\newif\ifhouches
\houchesfalse
\ifhouches
\documentclass{hch}
\else
\documentclass{article}
\fi
\usepackage{graphicx}
\newif\iffigs
\figstrue
\iffigs
\fi
\def\drawing #1 #2 #3 {
\begin{center}
\setlength{\unitlength}{1mm}
\begin{picture}(#1,#2)(0,0)
\put(0,0){\framebox(#1,#2){#3}}
\end{picture}
\end{center} }

\def\rf#1{(\ref{#1})}

\def\rset{{\rm I\kern -0.2em R}}
\def\un{\hbox{{1\kern -0.25em\raise 0.4ex\hbox{{\scriptsize $|$}}}}}

\def\cset{\hbox{{C\kern -0.55em\raise 0.5ex\hbox{{\tiny $|$}}}}}
\def\nset{\hbox{{I\kern -0.18em N}}}
\def\la{\left\langle}
\def\ra{\right\rangle} 

\def\v{{\bf v}}
\def\f{{\bf  f}}
\def\u{{\bf u}}
\def\x{{\bf x}}
\def\r{{\bf r}}

\def\a{{\bf a}}

\begin{document}
\ifhouches
\title{``Burgulence''}
\author{U.~Frisch \lowercase{and} J.~Bec}
\address{Laboratoire G.D. Cassini, UMR~6529\\
Observatoire 
de la C\^ote d'Azur \\ BP4229, 06304 Nice Cedex 4 \\ 
e-mail: uriel@obs-nice.fr}
\else
\title{\vspace*{-3.2cm}``Burgulence'' \\ \vspace{0.5cm}
\normalsize to appear in\\
 {\it Les Houches 2000: New Trends in Turbulence}\\
M.~Lesieur, ed., Springer EDP-Sciences}
\date{}
\author{U.~Frisch \lowercase{and} J.~Bec\\ \\
{\small Laboratoire G.D. Cassini, UMR~6529}\\
{\small Observatoire de la C\^ote d'Azur} \\
{\small BP4229, 06304 Nice Cedex 4} \\ 
{\small e-mail: uriel@obs-nice.fr}}
\fi
\maketitle

\ifhouches
\else
\begin{abstract}
This is a review 
of selected work on the
one- and multi-dimensional random Burgers equation (burgulence) with
emphasis on questions generally asked for incompressible
Navier--Stokes turbulence, such as the law of decay of the energy and
the pdf of velocity gradients.  Most of the material is devoted to
decaying (unforced) burgulence.  For more details see the Table of
Contents.
\end{abstract}
\tableofcontents
\fi

\section{Introduction}
\label{s:intro}

These lectures are about the $d$-dimensional Burgers equation
\begin{equation}
\partial_t \v + \left ( \v \cdot \nabla \right ) \v  = \nu \nabla^2 \v,
\;\;\;\;\; \v = -\nabla \psi.
\label{ddburg}
\end{equation}
Note that the constraint that $\v$ be derived from a (velocity)
potential $\psi$ is trivially satisfied if $d=1$.  The word
``burgulence'', as we use it here, is a contraction of
``Burgers'' and ``turbulence''.  It means ``the study of random
solutions to the Burgers equation''.  The randomness may arise because
random initial conditions $\v_0=-\nabla \psi_0$ are given or because a
random driving force $\f = -\nabla F$ is added to the r.h.s. of
(\ref{ddburg}), or both. When $\f = 0$ one speaks about ``decaying
burgulence''.

In the thirties when the Dutch scientist J.M.~Burgers introduced the 
equation in the one-dimensional case, he hoped to contribute to 
the study of turbulence with a simple model  which, obviously, 
has a lot in common with the Navier--Stokes equation:
\begin{itemize}
\item same type of advective nonlinearity 
\item presence of a diffusion term from which a Reynolds number may 
be defined
\item many invariance and conservation laws in common: invariance under
translations in space and time, parity invariance, conservation of momentum
and energy (only for $\nu=0$ and $d=1$).
\end{itemize}

Such hopes appeared to be shattered when, in the fifties, Hopf 
\cite{h50} and Cole \cite{c51} discovered -- some say 
rediscovered -- that the Burgers equation can actually be integrated 
explicitly (we shall return to this matter later). Indeed, an important 
property of the Navier--Stokes equation, not shared by the Burgers 
equation, is the sensitivity to small changes in the initial 
conditions in the presence of boundaries or driving forces and at 
sufficiently high Reynolds numbers. Hence, the Burgers equation is not 
a good model for one of the most important aspects of turbulence: the 
spontaneous arise of randomness by chaotic dynamics.  

In spite of this there has been a strong renewal of interest in the Burgers
equation, starting in the eighties, for a variety of reasons which we shall
now explain.  As a quantitative measure of the current interest,
Table~\ref{table_google_lanl} shows some web-based statistical figures on the
number of hits as of August 2000 (Google is an all-purpose search engine and
``Los Alamos'' stands for the nlin (ex-chao-dyn) preprint archive):
\begin{table}[h]
\centerline{\begin{tabular}{|l|c|c|} \hline
 &Navier--Stokes equation&Burgers equation\\ \hline
 Google&15000&4000\\ \hline
 Los Alamos&100&75 \\ \hline \end{tabular} } 
\caption{Web-based statistical data}
\label{table_google_lanl}
\end{table}
The Burgers equation, which obviously describes a compressible flow (in one
dimension there exist only trivial incompressible flows), has found many
applications in {\em nonlinear acoustics} and other nonlinear wave problems.
A review may be found in Ref.~\cite{gms91}.

\subsection{The Burgers equation in cosmology}
\label{zeladh}

The Burgers equation has found interesting applications in cosmology,
where it is known, in one instance, as the ``Zel'dovich
approximation'' \cite{z70} and, in another instance, as the ``adhesion
model'' \cite{gs84}. Here, we shall give a brief introduction to how
the Burgers equation arises in cosmology.  More details may be found
in Refs.~\cite{saf92,vdfn94,p93,cl95}.  Just after the baryon-photon
decoupling in the early Universe, there may have been a rarefied
medium formed by collisionless dustlike particles without pressure,
interacting only via Newtonian gravity \cite{p93}. The gravitational
potential is then determined from the fluctuations in mass density by
a Poisson equation. Limiting ourselves to the case of a single type of
matter, we can schematically write the acceleration of a fluid
particle as follows:
$$
\begin{array}{cccccccc}
\!\!\!\!\hbox{acceleration}= &\mbox{pressure } &+&\hbox{viscous} &+&
\hbox{expansion} &+ &\hbox{gravit.} \\
\!\!\!\! & \hbox{term}&& \hbox{term}&& \hbox{term}&& \hbox{term}\\[1.6ex]
\!\!\!\!\partial_{t}\v +\v\cdot\nabla \v&\hbox{negligible}&&?&&
\stackrel{\propto  \v\,\,\,{\rm in}}{\rm comov.\,\, coord.}&&
\end{array}
$$
On the left hand side (l.h.s.) we
recognize the familiar terms of the Burgers equation. The pressure is
usually neglected because the matter is very cold. We shall come back
to the viscous term later. The expansion term, proportional to the
velocity, arises  because the equation is written in a frame
comoving with the expansion of the Universe.

It turns out that when the problem of self-gravitating gas in an expanding
universe is examined in the linear approximation (small density fluctuations)
an instability is obtained in which the dominant mode has the following
properties \cite{p93,cl95}:
\begin{itemize}
\item it is potential ($\v =-\nabla \psi$),   
\item the expansion and gravitational terms cancel.
\end{itemize} 

\ifhouches
\begin{figure}[h]
\iffigs
\centerline{\includegraphics[width=5.5cm,angle=-90]{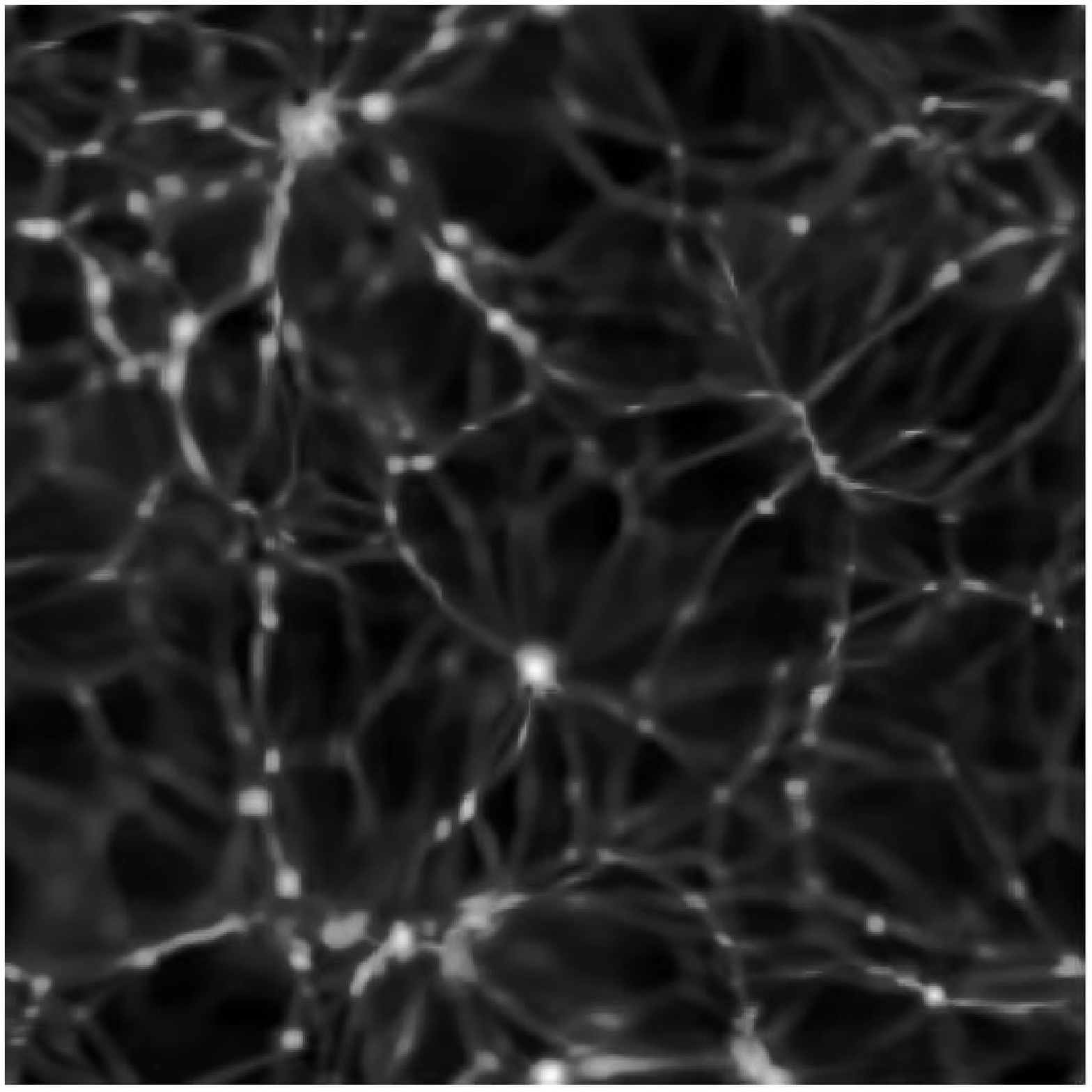}}
\else\drawing 100 10 {Simu N corps}
\fi
\caption{$N$-body simulation by the Virgo Consortium (see
\hbox{http://star-www.dur.ac.uk/$\sim$frazerp/virgo/virgo.html}).  The
simulation has $256^{3}$ particles and was done on two large Cray~T3D
parallel supercomputers at the computing centers Garching (D) and Edinburgh
(GB).  The brightness is proportional to the log of the density of the
particles.}
\label{f:virgo}
\end{figure}
\begin{figure}[h]
\iffigs
\centerline{\includegraphics[width=5.5cm]{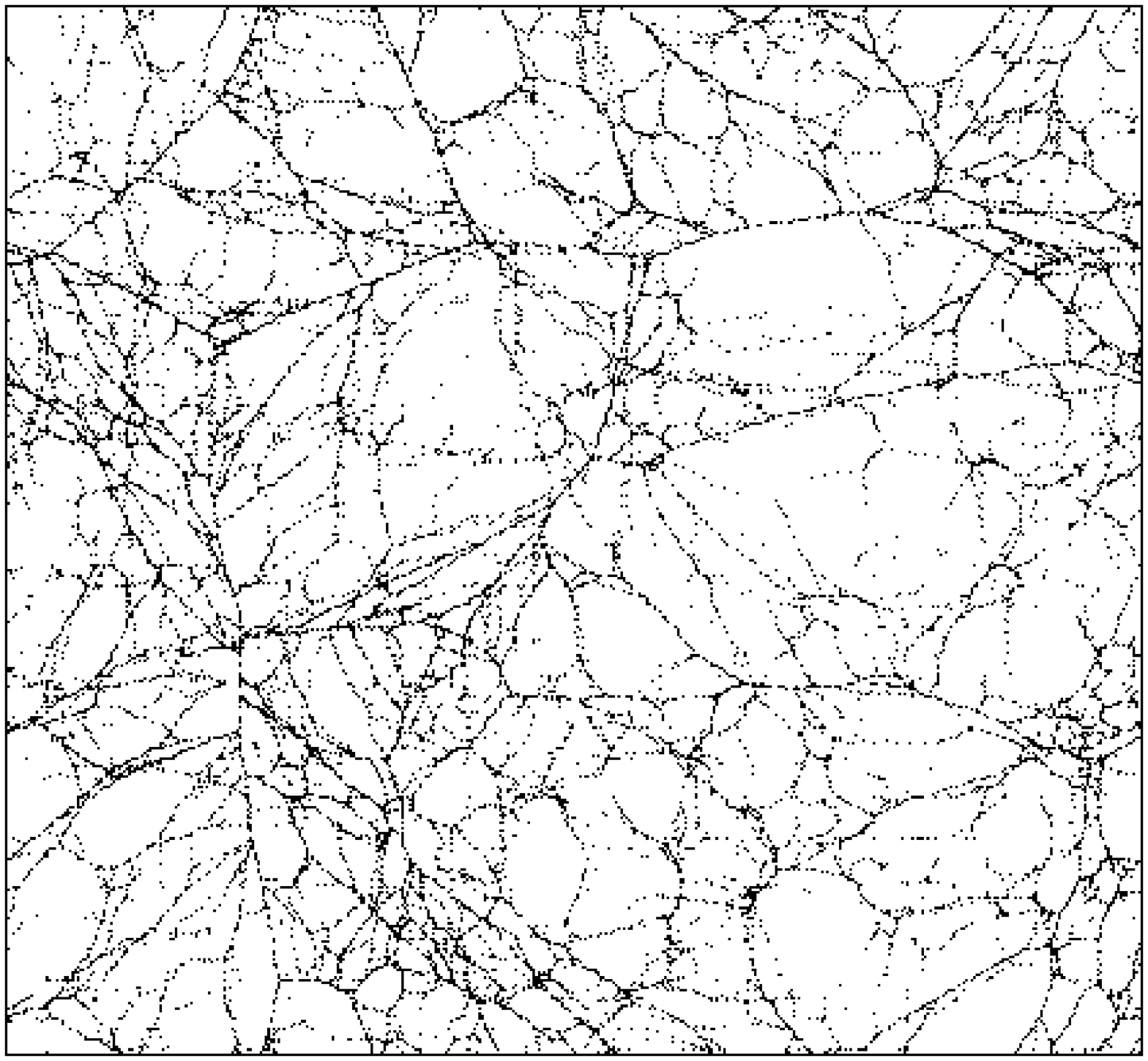}}
\else\drawing
  100 10 {Simu massimo 2d} \fi
\caption{Mass density field from a $512^2$ simulation of the Burgers equation
with random scale-free initial data (from Ref.~\cite{vdfn94}).}
\label{f:massimo}
\end{figure}
\fi

In 1970, Zel'dovich \cite{z70} proposed to extend these properties
into the nonlinear regime where density fluctuations become strong and
mass condensation forming large-scale structures appear.  Furthermore,
this ``Zel'dovich approximation'' is exact in one dimension,
irrespective of the strength of fluctuations. Clearly, in the
Zel'dovich approximation each fluid particle is just moving in a
straight line with constant velocity (after a suitable nonlinear
change of variable of the time). Just like a family of straight light
rays forms generally caustics along which the intensity is infinite,
the material particle lines form singular objects along which the
mass density is  infinite.  Arnold, Shandarin and Zel'dovich
\cite{asz82} studied the various kind of singularities which can form
in this way, to see if they could account for known large-scale
structures such as galaxies and clusters. Observations and numerical
simulations have now revealed that large-scale structures are much
simpler than the mathematical objects generated  in a caustic-type theory.

\ifhouches
\else 
\begin{figure}[p]
\iffigs
\centerline{\includegraphics[width=6.5cm,angle=-90]{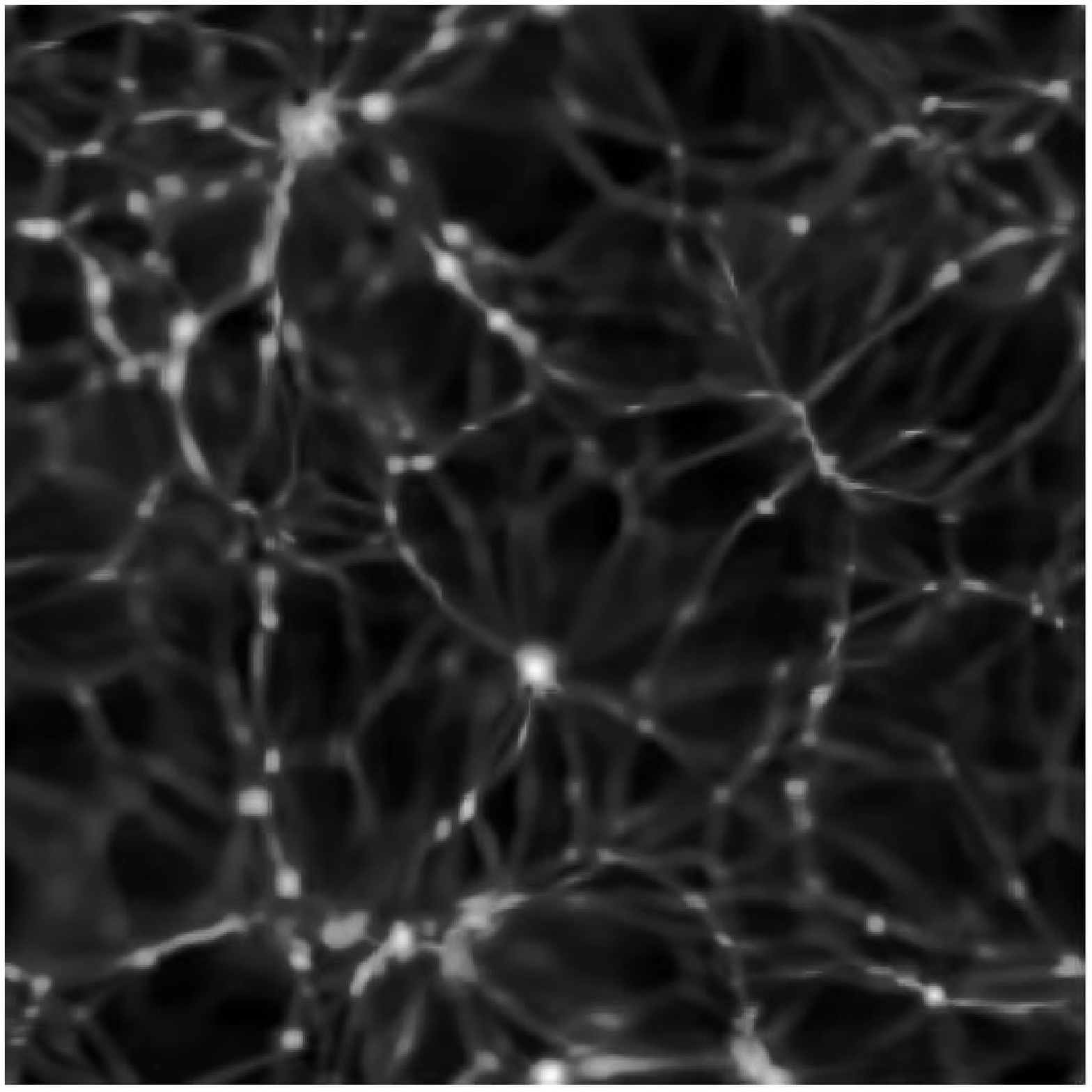}}
\else\drawing 100 10 {Simu N corps}
\fi
\caption{$N$-body simulation by the Virgo Consortium (see
\hbox{http://star-www.dur.ac.uk/$\sim$frazerp/virgo/virgo.html}).  The
simulation has $256^{3}$ particles and was done on two large Cray~T3D
parallel supercomputers at the computing centers Garching (D) and Edinburgh
(GB).  The brightness is proportional to the log of the density of the
particles.}
\label{f:virgo}
\end{figure}
\begin{figure}[p]
\iffigs 
\centerline{\includegraphics[width=6.5cm]{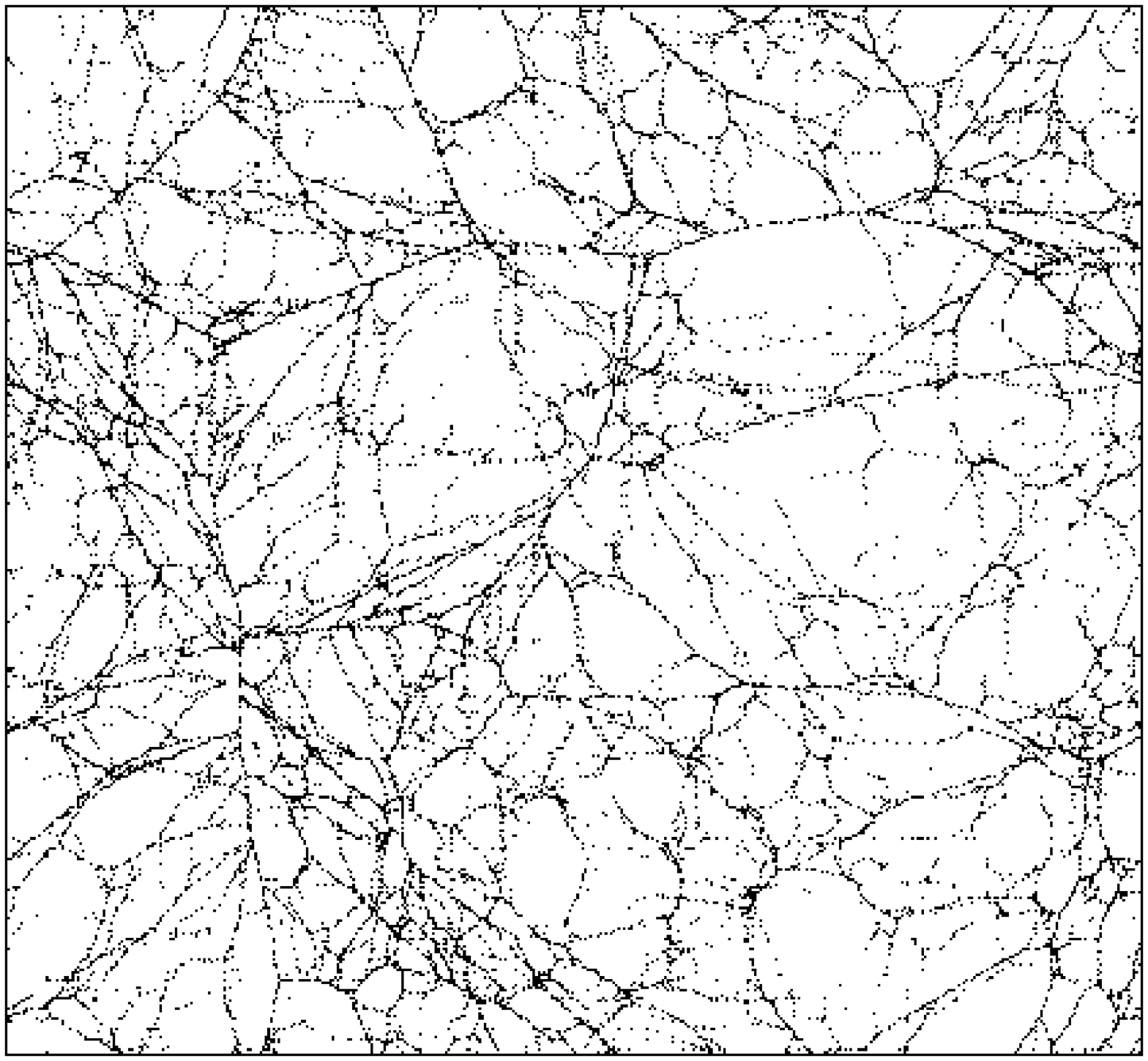}} 
\else\drawing
  100 10 {Simu massimo 2d} \fi
\caption{Mass density field from a two-dimensional 
$512^2$ simulation of the Burgers equation with random scale-free
initial data (from Ref.~\cite{vdfn94}).}
\label{f:massimo}
\end{figure}
\fi

Consider for example Fig.~\ref{f:virgo} which shows a thin slice of a
simulated universe using the so-called $\Lambda$CDM model (cold dark matter
with cosmological constant).  The line-like and node-like features on this
slice are actually sheets and filaments embedded in the three-dimensional
space. Sheets (walls), filaments and nodes (clusters) are the most common
structures observed in such simulations.  As was shown by Gurbatov and Saichev
\cite{gs84} these are precisely the structures obtained if one modifies the
Zel'dovich approximation by requiring that particles should not cross
but rather adhere. This {\em adhesion model\/} is just the
three-dimensional Burgers equation (\ref{ddburg}), taken in the limit
of vanishing viscosity.  Numerical experiments indicate that the
adhesion model reproduces quite well the early skeleton of large-scale
structures in N-body numerical simulations (see, for example, Figs.~6a
and 6b of Ref.~\cite{kbgn94}).  Since dark matter is essentially
collisionless, it is not clear at the moment what is the physics
behind this agreement which seems to require some viscosity-generating
mechanism to prevent -- or dramatically slow down -- particle
crossing. Furthermore, the adhesion model cannot cope with many
important aspects of gravitational dynamics. For example, in $N$-body
simulations, one frequently observes the collapse of a filament into
an isolated node (cluster). As we shall see, there is nothing of this
sort in Burgers dynamics.

\subsection{The Burgers  equation in condensed matter and statistical 
physics}
\label{condmat}
The Burgers equation arises in a number of condensed matter and statistical
physics problems and even in non-physics problems such as vehicular traffic
(for review see Ref.~\cite{css00}) A frequently studied problem is the
Kardar--Parisi--Zhang or KPZ equation \cite{kpz86} (see also
Ref.~\cite{bs95})
\begin{equation}
\partial_t \psi = {1 \over 2}\left| \nabla\psi \right|^2 + \nu
\nabla^2 \psi +F,
\label{KPZ}
\end{equation}
which appears in studying the motion of an interface under
deposition. Here, $\psi$ is the vertical displacement of the interface
as a function of $d-1$ horizontal coordinates and of the time. It is
immediately checked, by taking the horizontal gradient of (\ref{KPZ}),
that one obtains the Burgers equation (\ref{ddburg}) with an
additional forcing term ${\bf f}=-\nabla F$. Burgers equation also
arises in studying directed polymers (see, e.g.,
Refs.~\cite{kz87,bmp95}), but with the time variable now interpreted
as a space variable in the direction of main extension of the
polymers. On all these problems there is considerable literature which
it is not our purpose to review here.

\subsection{The Burgers equation as testing ground for Navier--Stokes}
\label{test}

The Burgers equation, because of its known solutions, is frequently
used for testing numerical schemes, particularly those intended for
compressible flow (many of the Google hits are of this kind). If one
is mostly interested in turbulence, as is the case for participants of
the present School, Burgers equation turns out to be quite useful for
testing -- and mostly discarding -- certain types of theories of
turbulence. Indeed, there have been many attempts to tackle the
problem of the statistical theory of turbulence by adapting to it
tools borrowed from field theory (for reviews, see
Refs.~\cite{o77,rs78,livreufcup95,l97}).  Such methods had little
impact on the field until recently when they have permitted a real
breakthrough in understanding the mechanism for intermittency and
anomalous scaling (see, e.g.  the lectures by G.~Falkovich {\it et
al.} in the same volume). In the past such field-theoretic methods
have frequently involved formal expansions in powers of the
nonlinearity, with Feynman graphs used for the bookkeeping of all the
terms generated after averaging over Gaussian initial conditions
and/or random forces.  Since the Burgers equation has the same type of
nonlinearity as the Navier--Stokes equation such methods are typically
also applicable to the Burgers equation. Hence it is possible to find
what they predict for the latter and to compare the results with those
obtained by more reliable methods. From this point of view, that is of
using the Burgers equation as testing ground, it is desirable to know
the answers to questions similar to those generally asked for
Navier--Stokes turbulence. For example, what are the scaling
properties of structure functions; what are the probability
distribution functions (pdf) of velocity increments and velocity
gradients?  Such questions will be at the center of these
lectures. Whenever possible we shall comment on the corresponding
Navier--Stoke issues. The emphasis will be exclusively on what happens
in the real space-time domain in the limit of vanishing viscosity,
which is of course not the same as naively putting the viscosity equal
to zero.  A number of interesting questions, requiring a finite
viscosity, such as the pole decomposition \cite{c75,fm81} will thus be
left out.

\section{Basic tools}
\label{s:tools}

In this section we introduce various analytical, geometrical and numerical
tools which are useful for constructing solutions to the decaying (unforced)
Burgers equation \rf{ddburg}. Mostly, we shall deal with the deterministic
equation, while making occasional comments on consequences for burgulence.

\subsection{The Hopf--Cole transformation and the maximum representation}
\label{subs:hopfcole}

If in \rf{KPZ} with $F=0$ we set $\psi = 2\nu\ln \theta$ we obtain the
$d$-dimensional heat equation \cite{h50,c51}
\begin{equation}
\partial_t \theta = \nu \nabla^2 \theta,
\label{heat}
\end{equation}
which can be solved explicitly if there are no boundaries. One thus obtains 
\begin{equation}
\psi(\r, t) = 2\nu \ln \left \{ {1 \over (4\pi\nu t)^{d/2}} \int_{\rset^d}
\exp \left [ {1 \over 2\nu} \left ( \psi_0 (\a) - {|\r-\a|^2 \over 2t}
\right ) \right ] d^da \right \},
\label{hopfcole}
\end{equation}
where $\psi_0 (\a)$ is the initial potential.
The limit of vanishing viscosity ($\nu\to 0$), obtained by steepest descent,
has the following ``maximum representation''
\begin{equation}
\psi(\r,t) = \max_{\a} \left (\psi_0 (\a) - {|\r-\a|^2 \over 2t}\right ).
\label{maximum}
\end{equation}
Note that the operation of taking a maximum is global in nature, whereas
the viscous Burgers equation is a local partial differential equation. 
If $\psi_0 (\a)$ is differentiable (i.e.\ the initial velocity $\u_0(\a)$
exists as an ordinary function rather than a distribution), the maximum in
\rf{maximum} will be achieved at one or several points $\a$ where the gradient
of the r.h.s. vanishes, that is, where 
\begin{equation}
\r = \a+t\v_0(\a)
\label{naive}
\end{equation}
In other words, $\r$ is the position at time $t$ of the fluid particle
starting at $\a$ and retaining its initial velocity $\v_0(\a)$. Hence,
we can interpret $\a$ and $\r$ as being, respectively, Lagrangian and Eulerian
coordinates. Along this Lagrangian trajectory, the velocity being
conserved, we have
\begin{equation}
\v(\r,t)=\v_0(\a).
\label{uuzero}
\end{equation}

The map $\a\mapsto \r$ defined by \rf{naive} is called the naive Lagrangian
map. It is not necessarily invertible: if there are several Lagrangian
locations satisfying \rf{naive} for a given $\r$ the only acceptable one
is that which maximizes the argument on the r.h.s. of \rf{maximum}. 
As long as the Jacobian of the naive Lagrangian map \rf{naive}
\begin{equation}
J(\a,t)= \det \left ( \delta_{ij} - t {\partial^2 \psi_0 \over
    \partial a_i \partial a_j } \right )
\label{jacobian}
\end{equation}
does not vanish the map is guaranteed to be invertible and the solution of
the Burgers equation cannot have a singularity.  For sufficiently
smooth initial data with bounded second derivatives of $\psi_0$ the
first singularity appears  at 
\begin{equation}
t_\star = {1\over \max_{\a}
\left [ \lambda(\a) \right ]},
\label{tstar}
\end{equation}
where $\lambda(\a)$ is the largest eigenvalue of the Hessian matrix 
$\partial^2   \psi_0 /\partial a_i \partial a_j$.

\begin{figure}[h]
\iffigs
\centerline{\includegraphics[width=3.5cm,angle=-90]{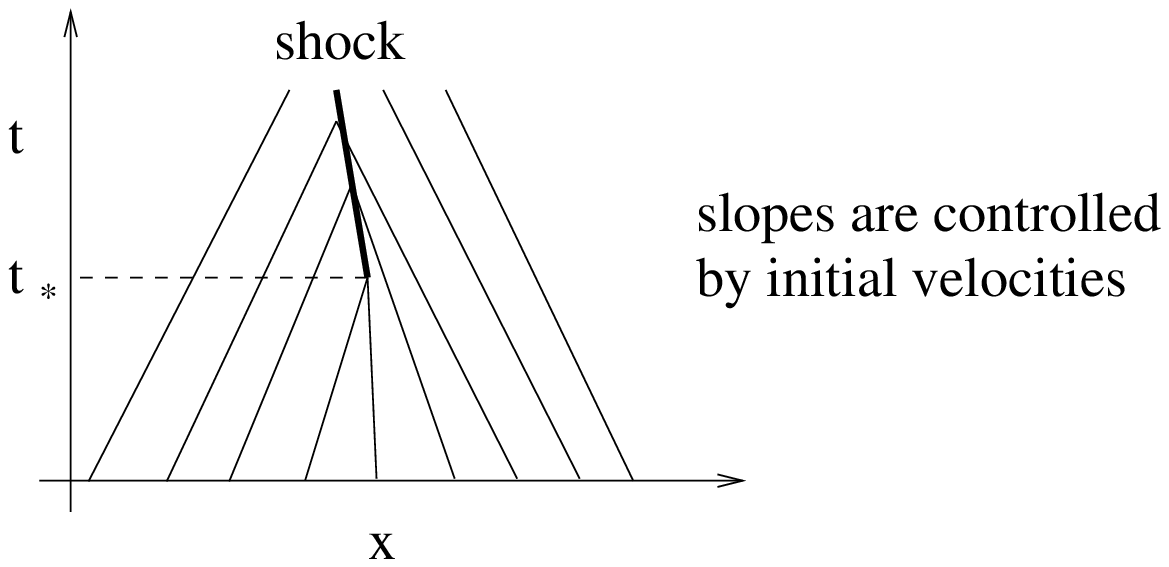}}
\fi
\caption{Characteristics for the unforced  one-dimensional Burgers equation in
the $(x,t)$ plane.}
\label{f:characteristics}
\end{figure}

In one dimension, we denote the velocity by $u$. Now, the time  $t_\star$ is the inverse of the absolute value of the
most negative initial velocity derivative $du_0(a)/da$. It is the first time
at which the characteristics $x=a+tu_0(a)$ of the hyperbolic inviscid Burgers 
equation are crossing (Fig.~\ref{f:characteristics}). 
The first singularity in one or more dimension, is known as a ``preshock''
\cite{ff83}  and plays an important role in the theory of pdf for velocity
gradients and densities (Sections~\ref{s:pdfgrad} and \ref{s:pdfdensity}).

Note that, for Gaussian random initial conditions, $t_\star$ is itself random
and can become arbitrarily small with very small but nonvanishing
probability. As a consequence, most averaged quantities (e.g. the two-point
correlation function) will have an essential singularity at $t=0$. Note also
that the distribution of eigenvalues of the Hessian matrix extends to infinite
values in any finite dimension $d$, but becomes compactly supported (on a
semi-circle) when dividing the eigenvalues by $\sqrt{d}$ and letting $d\to
\infty$. This is indeed a consequence of the properties of large random
symmetric matrices, called Wigner matrices (see, e.g., Ref.~\cite{fb70}).

\subsection{Shocks in one dimension}
\label{subs:shocks1d}

After the time $t_\star$ the Lagrangian map \rf{naive} ceases to be
invertible. More precisely, for a given Eulerian position $\r$ there is more
than one Lagrangian position $\a$ satisfying \rf{naive}. This implies the
presence of shocks in the Eulerian velocity field. In this subsection we
consider the one-dimensional case and give various geometrical
constructions of the solution (including shocks). 

\begin{figure}[h]
\iffigs
\centerline{\includegraphics[width=4.5cm]{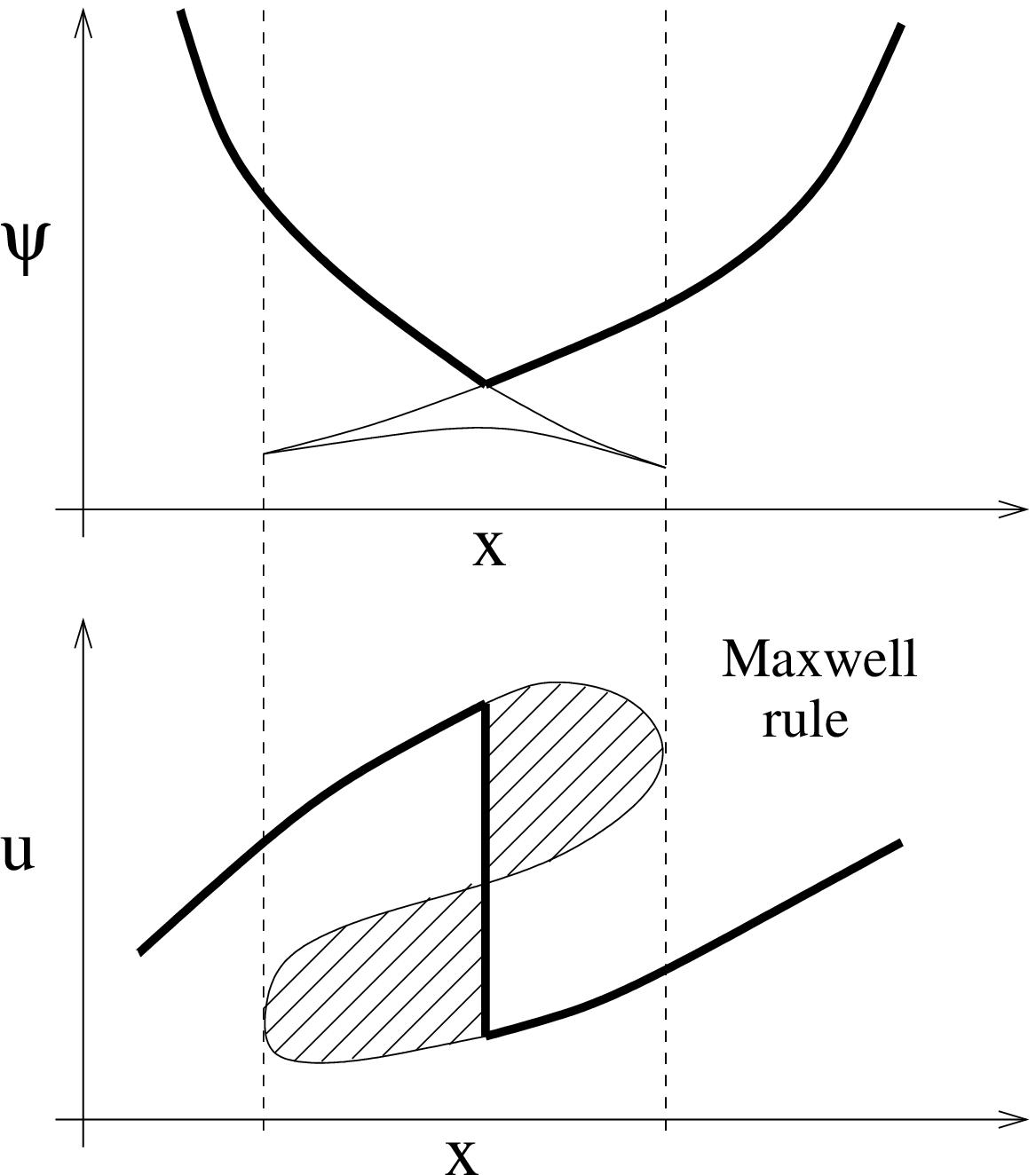}}

\fi
\caption{Lagrangian manifolds  for the potential in the $(x,\psi)$-plane
(upper) and the velocity in the $(x,u)$-plane (lower). The heavy lines
correspond to the correct Eulerian solutions. The vertical dashed lines
delineate the multivalued region.}
\label{f:mutivalue}
\end{figure}

First, let us define in the $(x,\psi)$-plane the Lagrangian manifold (a curve
in one dimension)
\begin{eqnarray}
x &\equiv& a+tu_0(a) \label{naiverep}\\
\psi  &\equiv& \psi_0(a)-{t\over 2} u_0^2(a),
\label{psilag}
\end{eqnarray}
where the second line is just the r.h.s.\ of \rf{maximum} without the maximum,
evaluated at the (naive) Eulerian position $a+tu_0(a)$. Fig.~\ref{f:mutivalue}
(upper) shows this Lagrangian manifold after the time $t_\star$. Hence, above
some Eulerian locations $x$ there is more than one branch and cusps are
present at Eulerian locations such that the number of branches changes.
Clearly, the correct Eulerian potential is obtained by taking the maximum,
i.e.\ always the highest branch. Note that this potential will have one or
several points with discontinuous slope, the right derivative being always
greater than the left one. Hence the velocity, which is the negative space
derivative of the potential (shown in the lower part of
Fig.~\ref{f:mutivalue}) will have discontinuities at shock locations with $u_-
>u_+$. It is also possible to directly construct the velocity starting
from the Lagrangian manifold in the $(x,u)$-plane
\begin{eqnarray}
x &\equiv& a+tu_0(a) \label{naivereprep}\\
u  &\equiv& u_0(a).
\label{ulag}
\end{eqnarray}
If there is a single shock present, it follows obviously that its position is
determined by a Maxwell rule: the hashed loops shown on Fig.~\ref{f:mutivalue}
(lower part) right and left of the shock should have equal areas. A Maxwell
rule construction can become very cumbersome if there are several shocks
present.  

\begin{figure}[h]
\iffigs
\centerline{\includegraphics[width=4cm,angle=-90]{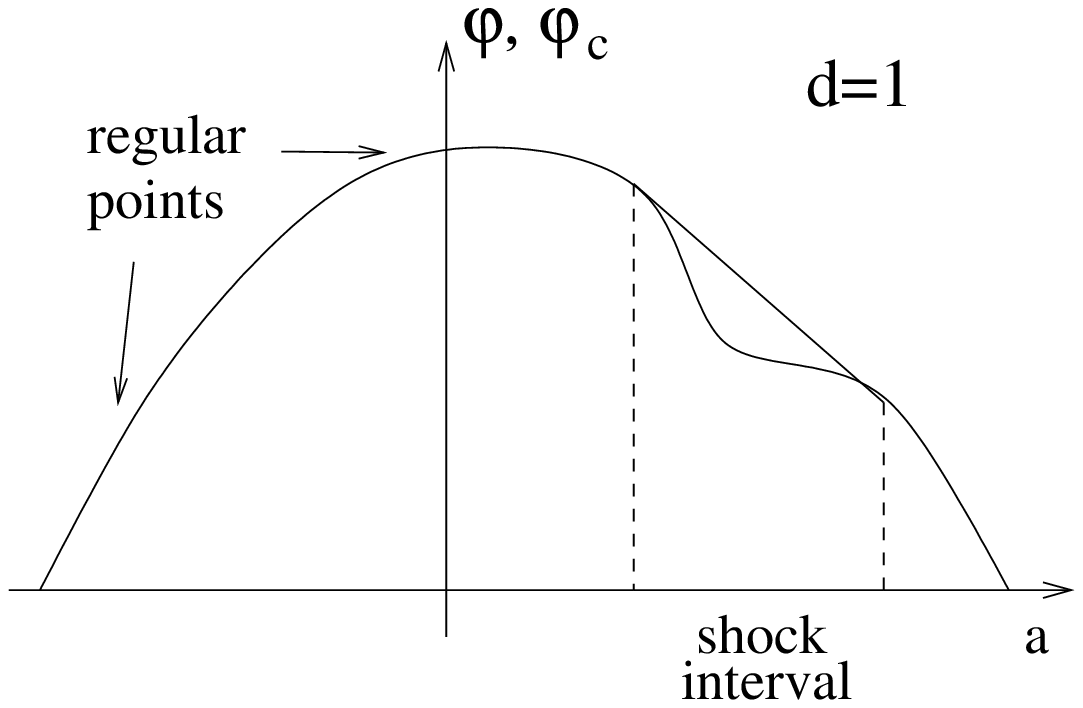}}
\fi
\caption{Convex hull construction in terms of the Lagrangian potential.}
\label{f:convexhull1d}
\end{figure}

\begin{figure}[h]
\iffigs
\centerline{\includegraphics[width=4cm,angle=-90]{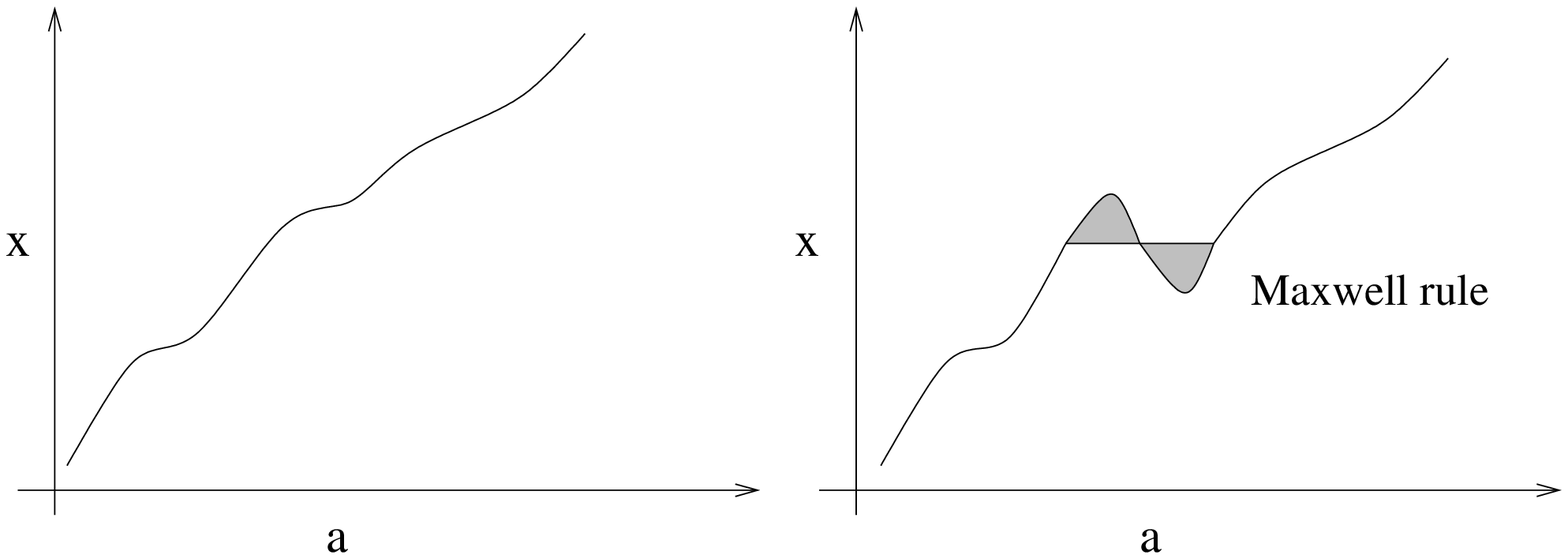}}
\fi
\caption{Naive Lagrangian map before (left) and after (right) appearance of
a shock. The correct Lagrangian map is obtained by a Maxwell rule.}
\label{f:lagmap}
\end{figure}

Another geometrical construction uses the {\em Lagrangian potential}
\begin{equation}
\varphi(a,t)\equiv t\psi_0(a) - {a^2 \over 2},
\label{deflagpot1d}
\end{equation}
whose negative gradient is obviously the naive Lagrangian map.
We can rewrite \rf{maximum} as
\begin{equation}
t\psi(x,t)+{x^2\over 2} = \max_{a}\left[ \varphi(a,t)+ax
\right ],
\label{legendre1d}
\end{equation}
which represents the potential as, basically, a Legendre transform of
the Lagrangian potential. (Note that the Legendre transformation is
also used in the theory of multifractals.) The r.h.s. of
\rf{legendre1d} is equivalent to finding the largest algebraic
vertical distance between the graph of the Lagrangian potential and
the line of slope $-x$ through the origin. If the graph is convex
(second derivative negative everywhere), the maximum is attained at
the unique point where the derivative has the value $-x$.  Otherwise,
it suffices to replace the graph of $\varphi$ by its convex hull
$\varphi_c$, that is the intersection of all half-planes containing
the graph. This is illustrated in Fig.~\ref{f:convexhull1d}, which
shows both regular points (Lagrangian points which have not fallen
into a shock) and one shock interval, situated below the segment which
is part of the convex hull. Again, it is possible to work directly
with the (negative) derivative of the Lagrangian potential, namely,
the naive Lagrangian map.
The convex hull construction becomes then 
a Maxwell rule as shown in Fig.~\ref{f:lagmap}. From this one can easily show 
that the speed of a shock is the half-sum
of the velocities immediately to the right and to the left.

\begin{figure}[h]
\iffigs
\centerline{\includegraphics[width=4.5cm,angle=-90]{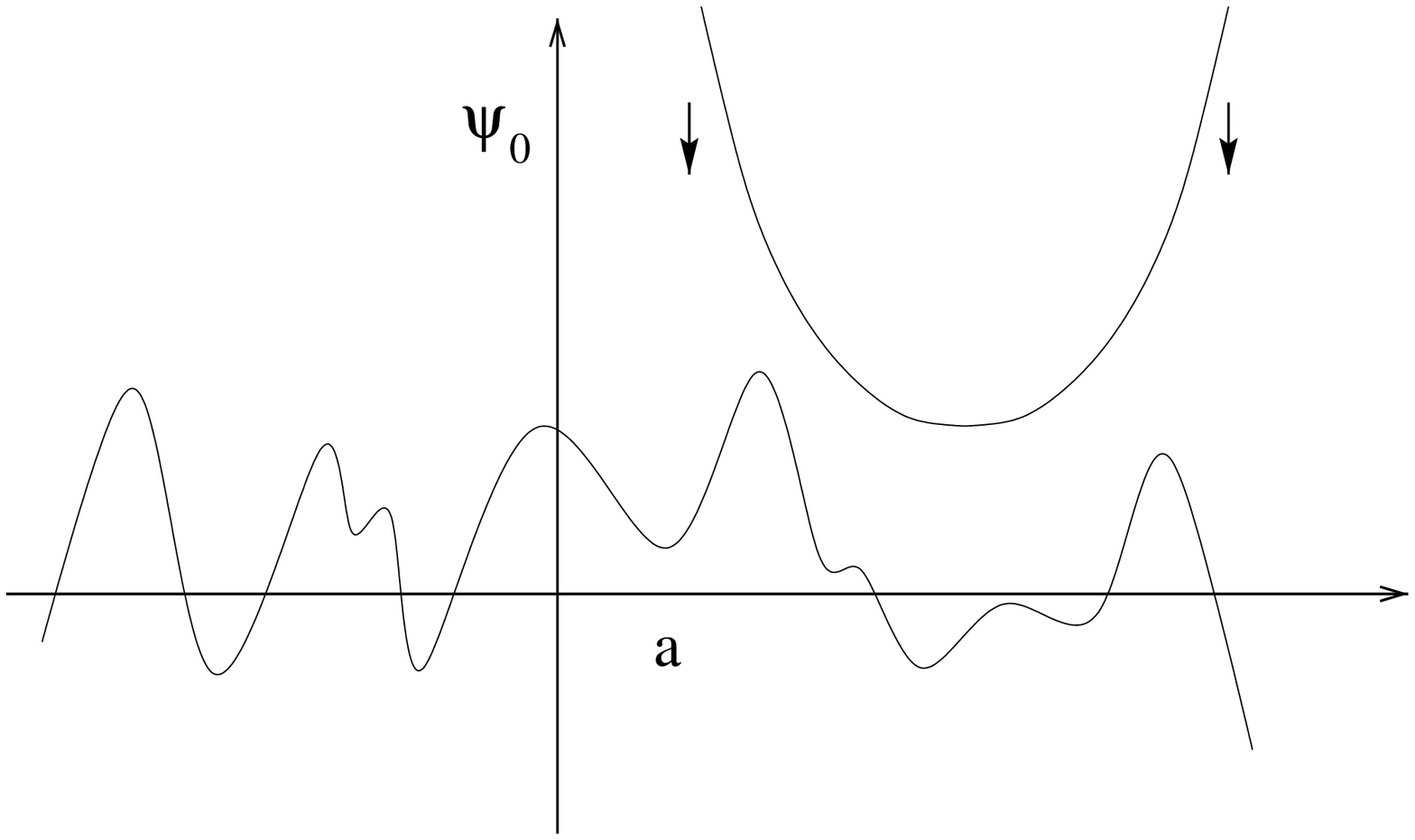}}
\fi
\caption{Parabola construction of the solution.}
\label{f:parabola}
\end{figure}

Finally, the maximum formula \rf{maximum} yields directly a 
``parabola construction'', illustrated in Fig.~\ref{f:parabola}: a parabola
with apex at $x$ and radius of curvature proportional to $t$ is moved down
until it touches the initial potential $\psi_0(a)$ at the Lagrangian location
associated to $x$ (or at two such locations if there is a shock).
Which of the five geometrical methods given is more convenient depends
on the application considered. The parabola construction is best for
understanding evolution in time (cf Section~\ref{s:kida}).  It may be
used, for example, to show that the long-time Eulerian solution has a
sawtooth structure with shocks separated by ramps of slope $1/t$ (see
Fig.~\ref{f:ramps}). The ramps are associated to high local maxima in
the potential $\psi_0$.

\begin{figure}[h]
\iffigs
\centerline{\includegraphics[width=2.5cm,angle=-90]{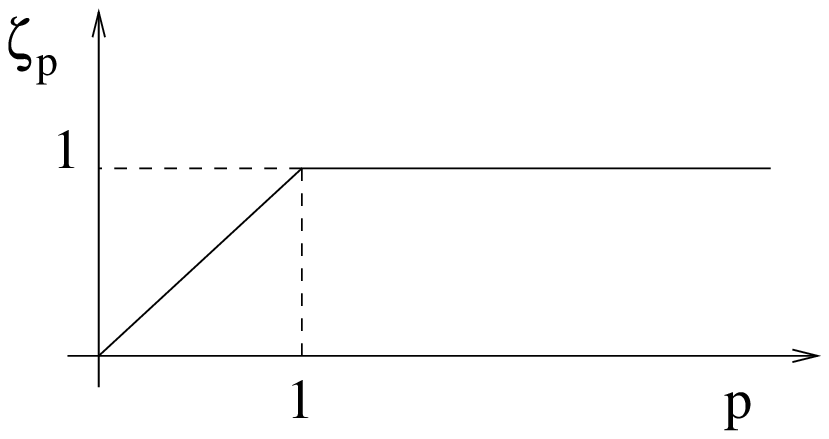}}
\fi
\caption{Exponent of the structure function of order $p$. Note the 
``phase transition'' at $p=1$.}
\label{f:zetap1d}
\end{figure}

With random and homogeneous initial conditions there will be shocks
(discontinuities) at random Eulerian locations which do not cluster
(unless we use non-smooth initial conditions as in Section~\ref{s:sinai}).
From this it is easily inferred that, for $p>0$, the structure functions
\begin{equation}
S_p(\Delta x,t) \equiv\la\left |u(x+\Delta x,t)-u(x,t)\right |^p\ra
\label{defstructfunc}
\end{equation}
behave, for small $\Delta x$, as
\begin{equation}
S_p(\Delta x,t) \sim C_p |\Delta x|^p +C_p' |\Delta x|, 
\label{smalldeltax}
\end{equation}
where the first term comes from regular (smooth) parts of the 
Eulerian velocity, while the second comes from the $O(|\Delta x|)$ probability
to have a shock somewhere in an interval of Eulerian length $|\Delta x|$.
For $0<p< 1$ the first term dominates as $\Delta x \to 0$, while, for 
$p>1$, it is the second. Hence, $S_p \sim |\Delta x|^{\zeta_p}$, with the
exponents $\zeta_p$ as shown in Fig.~\ref{f:zetap1d}. There are also
higher-order corrections to the simple scaling law given in \rf{smalldeltax}
which cannot be obtained by such simple arguments \cite{bfk00}. Note that a
second-order structure functions with a behavior $\propto |\Delta x|$ at small
distances implies an energy spectrum $E(k)\propto k^{-2}$ as $k \to \infty$.

The ``phase transition'' at $p=1$ seen in Fig.~\ref{f:zetap1d} is due
to the isolated character of the dissipative structures (the shocks), a
feature not present in incompressible three-dimensional
Navier--Stokes turbulence.

\subsection{Convex hull construction in more than one dimension}
\label{subs:convexhull}

Some of the methods used for the one-dimensional case are readily extended to
dimensions $d>1$, for example the construction from the Lagrangian manifold in
the $(d+1)$ dimensional space $(\x,\psi)$.  In Section~\ref{s:pdfdensity} we
shall use the multidimensional generalization of the convex hull construction,
which we now briefly outline. We define the Lagrangian potential
\begin{equation}
\varphi(\a,t)\equiv -{|\a|^2\over2}+t\psi_0(\a).
\label{lagpotdd}
\end{equation}
and find, from \rf{maximum}, that
\begin{equation}
t\psi(\r,t)+{|\r|^2\over2}
=\max_\a\left[\varphi(\a,t)+\r\cdot\a\right]. 
\label{hlegendredd}
\end{equation}
As before, this involves a (multidimensional) Legendre transformation which
leads us to the construction of the  convex hull in a $(d+1)$
dimensional space of the graph of the Lagrangian potential. In more than one
dimension, singularities of convex hulls are considerably more involved.
As a consequence, the equivalent of shocks are discontinuities across
$(d-1)$ manifolds, but there are many other singularities of higher codimension
(the codimension is $d$ minus the dimension of the object).

\begin{figure}[h]
\iffigs
\centerline{\includegraphics[width=8cm]{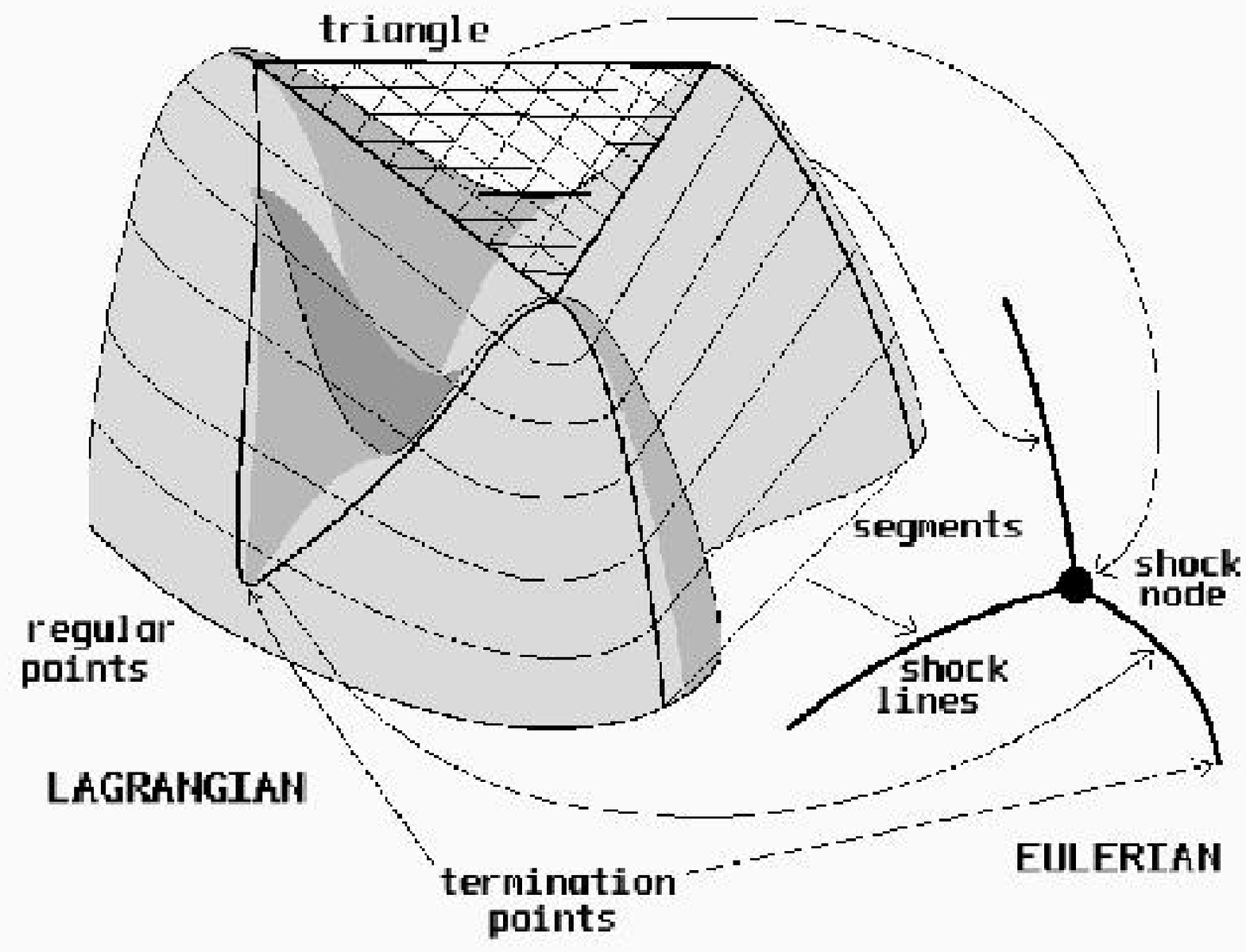}}
\fi
\caption{Construction of the convex hull for a two-dimensional Lagrangian
potential and associated Eulerian picture (Figure adapted from
Ref.~\cite{vdfn94}).}
\label{f:convex-hull-2d}
\end{figure}

In two dimensions the convex hull consists generically of four kinds of
objects: (i) parts of the original graph, (ii) pieces of ruled surfaces, (iii)
``kurtoparabolic points'', to which we shall come back, and (iv) triangles
(see Fig.~\ref{f:convex-hull-2d}). The associated Eulerian objects are,
respectively, (i) regular points, (ii) shock lines, (iii) end points of shocks
and (iv) shock nodes. Likewise, in three dimensions we have two-dimensional
shock surfaces meeting in triples at shock lines, meeting in quadruples at
shock nodes. (Nodes are always connected to shock lines and never isolated.)
Note that the Eulerian part of Fig.~\ref{f:convex-hull-2d} looks just like a
thermodynamic phase diagram, with the three shock lines playing the role of
the liquid-gas, liquid-solid and solid-gas transition lines, the node playing
the role of the triple point and the end point the role of the critical
point. This is not accidental. In thermodynamics, equilibrium states are
obtained by minimizing the Gibbs potential. This is equivalent to taking a
Legendre transform of the internal energy in which the pressure and the
temperature play the role of the Eulerian coordinates \cite{ll80}.  This
analogy holds also in higher dimensions: the classification of ``Legendrian
singularities'' can be used both for studying the Burgers equation \cite{b99}
and for studying multi-variable phase transitions \cite{a00}.

A more complete description of singularities is obtained by considering
the metamorphoses  of singularities as
time elapses. A complete classification in two and three dimensions may 
be found in the appendix (supplement 2) by V.I.~Arnold, Yu.M.~Baryshnikov
and I.A.~Bogaevski of Ref.~\cite{gms91}. 

For random initial conditions the structure functions scale the same way as 
in one dimension. For example, the probability of having a $(d-1)$-dimensional 
shock intercepting an Eulerian segment of small length $r$ is $O(r)$.
The higher codimension structures give only subdominant corrections.

\subsection{Remarks on numerical methods}
\label{subs:remnum}

Here, we give just some indications on how the (decaying) Burgers equation can
be solved numerically. (For more details, see Refs.~\cite{vdfn94,nv94,bfk00}
or any textbook on numerical methods for nonlinear hyperbolic equations; for
the case with forcing, see Refs.~\cite{kz87,bfk00} and references therein.)
First, one can of course, solve the Burgers equation with viscosity. This
should be avoided unless one is interested in what happens at dissipative
scales (e.g., inside shocks). For the inviscid limit and only in the decaying
case it is possible to construct the solution at time $t$ directly from the
initial condition without recourse to any time marching. One way is to
directly use the maximum representation \rf{maximum}, assuming that Lagrangian
and Eulerian locations have been discretized on the same grid. Then, for a
given Eulerian $\r$ one searches the Lagrangian $\a$ which maximizes the
r.h.s. If there are $N$ grid points, this seems to require $O(N^2)$
operations, but it can actually be done in $O(N\log_2 N)$ operations
\cite{vdfn94,nv94}.  Such a strategy must be combined with suitable
interpolations to increase accuracy and avoid getting complete garbage for
derivatives \cite{bfk00}. In one dimension one can also use Lagrangian
strategies with particle and/or shock tracking. To be consistent with the
inviscid limit, the particles must stick upon collisions.
for sufficiently smooth initial data, it may be possible to construct the
solution from the Lagrangian manifold \rf{naiverep}-\rf{psilag} or its
multidimensional generalization, by just searching the maximum, for a given
$\x$, of the finitely many branches present.
 
\section{The Fourier--Lagrange representation and artefacts}
\label{s:fournierfrisch}

In this section we show that formal manipulations of the inviscid Burgers
equation with random initial conditions, even though they include apparently
terms of all orders, can nevertheless lead to completely incorrect results,
e.g. for the energy spectrum. This section is entirely based on work 
by Fournier and Frisch \cite{ff83}. The theory is given in
one dimension but similar results can be established in higher dimensions.

In one dimension, it follows from \rf{naivereprep}-\rf{ulag}, that  the
Eulerian solution to the initial value problem for the decaying Burgers 
has the following implicit representation:
\begin{eqnarray}
u(x,t) = u_0(a) \nonumber\\
x = a+tu_0(a). 
\label{implicit}
\end{eqnarray}
This becomes explicit if, instead of working with $u(x,t)$, we use its
spatial Fourier transform ($2\pi$-periodicity is assumed for convenience)
\begin{equation}
\hat u(k,t) \equiv {1\over 2\pi} \int_0^{ 2\pi} {\rm e}^{-ikx}u(x,t)\,dx
\label{defuchap}
\end{equation}
and make the change of variables $x\mapsto a$, to obtain
\begin{equation}
\hat u(k,t) = {1\over 2\pi} \int_0^{2\pi} {\rm e}^{-ikx(a,t)}\,u_0(a){\partial
  x \over \partial a} \,da, \quad x(a,t) \equiv a+tu_0(a).
\label{explicit}
\end{equation}
Eq.~\rf{explicit} is called the {\em Fourier--Lagrangian\/} representation.
A first integration by parts yields
\begin{equation}
\hat u(k,t) = {1\over 2\pi} {1\over ik} \int_0^{2\pi} {\rm
  e}^{-ik\left(a+tu_0(a)\right)}\,u_0'(a)\,da.
\label{explicitfirst}
\end{equation}
A second integration by parts leads then to  
\begin{equation}
\hat u(k,t) = {1\over 2\pi} {1\over ikt} \int_0^{2\pi} {\rm
  e}^{-ik\left(a+tu_0(a)\right)}\,da, \quad k\neq 0.
\label{explicitsecond}
\end{equation}
If we now take random homogeneous Gaussian initial conditions, we can easily
calculate moments of $\hat u(k,t)$ because they just involve averages 
of exponentials having the Gaussian initial velocity in their arguments.
For example, the energy spectrum, related to the correlation function by
\begin{equation}
\la\hat u(k,t)\hat u(k',t)\ra = E(k,t) \delta_{k,k'}, 
\label{defspectrum}
\end{equation}
where $\delta_{k,k'}$ is a Kronecker delta, has the following expression
\begin{equation}
E(k,t) = {1\over 2\pi} {1\over k^2t^2} \int_0^{2\pi} {\rm e}^{-ikh} {\rm e}^{-{1\over
      2}k^2 t^2 S_2(h,0)}\, dh,
\label{ekfoncs2}
\end{equation}
where $S_2(h,0) \equiv \la \left [ u_0(h) - u_0(0) \right ]^2 \ra$ is the
second-order structure function of the initial velocity field.  If the latter
is smooth, as we shall assume, we have $S_2(h,0)\propto h^2$ for $h \to 0
$. It then follows by a simple Laplace-type asymptotic expansion of
\rf{ekfoncs2} that
\begin{equation}
E(k,t) \propto k^{-3} \,\,\,\mbox{ when } \,\,k\to\infty.
\label{kminus3}
\end{equation}
This is obviously the wrong answer: for Gaussian initial conditions
there will be shocks with a non-vanishing probability for any
$t>0$. Their signature is a $k^{-2}$ law in the energy spectrum at
high wavenumbers, as shown in Section~\ref{subs:shocks1d}.

\begin{figure}[h]
\iffigs
\centerline{\includegraphics[width=5cm]{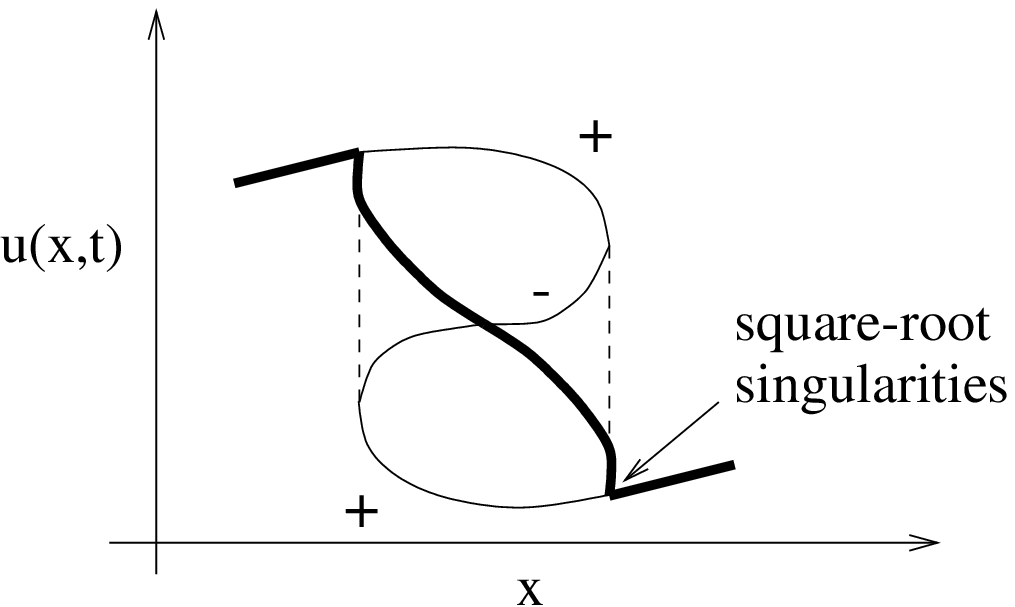}}
\fi
\caption{Spurious solution of Burgers equation when three branches
of a multi-valued solution are combined into one.}
\label{f:3branches}
\end{figure}

What went wrong? After the appearance of the first shock the Lagrangian
map $a\mapsto x$ is not monotonic and the change of variable from
\rf{defuchap} to \rf{explicit} is valid only outside of the Lagrangian
shock interval. Hence, in \rf{explicit} we should excise this interval
from the domain of integration. If we do not remove it, we are actually
calculating the Fourier transform of a function obtained by superposing
the threee branches shown in Fig.~\ref{f:3branches} with a plus sign
for the two direct branches and a minus sign for the retrograde branch
(the sign comes from the lack of an absolute value on the Jacobian
$\partial x/\partial a$ in \rf{explicit}). Obviously, this superposition
has two square-root cusps as shown in Fig.~\ref{f:3branches}. This produces
$k^{-3/2}$ tails in the Fourier transform and, hence, explains the
spurious $k^{-3}$ energy spectrum. Note also that this superposition
of three branches is not a solution to the Burgers equation, the latter being
nonlinear. This phenomenon is not related to the well known non-uniqueness
of the solution to the Burgers equation with zero viscosity without proper
additional conditions \cite{lax}.

The problem is actually worse than suggested so far. It is easily shown that
if the the initial velocity is deterministic and smooth, the function of the
time defined by \rf{explicit}, for {\em fixed\/} wavenumber $k$, is entire,
that is, its Taylor series around $t=0$ has an infinite radius of convergence.
There is no way to see the time $t_\star$ of the first preshock from this
function. A preshock is indeed an ``ultraviolet'' singularity which
is not seen in the temporal behavior of a single spatial Fourier
component. This result has an important consequence for the case of random 
Gaussian initial conditions. Suppose we simply ignore the viscosity in the
Burgers equation and expand the solution to all orders
in a temporal Taylor series
around $t=0$ and then calculate various correlation functions
and use Feynman graphs for bookkeeping of all the terms generated from
averaging. We then find that the whole set can be resummed exactly
and gives a spectrum with a $k^{-3}$ tail. Of course, the origin
of the ``resummation miracle'' is the Fourier--Lagrangian representation.

\section{The law of energy decay}
\label{s:kida}

An important issue in burgulence and turbulence is that of the law of
decay at long times when the viscosity is very small. Before turning
to the Burgers equation let us recall a few things about the
Navier--Stokes case.  It is generally believed that high-Reynolds
number turbulence has universal and non-trivial small-scale
properties. In contrast, large scales, important for practical
applications such as transport of heat or pollutants, are believed to
be non-universal. This is however so only for the toy model of
turbulence maintained by prescribed large-scale random forces.  Very
high-Reynolds number turbulence, decaying away from its production
source, and far from boundaries can relax under its internal nonlinear
dynamics to a (self-similarly evolving) state with universal and
non-trivial statistical properties {\it at all scales}.  K\'arm\'an
and Howarth \cite{kh38}, investigating the decay of high-Reynolds
number, homogeneous isotropic three-dimensional turbulence, proposed a
self-preservation (self-similarity) ansatz for the spatial correlation
function of the velocity: the correlation function keeps a fixed
functional shape; the integral scale $L(t)$, characteristic of the
energy-carrying eddies, grows in time and the mean kinetic energy
$E(t)=u^2(t)$ decays, both following power laws; there are two
exponents which can be related by the condition that the energy
dissipation per unit mass $|\dot E(t)|$ should be proportional to
$u^3/L$.  But {\em an additional relation\/} is needed to actually
determine the exponents.  The invariance in time of the energy
spectrum at low wavenumbers, known as the ``permanence of large
eddies'' \cite{livreufcup95,l97,burgatov} can be used to derive the
law of self-similar decay when the initial spectrum $E_0(k)\propto
k^n$ at small wavenumbers $k$, with $n$ below a critical value equal to
3 or 4, the actual value being disputed because of the ``Gurbatov
phenomenon'' (see the end of this section). One then obtains a law of
decay $E(t) \propto t^{-2(n+1)/(3+n)}$. (Kolmogorov \cite{k41decay}
proposed a law of energy decay $u^2(t) \propto t^{-10/7}$, which
corresponds to $n=4$ and used in its derivation the so-called
``Loitsyansky invariant'', a quantity actually not conserved, as shown
by Proudman and Reid \cite{pr54}.) When the initial energy spectrum at
low wavenumbers goes to zero too quickly, the permanence of large
eddies cannot be used, because the energy gets backscattered to low
wavenumbers by nonlinear interactions. For Navier--Stokes turbulence
the true law of decay is then known only within the framework of
closure theories (see, e.g., Ref.~\cite{l97}).

\begin{figure}[h]
\iffigs
\centerline{\includegraphics[width=3.5cm]{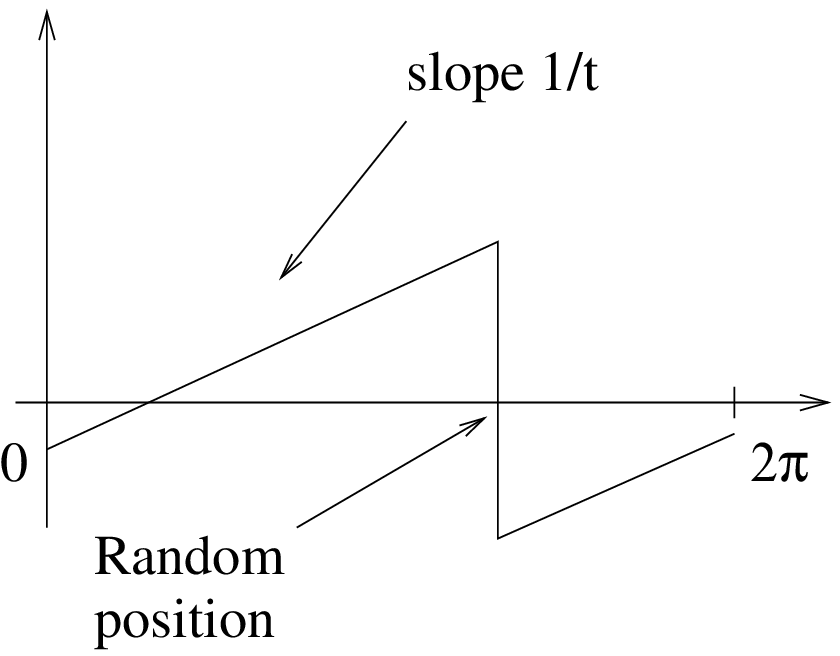}}
\fi
\caption{Snapshot of solution of decaying burgulence at long times when
spatial periodicity is assumed.}
\label{f:declinper}
\end{figure}

For one-dimensional burgulence, many of these questions are completely
settled. First, we observe that the problem of decay is quite simple if a
finite spatial periodicity is assumed. Indeed, eventually, all the shocks
produced will merge into a single shock per period, as shown in
Fig.~\ref{f:declinper}.  The position of the shock is random and the two ramps
have slope $1/t$, as is easily shown using the parabola construction of
Section~\ref{subs:shocks1d}. Hence, the law of decay is simply $E(t) \propto
t^{-2}$.
Nontrivial laws of decay are obtained if the burgulence is homogeneous
in an unbounded domain and has the ``mixing'' property (which means,
roughly, that correlations are decreasing with separation). The number
of shocks is then typically infinite but their density per unit length
is finite and decreases in time because shocks are constantly
merging. The $E(t) \propto t^{-2(n+1)/(3+n)}$ law mentioned above can
be derived for burgulence from the permanence of large eddies when
$n\le 1$ \cite{burgatov}. For $n=0$, this $t^{-2/3}$ law was actually
derived by Burgers himself \cite{livreburgers}.

The hardest problem is again when permanence of large eddies does not
determine the outcome, namely for $n>1$. This problem was solved by Kida
\cite{k79} (see also Refs.~\cite{gms91,ff83,burgatov}).

We now give some key ideas regarding the derivation of Kida's law of energy
decay. We assume Gaussian, homogeneous smooth initial conditions, such that the
potential is homogeneous. Since a homogeneous function is not, in general, the
derivative of another homogeneous function, we assume that the initial energy
spectrum
\begin{equation}
E_0(k) \propto k^n,\quad n>1; \quad k \to 0.
\label{ezeronpg1}
\end{equation}
This condition implies that the  mean square initial potential $\int k^{-2}
E_0(k)\,dk$ has no infrared (small-$k$) divergence (the absence 
of an ultraviolet divergence is guaranteed by the assumed smoothness). 

\begin{figure}[h]
\iffigs
\centerline{\includegraphics[width=4cm,angle=-90]{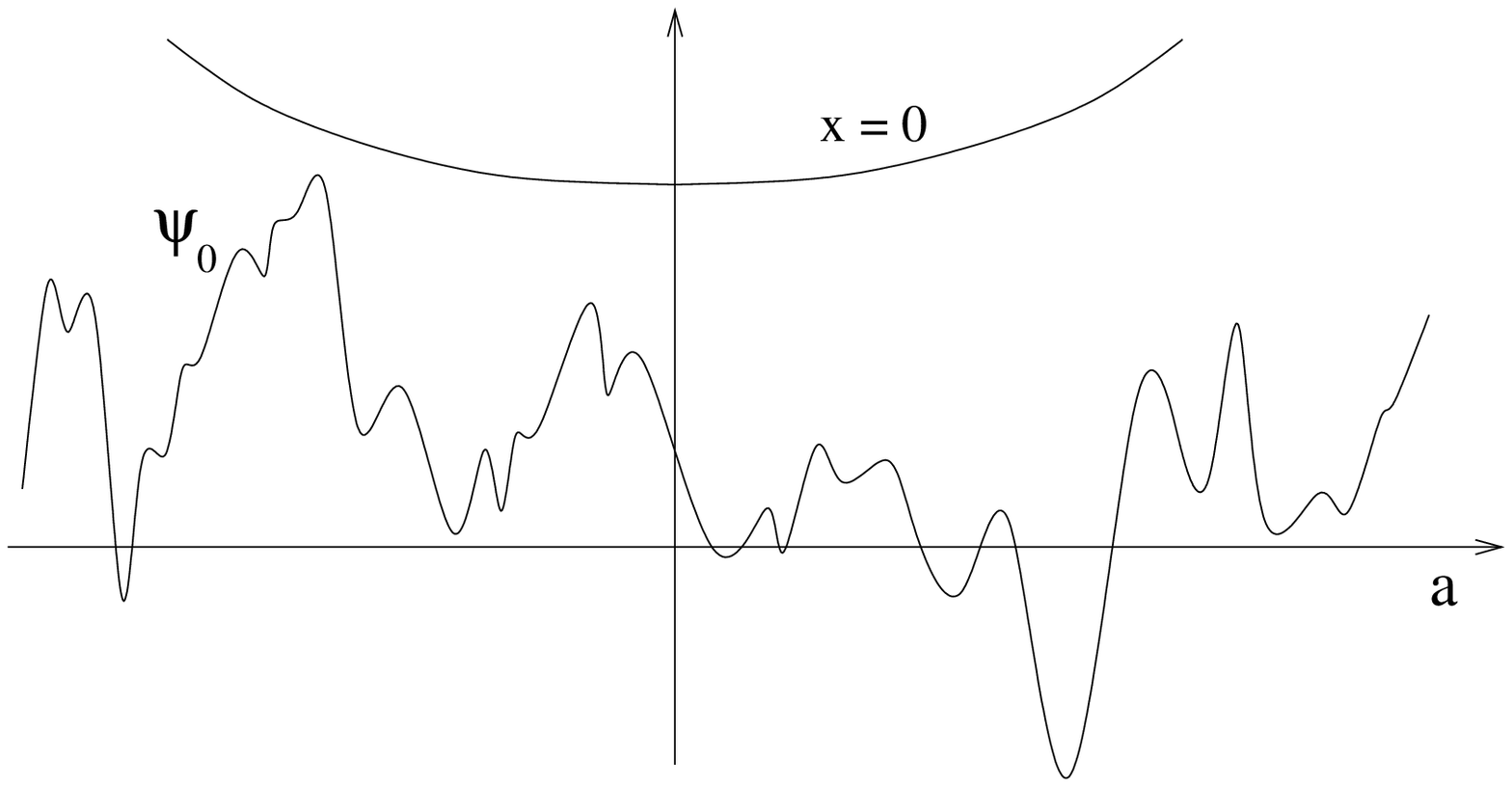}}
\fi
\caption{An initial potential which is everywhere below the parabola $a^2/(2t)
+ \psi$. The probability of such events gives the cumulative probability to
have a potential at time $t$ less than $\psi$.}
\label{f:parabkida}
\end{figure}

\begin{figure}[h]
\iffigs
\centerline{\includegraphics[width=3cm,angle=-90]{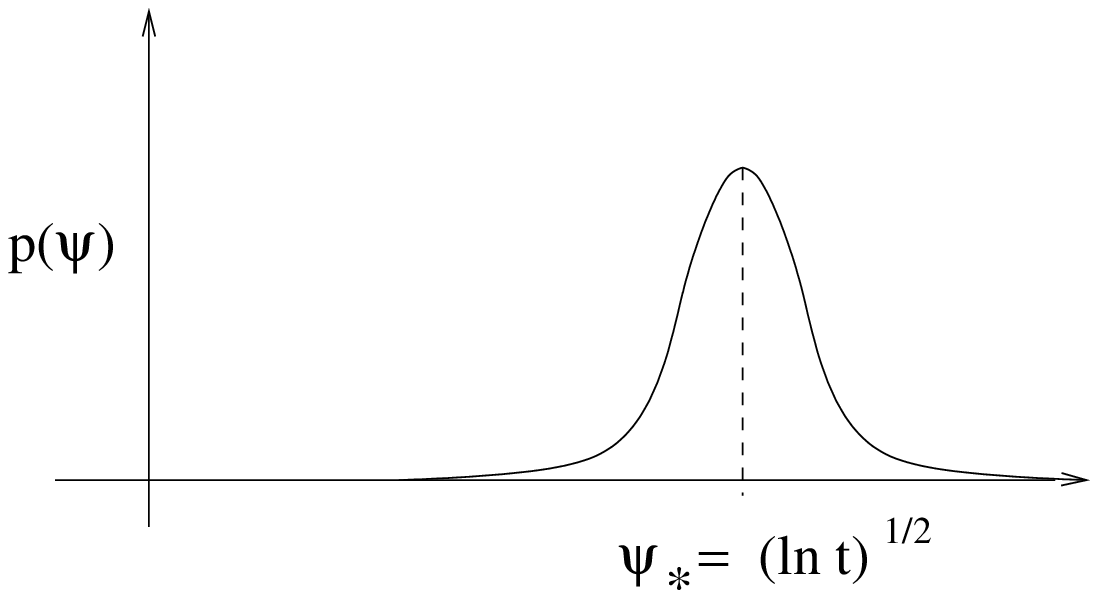}}
\fi
\caption{A sketch of the pdf of the potential at long times.}
\label{f:probapsigrand}
\end{figure}

\begin{figure}[h]
\iffigs
\centerline{\includegraphics[width=9cm]{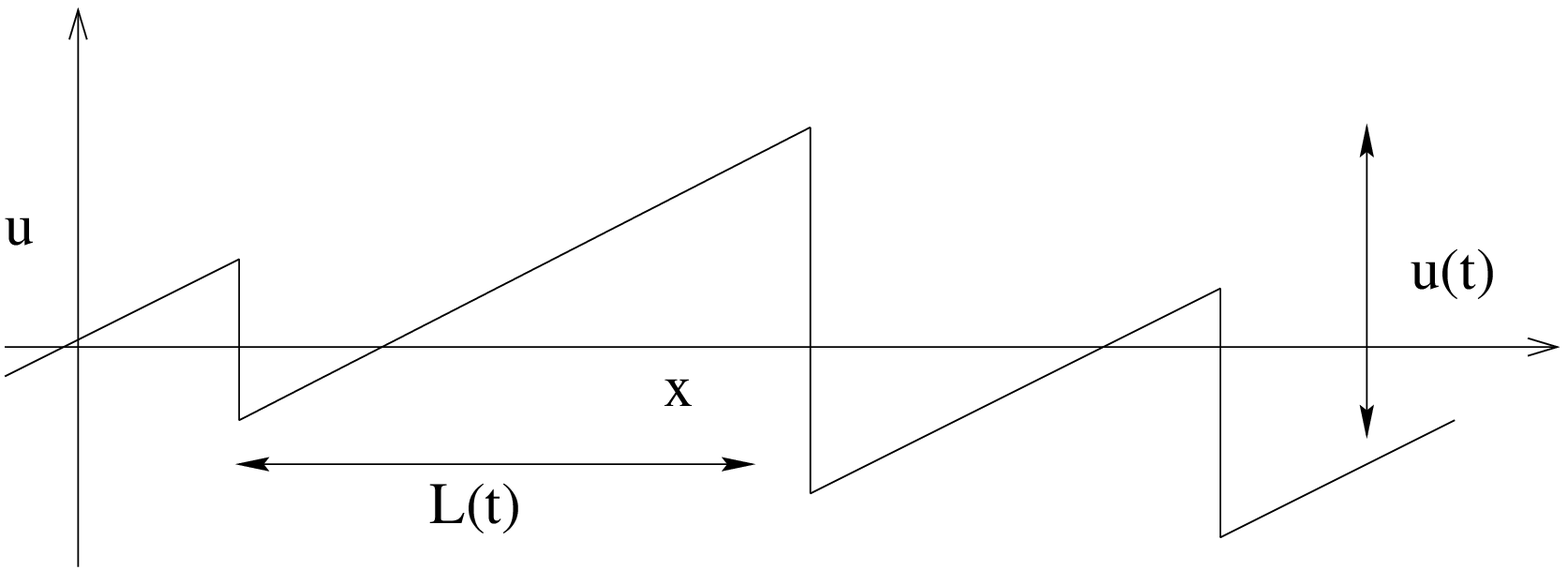}}
\fi
\caption{The Eulerian solution at long times $t$. The ramps have slope $1/t$. 
In time-independent scales, the  figure would be stretched horizontally and
squeezed vertically by a factor proportional to $t$.}
\label{f:ramps}
\end{figure}

A very useful property of decaying burgulence, with no known counterpart
for Navier--Stokes turbulence, is the relation
\begin{equation}
E(t) = {\partial \over \partial t} \la \psi \ra,
\label{epsider}
\end{equation}
which follows by taking the mean of \rf{KPZ} in the absence of a driving force.
Hence, the law of energy decay can be obtained from the law for the mean
potential. The latter can be derived from the cumulative probability of the
potential which, by homogeneity, does not depend on the position. By
\rf{maximum}, its expression at $x=0$ is 
\begin{equation}
  \mbox{Prob}\left\{\mbox{Potential}<\psi\right\} = \mbox{Prob}\left\{\forall
    a,\;\; \psi_0(a)<{a^2 \over 2t} + \psi \right\}.
\label{cumulpot}
\end{equation}
Expressed in words, we want to find the probability that the initial potential
does not cross the parabola $a^2/(2t) + \psi$ (see Fig.~\ref{f:parabkida}).
Since, at long times $t$, the relevant $\psi$ is going to be large,
the problem becomes that of not crossing a parabola with small
curvature and very high apex.  Such crossings, more precisely the
upcrossings, are spatially quite rare. As a
consequence of the mixing property,  for long $t$, they form a Poisson process
\cite{msw95} for which
\begin{equation}
\hbox{Prob}\, \{\hbox{no crossing}\} \simeq e^{-\la N(t)\ra},
\label{poisson}
\end{equation}
where $\la N(t)\ra$ is the mean number of upcrossings. By the Rice formula (a
consequence of the identity $\delta(\lambda x) = (1/|\lambda|) \delta(x)$,
\begin{equation}
\la N(t)\ra = \la \int_{-\infty}^{+\infty} da\; \delta \left ( m(a) - \psi
\right ) {dm \over da} \; H\!\left({dm \over da}\right) \ra,
\label{rice}
\end{equation}
where $H$ is the Heaviside function and
\begin{equation}
m(a)\equiv \psi_0(a) - {a^2 \over 2t}.
\label{defma}
\end{equation}
Since $\psi_0(a)$ is Gaussian, the r.h.s. of \rf{rice} can be easily
expressed in terms of integrals over the probability densities of $\psi_0(a)$
and of $d\psi_0(a)/da$ (as a consequence of homogeneity these variables are
uncorrelated and, hence, independent). The resulting integral can then be
expanded  by Laplace's method for large $t$, yielding
\begin{equation}
\la N(t)\ra \sim t^{1/2} \psi^{-1/2} e^{-\psi^2}, \quad t
\to \infty.
\label{asymptN}
\end{equation}
When this expression is used in \rf{poisson} and the result is
differentiated with respect to $\psi$ to obtain the pdf of $p(\psi)$, the
latter is found to be concentrated around $\psi_\star =(\ln t)^{1/2}$
(see Fig.~\ref{f:probapsigrand}).
It then follows that, at large times, we have Kida's log-corrected 
$1/t$ law for the energy decay
\begin{equation}
\la \psi \ra \sim (\ln t)^{1/2}, \quad E(t)\sim{1 \over t(\ln t)^{1/2}},
 \quad L(t)\sim{t^{1/4} \over (\ln t)^{1/4}}.
\label{psienLkida}
\end{equation}
The Eulerian solution, at long times, has the ramp structure shown in
Fig.~\ref{f:ramps} with shocks of typical strength $u(t)=E^{1/2}(t)$,
separated typically by a distance $L(t)$. The growth in time of $L(t)$
takes place because correlated particles, which initially cannot be
much apart, may propagate to far-apart locations at long times.

The fact that Kida's law is valid for any $n>1$, and not just for
$n\ge 2$ as thought originally, gives rise to an interesting
phenomenon now known as the ``Gurbatov effect'': if $1<n<2$ the
long-time evolution of the energy spectrum cannot be globally
self-similar. Indeed, the permanence of large eddies, which is valid
for any $n<2$ dictates that the spectrum should preserve exactly its
initial $C_nk^n$ behavior at small wavenumbers $k$, with a
constant-in-time $C_n$. Global self-similarity would then imply a
$t^{-2(n+1)/(3+n)}$ law for the energy decay, which would contradict
Kida's law. Actually, as shown in Ref.~\cite{burgatov}, for $1<n<2$
there are two characteristic wavenumbers with different time
dependences, the integral wavenumber $k_L(t) \sim (L(t))^{-1}$ and a
switching wavenumber $k_s(t)\ll k_L(t)$ below which holds the permanence of
large eddies. It was shown that the same phenomenon is also present in
the decay of a passive scalar \cite{ex00}. Whether or not a similar
phenomenon is present in three-dimensional Navier--Stokes
incompressible turbulence or closure models thereto is a controversial
matter \cite{et00,ol00}.

For decaying burgulence, if we leave aside the Gurbatov phenomenon which does
not affect energy-carrying scales, the following may be shown. If we rescale
distances by a factor $L(t)$ and velocities amplitudes by a factor
$u(t)=E^{1/2}(t)$ and then let $t\to \infty$, the spatial (single-time)
statistical properties of the whole random velocity field become
time-independent. In other words, there is a self-similar evolution at long
times. Hence, dimensionless ratios such as the velocity flatness
\begin{equation}
F(t)\equiv {\la u ^4\ra (t)\over \left[ \la u ^2\ra (t)\right]^2}
\label{defuflatness}
\end{equation}
have a finite limit as $t\to \infty$.  A similar property holds for the the
decay of passive scalars \cite{cefv00}.  We do not know if this property holds
also for Navier--Stokes incompressible turbulence or if, say, the velocity
flatness grows without bound at long times.

\section{One-dimensional case with Brownian initial velocity}
\label{s:sinai}

Burgers equation, when the initial velocity is Gaussian with a power-law
spectrum $\propto k^{-n}$, is what cosmologists call {\em scale-free\/}
initial conditions (see Refs.~\cite{p93,cl95}). Here, we consider the
one-dimensional case with Brownian motion (in the space variable) as initial
velocity, corresponding to $n=2$. The general case, including higher
dimensions, is discussed in Ref.~\cite{vdfn94} (an example of a 2-D simulation
with scale-free initial data is shown in Fig.~\ref{f:massimo}).

\begin{figure}[b]
\iffigs
\centerline{\includegraphics[width=4cm,angle=-90]{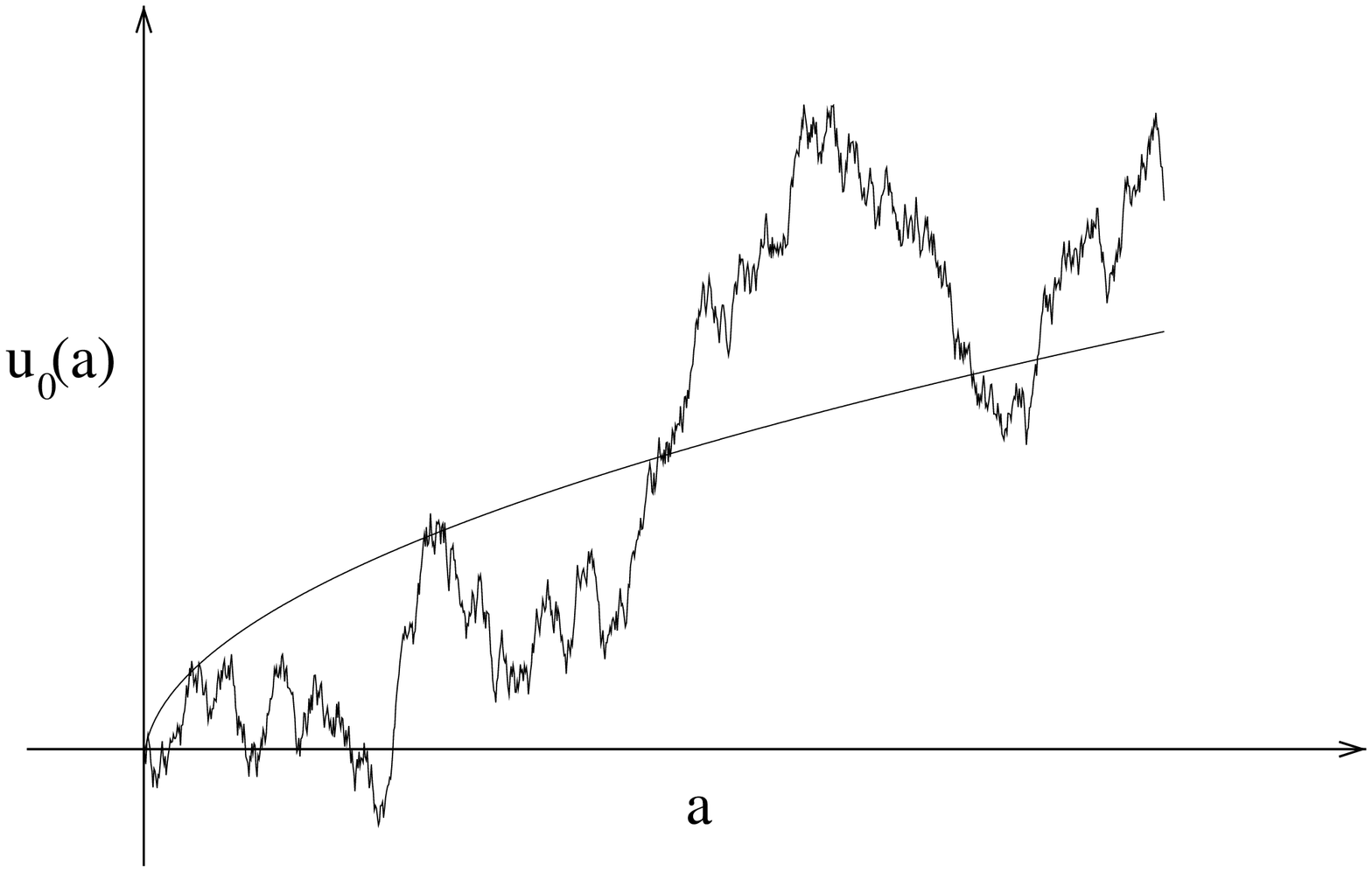}}
\fi
\caption{A realization of the Brownian motion curve. The parabola shows the
root-mean-square velocity $\propto a^{1/2}$.}
\label{f:brown}
\end{figure}
\begin{figure}[h]
\iffigs
\centerline{\includegraphics[width=5cm]{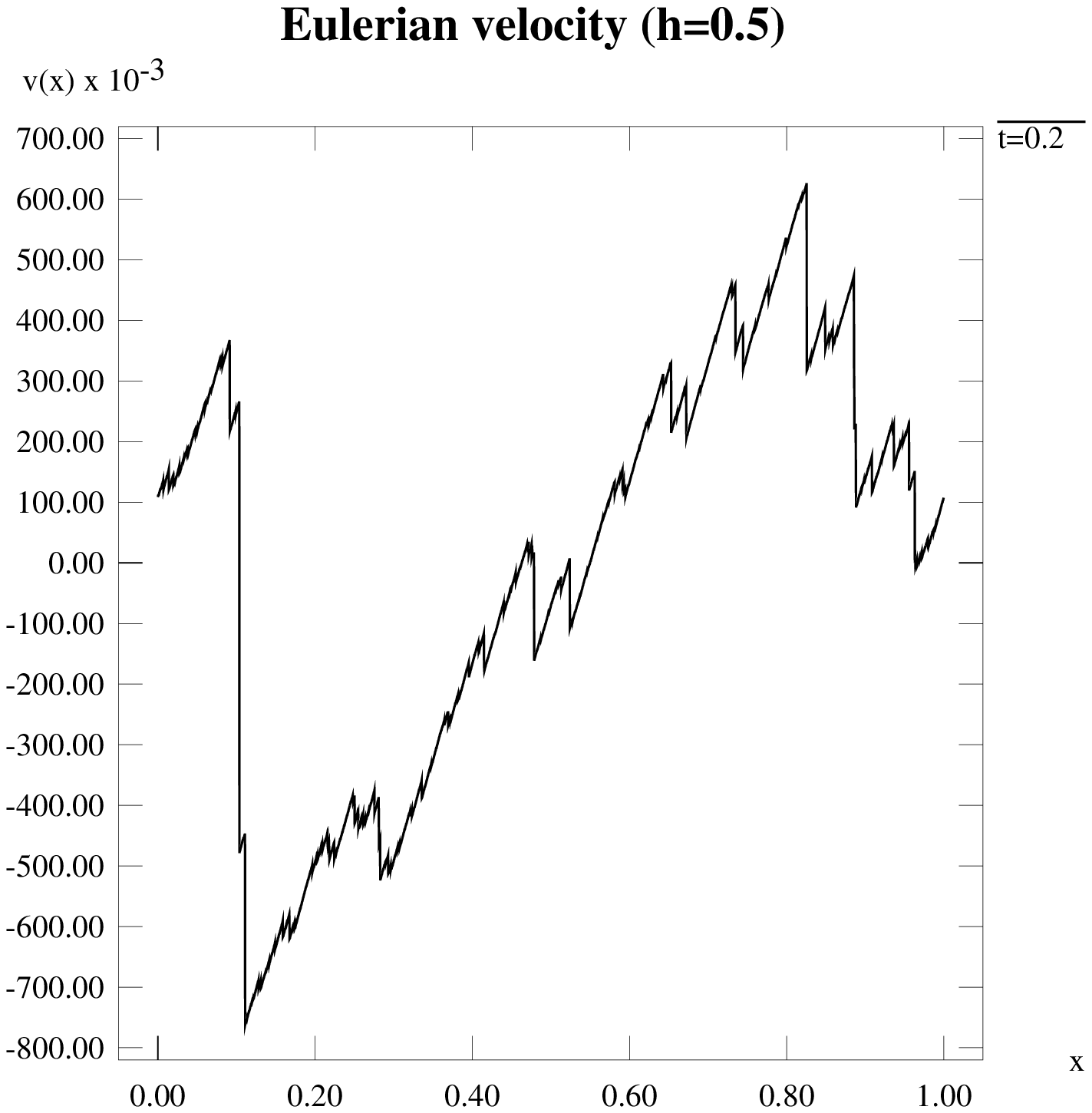}}
\fi
\caption{Snapshot of the velocity resulting from Brownian initial data.
Notice the dense proliferation of shocks (from Ref.~\cite{vdfn94}).}
\label{f:vitesse-cibr}
\end{figure}
\begin{figure}[h]
\iffigs
\centerline{\includegraphics[width=7cm]{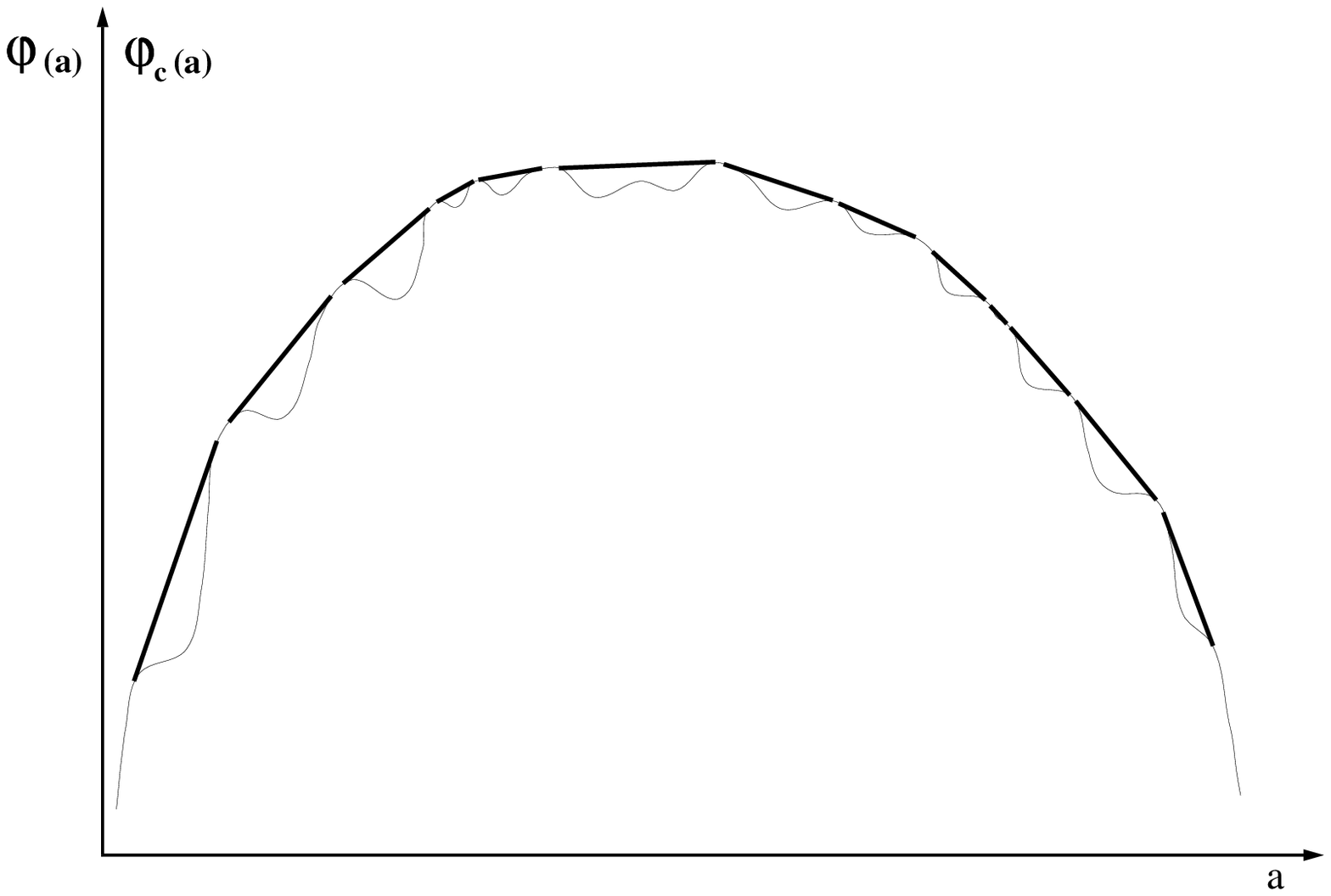}}
\fi
\caption{Sketch of the Lagrangian potential together with its convex hull
(from Ref.~\cite{vdfn94}).}
\label{f:envconvexe-cibr}
\end{figure}
\begin{figure}[h]
\iffigs
\centerline{\includegraphics[width=4cm]{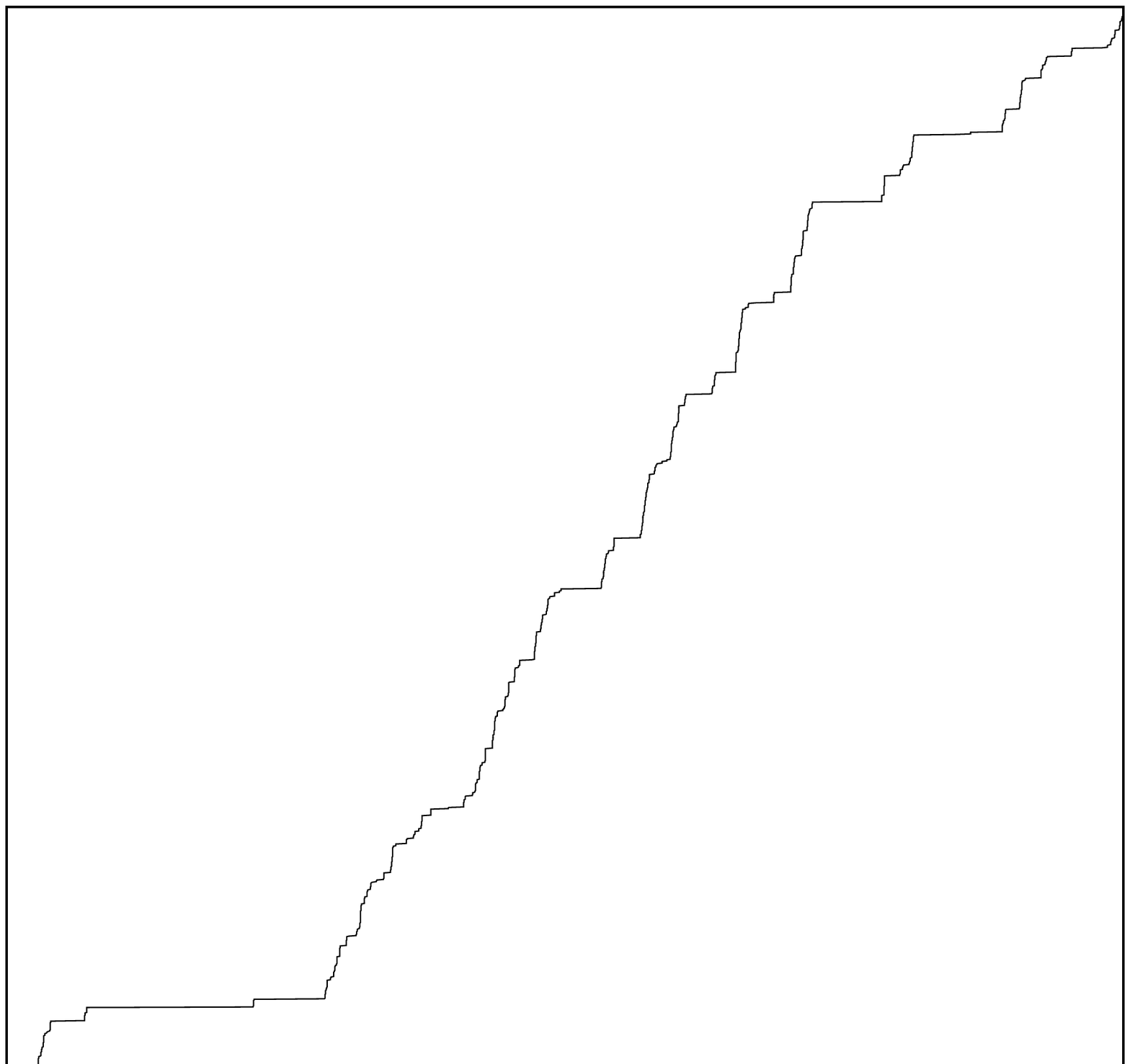}\hspace{1cm}
  \includegraphics[width=4cm]{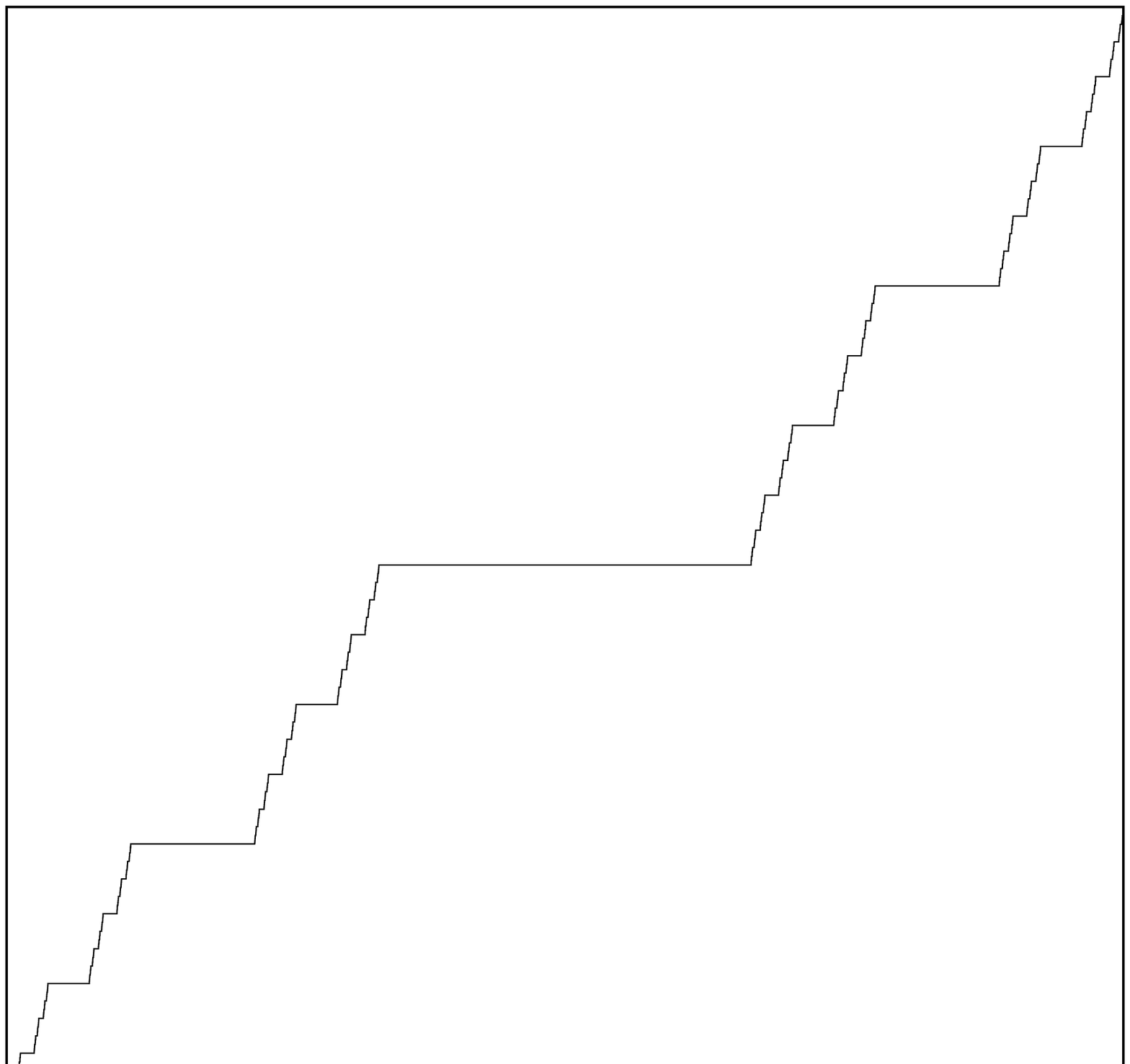}}
\fi
\caption{Left: the Lagrangian map looks like a devil's staircase. Right:
  standard devil's staircase over the triadic Cantor set, which is constant
  almost everywhere, except on a fractal (from Ref.~\cite{vdfn94}).}
\label{f:escaliers}
\end{figure}

Brownian motion is continuous but not differentiable (see Fig.~\ref{f:brown});
hence, shocks appear after arbitrarily short times and are actually {\em
dense} (see Fig.~\ref{f:vitesse-cibr}). Numerically supported conjectures made
in Ref.~\cite{saf92}, have led to a proof by Sinai \cite{s92} of the following
result: in Lagrangian coordinates, the regular points, that is fluid particles
which have not yet fallen into shocks, form a fractal set of Hausdorff
dimension $1/2$. This implies that there is a Devil's staircase of dimension
$1/2$ in the Lagrangian map (see Fig.~\ref{f:escaliers}). Note that when the
initial velocity is Brownian, the Lagrangian potential has a second space
derivative which is delta-correlated in space; this can be approximately
pictured as a situation where the Lagrangian potential has very strong
oscillations in curvature. Hence, it is not surprising that very few points of
its graph can belong to its convex hull (see Fig.~\ref{f:envconvexe-cibr}).
  
\begin{figure}[h]
\iffigs
\centerline{\includegraphics[width=7.5cm]{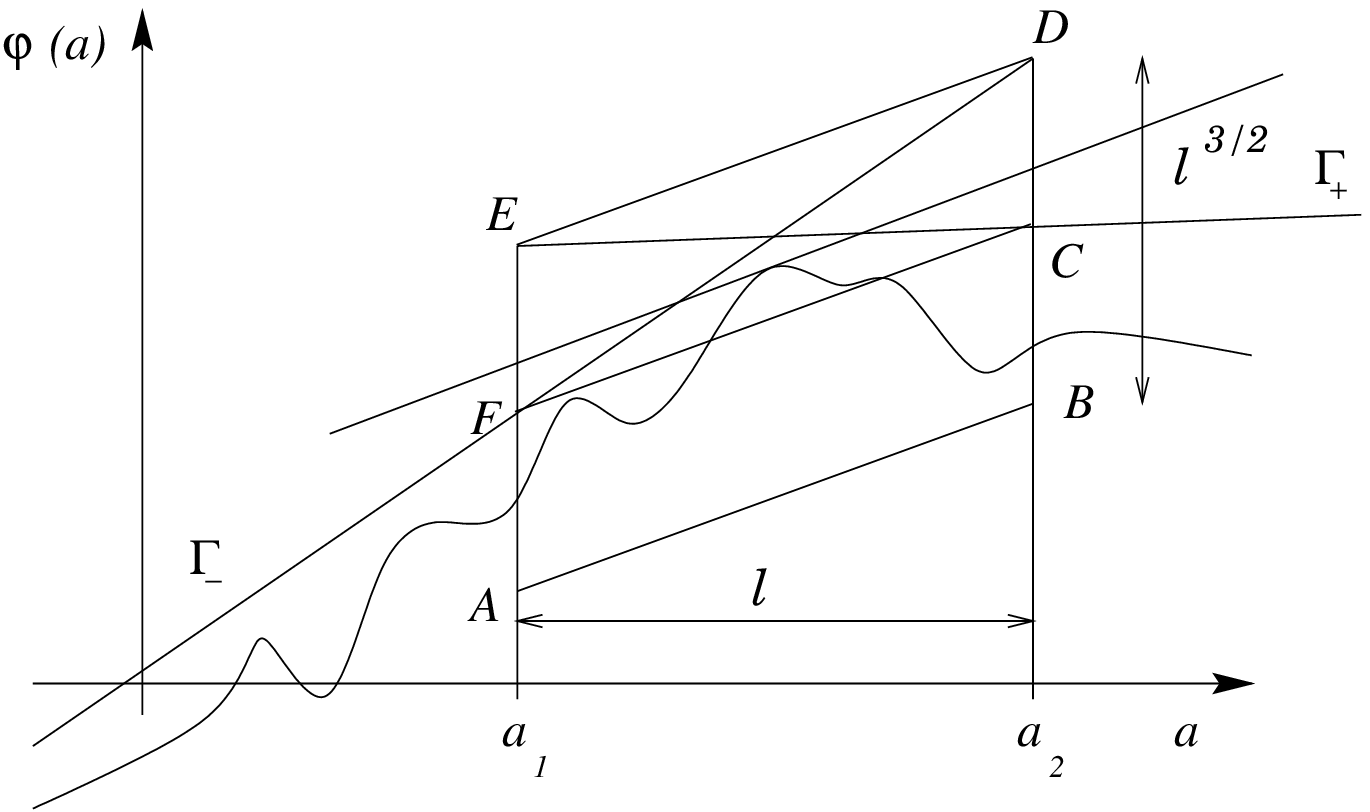}}
\fi
\caption{The box construction used to find a regular point $R$ (point of 
tangency with the graph entirely on one side of the tangent) within a
Lagrangian interval of length $\ell$ (from   Refs.~\cite{s92,vdfn94}).}
\label{f:leftboxright}
\end{figure}

We will now give some highlights of Sinai's proof of this result. For
this problem, it turns out that the Hausdorff dimension of the regular
points (determined in Ref.~\cite{s92}) is also equal to its
box-counting dimension, which is easier to determine. One obtains the latter
by finding the probability that a small Lagrangian interval of length
$\ell$ contains at least one regular point which belongs
simultaneously to the graph of the Lagrangian potential $\varphi$ and
to its convex hull. In other words, one looks for points, such as $R$,
with the property that the graph of $\varphi$ lies below its tangent
at~$R$ (see Fig.~\ref{f:leftboxright}).  Sinai does this by the box
construction with the following constraints on the graph:
\begin{description}
\item[{\sl Left}:] the graph of the potential should be below the half line
  $\Gamma_-$,
\item[{\sl Right}:] the graph of the potential should be below the half line
  $\Gamma_+$,
\item[{\sl Box}:] $\left\{\begin{array}{l} \mbox{1: enter } (AF) \mbox{
        with a slope larger than that of } \Gamma_- \mbox{ by } O(\ell^{1/2})\\
      \mbox{2: exit } (CB) \mbox{ with a slope less than that of }
      \Gamma_+ \mbox{ by } O(\ell^{1/2})\\
      \mbox{3: cross } (FC) \mbox{ and stay below } (ED).
    \end{array}\right.$
\end{description}
It is obvious that such conditions ensure the existence of at least one
regular point. (Move $(ED)$ down parallel to itself until it touches the
graph.)  Note that $A$ and the slope of $(AB)$ are prescribed. Hence, one is
calculating conditional probabilities; but it may be shown that the
conditioning is not affecting the scaling dependence on $\ell$.

As the Brownian motion $u_0(a)$ is a {\em Markov process}, the constraints
{\sl Left}, {\sl Box} and {\sl Right} are independent and hence,
\begin{eqnarray}
  P^{\rm reg.} (\ell) & \equiv & {\rm Prob} \left \{ \mbox{regular point in
  interval of length } \ell \right \}\nonumber \\ & = & {\rm Prob} \left \{
  \mbox{\sl Left} \right \} \times {\rm Prob} \left \{ \mbox{\sl Box} \right
  \} \times {\rm Prob} \left \{ \mbox{\sl Right} \right \}
\label{leftboxright}
\end{eqnarray}
The scales of the box were chosen so that ${\rm Prob} \left \{
  \mbox{\sl Box} \right \}$ is independent of $\ell$:
\begin{equation}
{\rm Prob} \left \{
  \mbox{\sl Box} \right \} \sim \ell^0.
\label{box0}
\end{equation}
Indeed, Brownian motion and its integral have scaling exponent $1/2$ and
$3/2$, respectively and the problem with $\ell << 1$ can be rescaled
into that with $\ell = 1$ without changing probabilities.

\begin{figure}[h]
  \iffigs \centerline{\includegraphics[width=9cm]{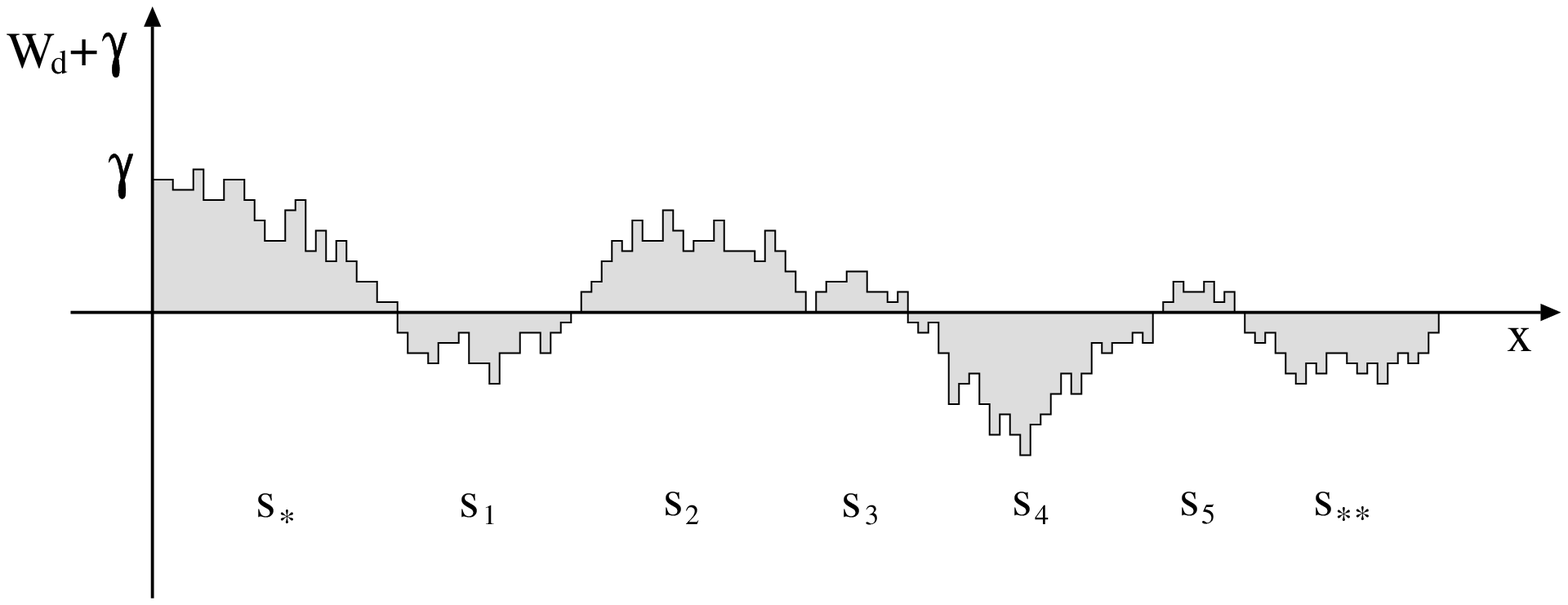}} \fi \caption{The arches construction which uses the
  zero-passings of a random walk to estimate the integral of Brownian motion
  (from Refs.~\cite{s92,vdfn94}).}  \label{f:arches}
\end{figure}

It is clear by symmetry that ${\rm Prob} \left \{ \mbox{\sl Left} \right \}$
and $ {\rm Prob} \left \{ \mbox{\sl Right} \right \}$ have the same scaling in
$\ell$.  Let us concentrate on ${\rm Prob} \left \{ \mbox{\sl Right} \right
\}$. We can write the equation for the half line $\Gamma_+$ in the form
\begin{equation}
\Gamma_+: a \mapsto \varphi(a_2) + \delta \ell^{3/2} +\left( \partial_a
  \varphi(a_2)+\gamma \ell^{1/2} \right ) (a-a_2),
\label{eqgammaplus}
\end{equation}
where $\gamma$ and $\delta$ are positive $O(1)$ quantities. Hence, introducing
$\alpha \equiv a-a_2$, the  condition {\sl Right} can be written to the
leading order as
\begin{equation}
\int_0^\alpha \left ( u_0(a) + \gamma \ell^{1/2}\right)da +\delta \ell^{3/2} +
{\alpha^2 \over 2} > 0, \mbox{ for all } \alpha>0.
\label{condright1}
\end{equation} 
By the  change of variable $\alpha = \beta\ell$ and use of the fact that the
Brownian motion has scaling exponent $1/2$, one can write the condition
{\sl Right} as 
\begin{equation}
  \int_0^\beta \left ( u_0(a) + \gamma \right) da >
    -\delta, \mbox{ for all } \beta  \in [0, \ell^{-1}]. 
\label{condright2}
\end{equation}
Without affecting the leading order, one can replace the Brownian motion by a
stepwise constant random walk with jumps of $\pm1$ at integer $a$'s.  The
integral in \rf{condright2} has a geometric interpretation, as
highlighted in Fig.~\ref{f:arches}, which shows a random walk starting from the
ordinate $\gamma$ and the arches determined by successive zero-passings. The
areas of these arches are denoted $S_\star, S_1, ... S_n, S_{\star\star}$.  It
is easily seen that
\begin{equation}
{\rm Prob} \left \{ \mbox{\sl Right} \right \} \sim {\rm Prob} \left \{ 
S_1>0,S_1+S_2>0,...,S_1+...+S_n>0 \right \},
\label{condright3}
\end{equation}
where $n=O(\ell^{-1/2})$ is the number of zero-passings of the random walk in 
the interval 
$[0, \ell^{-1}]$. The probability \rf{condright3} can be evaluated by
random walk methods (see, e.g., Ref.~\cite{fellervol2}, Chap.~12, Sec.~7),
yielding 
\begin{equation}
  {\rm Prob} \left \{ \mbox{\sl Right} \right \} \sim {\rm Prob}\left\{ n \mbox{ first sums} > 0 \right \} \propto n^{-1/2}
  \propto \ell^{1/4}.
  \label{condrightfin}
\end{equation}
By (\ref{leftboxright}), (\ref{box0}) and (\ref{condrightfin}), the
probability to have a regular point in a small interval of length $\ell$
behaves as $\ell^{1/2}$ when $\ell\to0$. Thus, the regular points have a
box-counting dimension $1/2$.

\section{Preshocks and the pdf of velocity gradients in one dimension}
\label{s:pdfgrad}

In this section we shall determine the tail behavior of the probability
density function (pdf) of the velocity gradient for one-dimensional decaying
burgulence. To explain some of the motivations for this study, it is useful to
make a digression concerning the forced one-dimensional Burgers equation:
\begin{eqnarray}
\partial_t u + u\partial_x u = \nu \partial_{xx} u +f(x,t),
\label{burgforce1d}\\
u(x,t_0) = u_0(x).
\label{condinitu}
\end{eqnarray}
The latter displays much richer features than the unforced problem. The case
where the force is random has often been studied as a prototype for a wide
range of problems in non-equilibrium statistical physics (see
Section~\ref{condmat}).

Eq.~(\ref{burgforce1d}) can also be used in the same spirit as the
forced Navier--Stokes equation, namely to investigate universality of
various statistical properties with respect to the forcing. For
Navier--Stokes turbulence, when the force is confined to large spatial
scales and the Reynolds number is very high, small-scale (inertial
range) statistical scaling properties are generally conjectured not to
depend on the forcing, except through overall numerical
factors. Similar conjectures have been made for burgulence
with large-scale forcing.  For example, there is little doubt that,
because of the presence of shocks, structure functions of order $p>1$
have universal exponents equal to unity, as in the decaying case (see,
e.g., Refs.~\cite{bmp95,ekms97}).  More controversial is the tail
behavior of the probability density function (pdf) of velocity
gradients and velocity increments in the limit of zero viscosity when
the force is a white-noise process in time. For increments, the
problem was addressed for the first time by Chekhlov and Yakhot
\cite{cy95}, who considered a force with a power-law spectrum, acting
both at large and at small scales. Concerning the pdf $p(\xi)$ at
large negative gradients $\xi$, it is generally believed that it
follows a power law
\begin{equation}
p(\xi) \propto |\xi|^{\alpha}, \quad {\rm for}\,\, \xi \to -\infty,
\label{pdfalpha}
\end{equation}
but the conjectured values of $\alpha$ differ markedly.  Polyakov
\cite{p95} and Boldyrev \cite{b97}, using a field-theoretical operator
product expansion, predicted $\alpha=-5/2$; E {\it et al}.\
\cite{ekms97}, using a semi-heuristic approach in which preshocks
(nascent shocks) are key, predicted $\alpha=-7/2$; Gotoh and Kraichnan
\cite{gk98}, using a Fokker--Planck equation approach, predicted
$\alpha=-3$; more recent work by Kraichnan \cite{k99} favored
$\alpha=-7/2$. E and Vanden Eijnden \cite{eve99,eve00} developed a
probabilistic formalism adapted to solutions with shocks and giving
insight into many aspects of the problem; they proved that
$\alpha<-3$, and made a good case for $\alpha=-7/2$.  The question of
the correct law for the case of white-noise forcing remains however
open (we shall come back to this later).

\begin{figure}[h]
\iffigs
\centerline{\includegraphics[height=0.3\hsize]{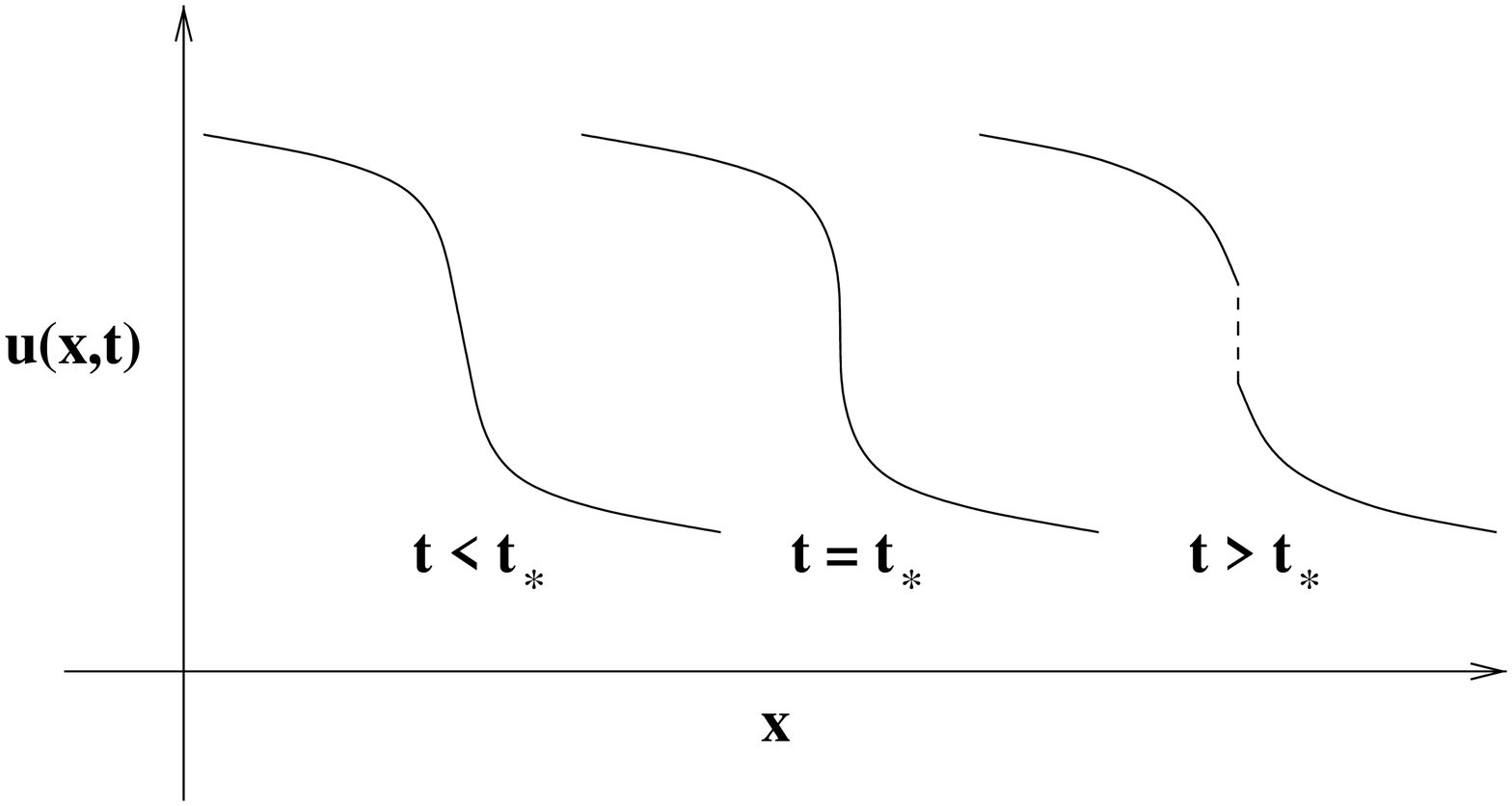}}
\fi
\caption{Eulerian structure of the solution just before a preshock, at the
time $t_\star$ of a preshock and just after.}
\label{f:preshock}
\end{figure}

Actually, there is a situation much simpler than that originally considered in
Ref.~\cite{ekms97}, for which the argument in favor of $\alpha=-7/2$ can be
made rigorous, namely decaying burgulence. This closes our
digression; in the remainder of this section we concentrate
on the the unforced problem
\begin{equation}
\partial_t u + u \partial_x u = \nu \partial_{x}^2 u,
\label{eq-burg-declin}
\end{equation}
in the limit of vanishing viscosity $\nu\rightarrow0$ and we follow
Refs.~\cite{bf00,fbv00}. We assume a random initial velocity
$u_0=-(d\psi_0/da)$, deriving from a smooth initial
potential. Homogeneity is not required. The value $\alpha = -7/2$ for
the exponent of the pdf at large negative gradients is easily
understood in this case. It is just the signature of the preshocks,
the cubic root singularities in Eulerian coordinates, which appear
when new shocks are created (see Fig.~\ref{f:preshock}). Preshocks
constitute discrete events in space-time, contrary to shocks which persist
in time (until they merge).
These preshocks are the only structures giving large {\em
finite\/} negative gradients: shocks give infinite negative gradients (unless a
finite viscosity is introduced) and the gradients in the immediate spatial
neighborhood of a mature shock are not particularly large. A simplified
presentation is given hereafter for the case of a single preshock; the
contributions of several preshocks to the pdf are just additive.

Let us suppose that the initial gradient $du_0/da$ has a minimum at $a=0$
(corresponding to an inflection point with negative derivative of the initial
velocity), so that a shock will appear at time $t=t_\star = -1/((du_0/da)(0))$
and at $x= t_\star u_0(0)$. Without loss of generality, we assume $u_0(0)=0$
(otherwise we perform a Galilean transformation to bring it to zero). As the
initial velocity is supposed to be sufficiently smooth, we can perform a
Taylor expansion of the initial potential in the neighborhood of $a=0$. We
then have, locally,
\begin{equation}
\psi_0(a)= c_1 a^2 - c_2 a ^4 +{\rm h.o.t.},
\label{taylorpot1d}
\end{equation}
where $c_1$ and $c_2$ are positive (random) constants and ``h.o.t.'' stands for
higher-order terms. The Lagrangian potential is locally
\begin{equation}
\varphi = -{a^2\over 2} + t \psi_0(a)= {\tau \over 2}  a^2 - t c_2 a^4 + {\rm
  h.o.t.},
\label{eq:lagpot1d}
\end{equation}
where $\tau = (t-t_*)/t_*$. The Lagrangian map outside the shock is thus
\begin{equation}
x(a,t) = -\partial_a \varphi(a,t) = -\tau a + 4t c_2 a^3+ {\rm h.o.t.}
\label{eq:lagmap1d}
\end{equation}

\begin{figure}[h]
\iffigs
\centerline{\includegraphics[height=0.3\hsize]{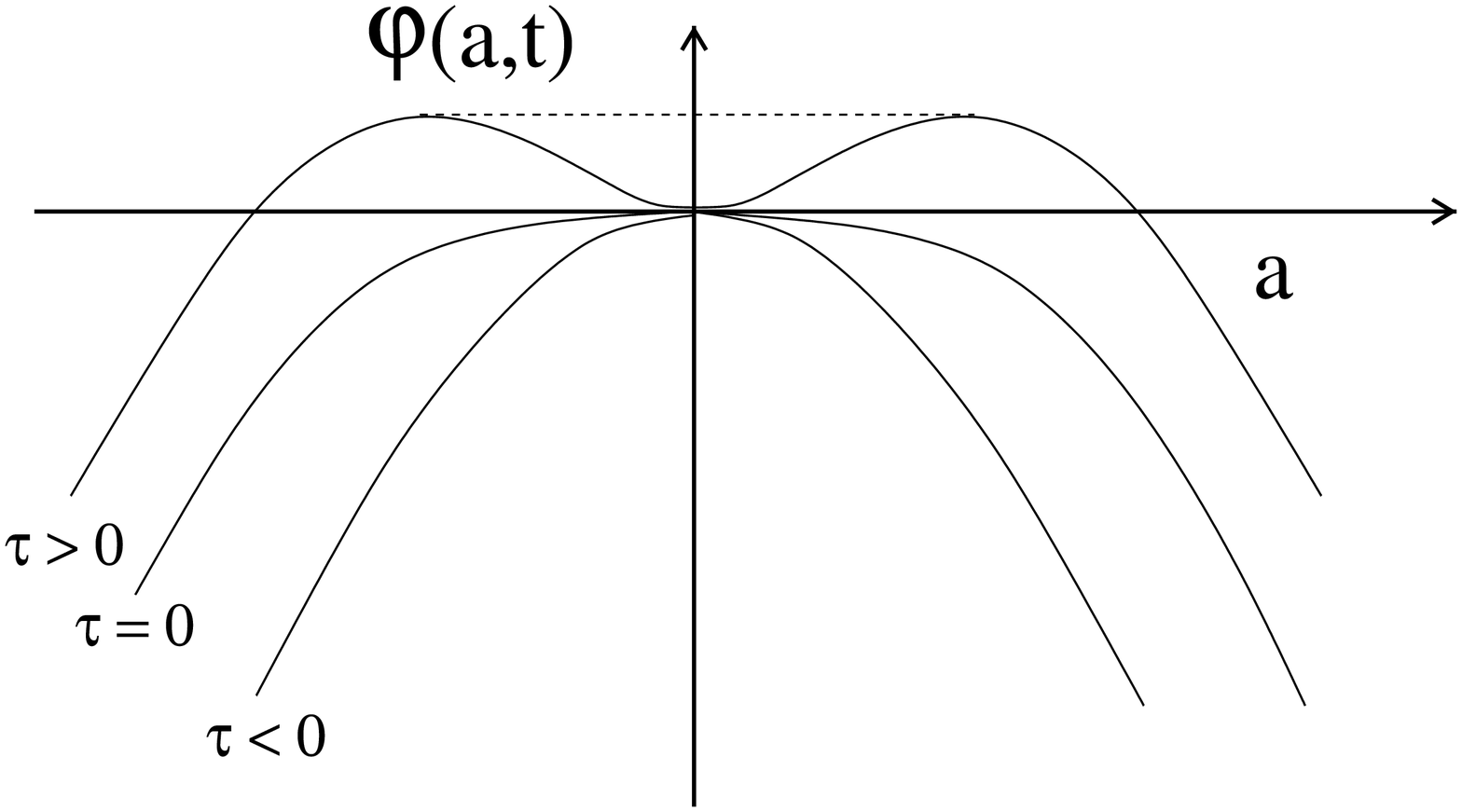}}
\fi
\caption{Normal form of the Lagrangian potential in the neighborhood
of a preshock in one dimension. At the time of the preshock
($\tau=(t-t_*)/t_*= 0$), the Lagrangian potential changes from a
single extremum to three extrema and develops a non-trivial convex
hull (shown as a dashed line).}
\label{f:lagpot1d}
\end{figure}

The Lagrangian potential, together with its convex hull, are shown in
Fig.~\ref{f:lagpot1d}. It is convex for $t\leq t_\star$. At $t=t_\star$, there
is a degenerate maximum with quartic behavior, and, immediately after $t_*$
(for $\tau>0$), convexity is lost and a shock interval is born. Given the
symmetry, resulting from our choice of coordinates, the convex hull contains a
horizontal segment extending between the two maxima $a_\pm = \pm
(\tau/(4c_2))^{1/2}$.
The velocity gradient can be
written locally as
\begin{equation}
\partial_x u (x,t) = {(du_0/da)(a) \over \partial_a x(a,t)} = {2/t_\star \over
- \tau  + 12 c_2 a^2},
\label{eq:gradvit}
\end{equation}
where $a$ is the unique preimage of $x$ by the naive Lagrangian map outside of
the shock interval $]a_-, a_+[$. Since, by~(\ref{eq:lagmap1d}), the relation
between $x$ and $a$ is cubic at $\tau = 0$, the velocity gradient $\partial_x
u (x,t_\star) \propto |x|^{-2/3}$, which is unbounded. For any $t\neq
t_\star$, the gradient remains bounded, except at the shock location. For
$\tau<0$, just before creation of the shock, the cubic relation between $x$
and $a$ still holds, except in a region of Lagrangian width of the order of
$\tau^{1/2}$, and hence of Eulerian width $\sim\tau^{3/2}$, where the relation
becomes linear to leading order.

The question is now: what is the fraction of Eulerian space-time where
$\partial_x u < \xi$, with $\xi$ a large negative number\,? Because of the
cubic root structure, $x$ must be in a small interval of width $\sim
|\xi|^{-3/2}$. The time must be sufficiently close to $t_\star$ for this
interval still to be in the region of validity of the cubic relation, that is,
within $\sim |x|^{2/3} \sim |\xi|^{-1}$. Hence, the relevant space-time
fraction or, in other words, the cumulative probability to have $\partial_x u
< \xi$ is proportional to $|\xi|^{-5/2}$.  This gives a pdf $p(\xi)\propto
|\xi|^{-7/2}$ at large negative $\xi$'s.

Actually, there is another contribution, also proportional to $|\xi|^{-7/2}$
stemming from a small time interval $\tau \sim |x|^{2/3} \sim |\xi|^{-1}$ just
{\em after\/} $t_\star$ when small-amplitude shocks are present which have not
yet completely destroyed the cubic root structure (see Ref. \cite{bf00}). 
Similar arguments can be used to show that there are power-law ranges
with exponent $-7/2$ and $+1$ in the pdf of velocity increments for decaying
burgulence \cite{bf00}.

The preshock argument has first been introduced phenomenologically in
Ref.~\cite{ekms97} to predict pdf's of velocity gradients and increments for
the case of white-noise in time forcing at large scales. In principle, in the
presence of forcing, spatio-temporal accumulations of preshocks, invalidating
the $-7/2$ law, cannot be ruled out. Nevertheless, numerical evidence in favor
of the $-7/2$ law has been recently obtained by one of us (JB), using particle
tracking simulations with a shot-noise approximation to white noise.

\section{The pdf of density}
\label{s:pdfdensity}

In cosmological applications of the adhesion model/Burgers equation, it is of
special interest to analyze the behavior of the density of matter, since the
large-scale structures may also be characterized as mass condensations. In
Eulerian coordinates, the mass density $\rho$ satisfies the continuity
equation
\begin{equation}
\partial_t \rho + \nabla \cdot \left( \rho \v \right) =0.
\label{ddrho}
\end{equation}
The initial density is denoted by $\rho_0(\a)$.

The question we intend to address here is the behavior, in the limit
of vanishing viscosity and at large $\rho$'s, of the pdf of mass
density $p(\rho)$, when the initial velocity is random and smooth (and
not necessarily homogeneous). This problem was studied in
Ref.~\cite{fbv00}, where it was shown that density pdf's have
universal power-law tails with exponent $-7/2$ in any dimension.  This
behavior stems from singularities, other than shocks, whose nature is
quite different in one and several dimensions. (Similar results can in
principle be obtained for velocity gradients and increments which are,
however, not scalars in more than one dimension.)

In one dimension, the pdf of the mass density at large arguments is
basically the same as the pdf of gradients at large negative
arguments. Indeed, it is easy to show that, for any $x$ not at a shock
location,
\begin{equation}
\rho(x,t) = \rho_0(a)\left(1-t\partial_x u(x,t)\right),
\label{formulegurbatov}
\end{equation}
where $a$ is the preimage of $x$ by the Lagrangian map \cite{gms91}. If now
$\rho_0$ is bounded from below and above (e.g., for uniform $\rho_0$), the
result of the previous section  implies that, for $\rho\to \infty$, the pdf 
$p(\rho)$ of the mass density satisfies a $\rho^{-7/2}$ law, which is again the
signature of preshocks.

\begin{figure}[h]
\iffigs
\centerline{\includegraphics[width=5.5cm]{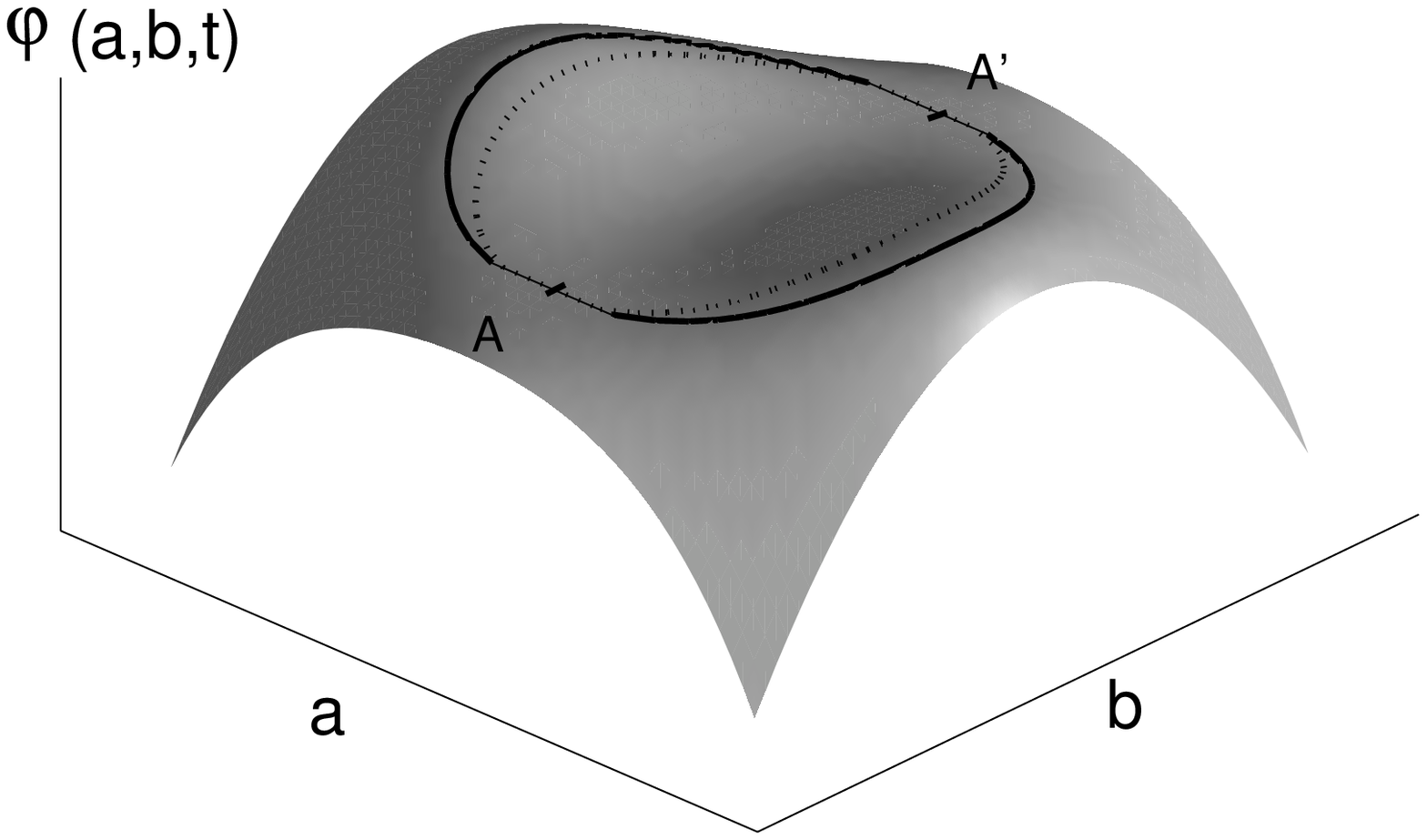}
\includegraphics[width=5.5cm]{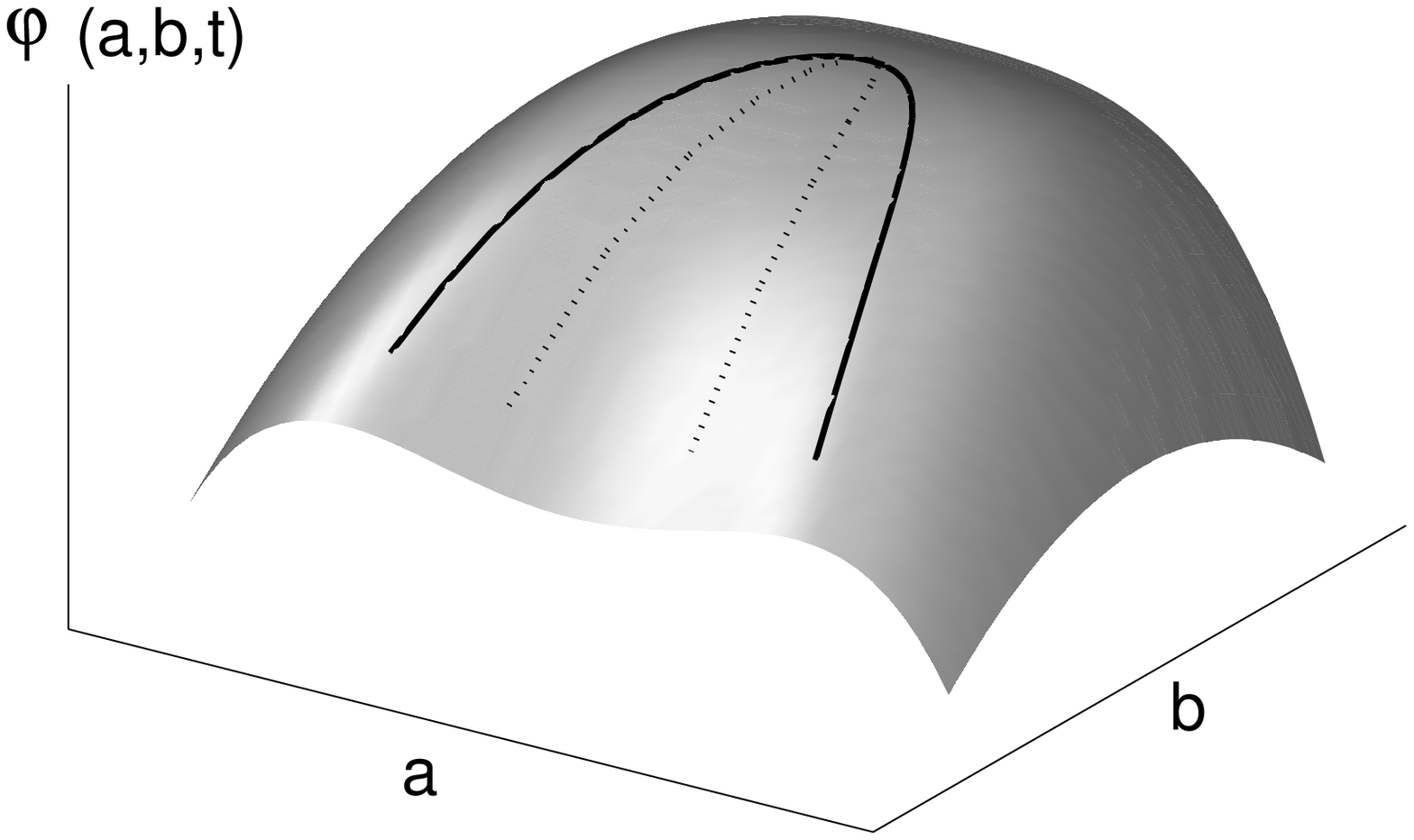}}
\fi
\label{f:prechetkurto}
\caption{Lagrangian potential in two dimensions with $(a,b)$ coordinates, just
after a preshock (left) and in the immediate neighborhood of a kurtoparabolic
point (right).  Continuous lines: separatrices between the regular part and
the ruled surfaces of the convex hull; dotted-lines: vanishing of the Jacobian
of the Lagrangian map. A and A' are a pair of kurtoparabolic points born with
the shock.}
\end{figure}

The key to studying this problem in more than one dimension is the geometric
construction of the solution via the convex hull of the Lagrangian potential
(see Section~\ref{subs:convexhull}). Conservation of mass (\ref{ddrho})
implies that the density is given at regular points by
\begin{equation}
\rho(\x,t) = {\rho_0(\a) \over J(\a, t)},
\label{rholag}
\end{equation}
where $J$ is the Jacobian of the Lagrangian map. (The density is infinite
within shocks.)  Since the Jacobian is (up to a factor $(-1)^d$) equal to the
Hessian of the Lagrangian potential (determinant of the matrix of second space
derivatives), it follows that large densities are typically obtained only near
parabolic points (where the Hessian vanishes). However, arbitrarily close to a
parabolic point there are generically hyperbolic points where the surface
defined by $\varphi$ crosses its tangent (hyper)plane and which, therefore, do
not belong to its convex hull.  Yet, there exist in general exceptional
``kurtoparabolic'' points which are parabolic and belong to the boundary of
the set of regular points ({\it kurtos} means convex in Greek). Near such
points, arbitrarily large densities are obtained.  In one dimension, the only
kurtoparabolic points are the preshocks which are discrete space-time events
in both Eulerian and Lagrangian coordinates. In two and more dimensions,
kurtoparabolic points are also born at preshocks but live in general for a
finite time; they reside on manifolds of spatial dimension $(d-1)$ (see
Fig.~\ref{f:prechetkurto}). In Eulerian space, they are associated to
boundaries of shocks (e.g. end points of shock lines for $d=2$).

\begin{figure}[h]
\iffigs
\centerline{\includegraphics[width=7cm]{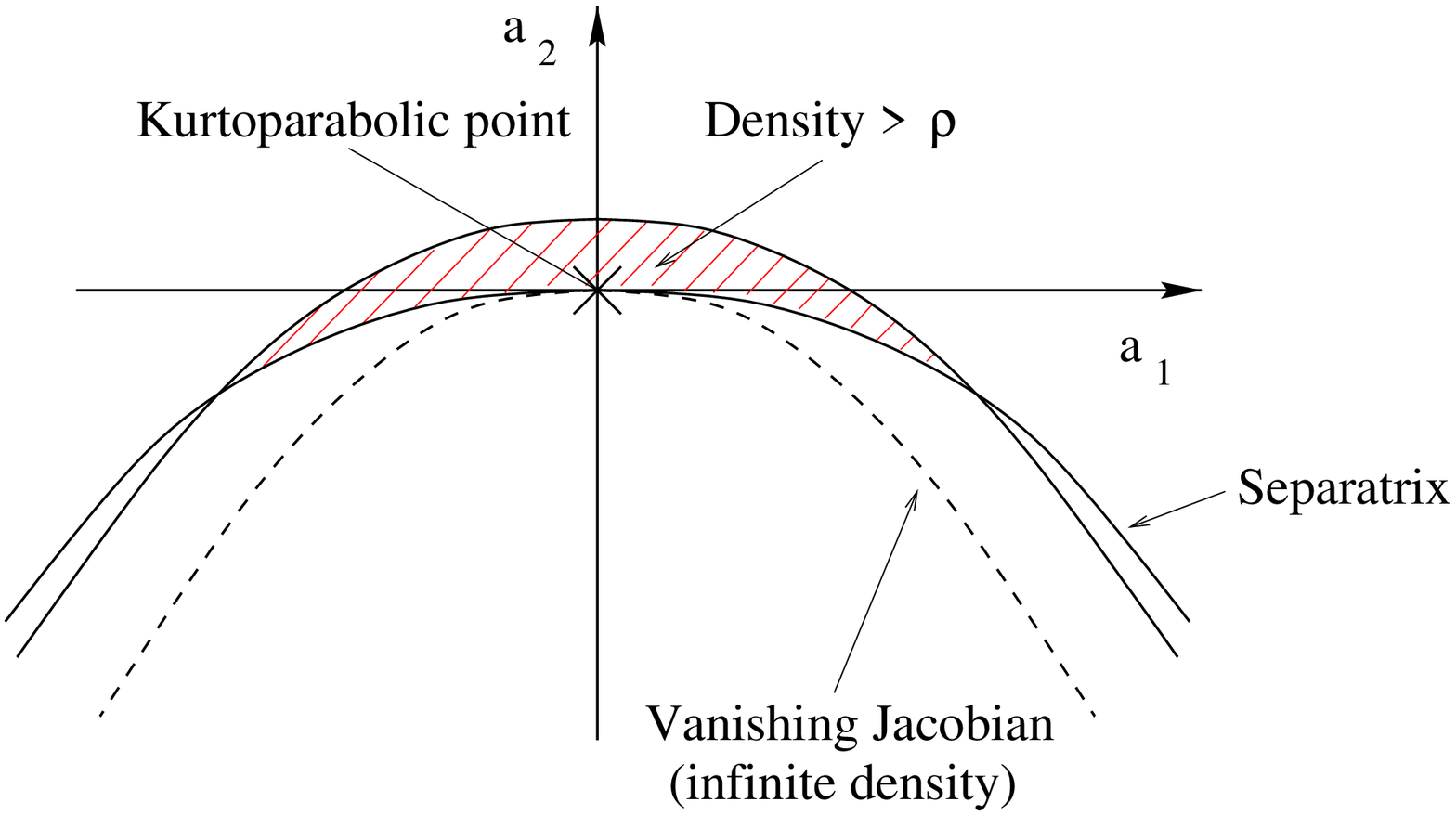}}
\fi
\caption{Projection in the two-dimensional case of the neighborhood of a
kurtoparabolic point.}
\label{f:kurto2d}
\end{figure}

The determination of the large-$\rho$ tail of the cumulative probability
distribution of the density, $P^>(\rho)$, is equivalent to finding the
fraction of Eulerian space-time where $\rho$ exceeds a given value (see Ref.
\cite{fbv00} for details). The latter is determined by changing from Eulerian
to Lagrangian coordinates and Taylor-expanding to the relevant order the
Lagrangian potential near a kurtoparabolic point, in a suitable coordinate
frame:
\begin{equation}
\varphi(\a,t) \simeq \zeta a_1^4 + \sum_{j>1} \left [ -{\mu_j\over 2} a_j^2 +
  \beta_j a_1^2 a_j \right ].
\label{normalformdd}
\end{equation}
From (\ref{normalformdd}), it is then easy to determine explicitly the
line of vanishing Jacobian, the separatrix of the convex hull and the area
where the density exceeds the value $\rho$ (as illustrated in
Fig.~\ref{f:kurto2d} for the 2-D case).

When $\rho\to\infty$, the cumulative probability can be estimated as follows
\begin{eqnarray}
P^>(\rho) \propto \underbrace{\rho^{-3/2}}_{\mbox{\scriptsize from }a_1} \times
\underbrace{\rho^{-1}}_{\mbox{\scriptsize from }a_2} \times \underbrace{1
  \times ... \times 1}_{\mbox{\scriptsize from }a_3...a_d} \times
\underbrace{1}_{\mbox{\scriptsize from time}}.
\label{estimrho}
\end{eqnarray}
Hence, the cumulative probability $P^>(\rho) \propto \rho^{-5/2}$ in any
dimension; so that the pdf of the mass density has a universal
power-law behavior with exponent $-7/2$. 
We have seen that the theory is rather different in one dimension and higher
dimensions, because kurtoparabolic points are persistent only in the latter
case.  However, the scaling law for the resulting pdf is the same in all
dimensions. Actually, two orthogonal spatial directions, $a_1$ and $a_2$ in
(\ref{estimrho}), play the same role as space and time in one
dimension.

It is now clear that, for burgulence, the algebraic tails of the pdf of
velocity gradients or of the density stem from singularities. Turning briefly
to {\em incompressible three-dimensional Navier--Stokes turbulence}, we note
that measurements of pdf's for space or time derivatives of Eulerian
velocities have not yet revealed power-law tails, but such tails may just have
been, so far, ``lost in the experimental noise''.  There has indeed been
considerable speculations about singularities of the Navier--Stokes equations
in the inviscid limit \cite{livreufcup95}. If singularities with divergent
gradients are present, they will give power-law tails, at least as
intermediate asymptotics when the viscosity is small (the converse is however
not true, since statistical effects not related to singularities can also give
power laws). The confirmed absence of power laws would probably rule out
singularities.

\section{Kicked burgulence}
\label{s:kicked}

\subsection{Forced Burgers equation and variational formulation}
\label{subs:varform}

\begin{figure}
\iffigs
\centerline{\includegraphics[height=0.3\hsize]{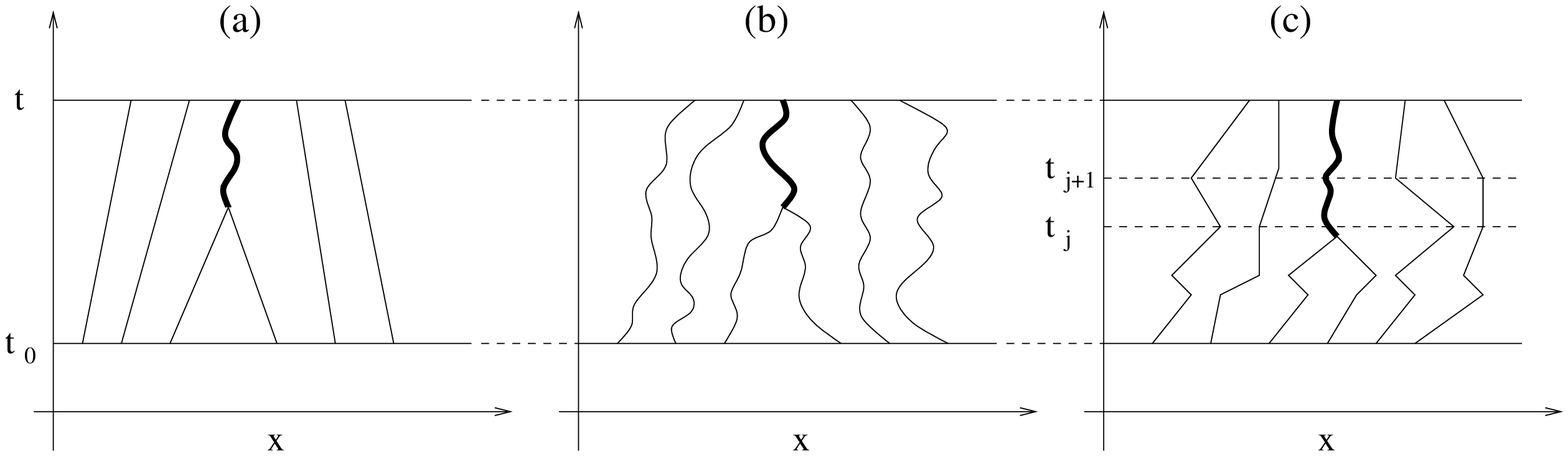}}
\fi
\caption{Trajectories of particles in the decaying case (a), in the
continuously forced
  case (b) and in the kicked case (c). The thick lines are shock trajectories.}
\label{figcarac}
\end{figure}

In the limit of vanishing viscosity and when no force is applied, the Burgers
equation just means that fluid particles keep their initial velocity until
they stick together in a shock. So, until merger, the position $X(t)$ of a
given fluid particle will depend linearly on time:
\begin{equation}
X(t) = X(t_0) + (t-t_0)u_0(X(t_0)), \quad u(X(t),t)=u_0(X(t_0)).
\label{caracdecay}
\end{equation}
When a force is applied, fluid particles trajectories, before merger with a
shock, follow forcing-dependent continuous trajectories governed by
\begin{equation}
{d^2 \over dt^2} X(t) = f(X(t),t), \quad {d \over dt} X(t) = u(X(t),t);
\label{caracforce}
\end{equation}
thus their dynamics can be rather complex (see Fig.~\ref{figcarac}).

Following Ref.~\cite{bfk00}, we shall be concerned here with the
one-dimensional case where the force is a sum of impulses (or kicks),
concentrated at discrete times $t_j$'s:
\begin{equation}
f(x,t)=\sum_{j>j_0} f_j(x) \delta(t-t_j),
\label{defkick}
\end{equation}
where $t_{j_0} = t_0$ is the initial time, and $t_{j_0+1}$ is the time of the
first kick. The kicking times $t_j$'s and the kick $f_j(\cdot)$'s are
prescribed. They can be either fixed or random. The meaning of such a forcing
is that, between kicks, we let the solution evolve as a solution of the
unforced problem. At each kicking time $t_j$, we discontinuously change the
velocity field by the amount $f_j(x)$:
\begin{equation}
u(x,t_{j+}) = u(x,t_{j-})+f_j(x).
\label{ukick}
\end{equation}
This is an intermediate case between decay and time-continuous forcing. Such
forcing implies a piecewise-linear time dependence of the position of
a given fluid particle (see Fig.~\ref{figcarac}).

It is of interest to notice that this kind of discrete-in-time forcing can be
applied also to the Navier--Stokes equations, with features of decaying
turbulence still present to some extent. The original motivation for
introducing such a forcing was to approximate white-noise-in-time forcing by
discrete random noise, also called shot noise.  But actually, the kicked case
displays interesting features of its own.  As will be shown later, the
problem can be understood in terms of area-preserving mappings to which we can
apply KAM theory (see Ref.~\cite{js98} and references therein) and
Aubry--Mather theory \cite{a83,m82}.

We will focus on the space-periodic case. Namely, we assume that both the
initial condition $u_0(\cdot)$ and the kicks $f_j(\cdot)$ are periodic with
period $1$, with respect to the space variable. For the moment, let us also
assume that the initial velocity and the kicks both have zero spatial mean
value over the space period $[0,1[$. Since the mean velocity is conserved by
Burgers dynamics,  we have
\begin{equation}
\int_0^1 u(x,t) dx = 0
\label{aegal0}
\end{equation}
at all times.  This constraint implies that the velocity potential $\psi(x,t)$,
defined by $u(x,t)=-\partial_x \psi(x,t)$, is itself periodic in space. Let us
define the kicking potentials $F_j(\cdot)$, so that
\begin{equation}
f_j(x) = -{d\over dx} F_j(x).
\label{defkickpot}
\end{equation}
It is then easy to write the potential at any time $t$, using between
successive kicks the standard maximum representation \rf{maximum} for decaying
solution in the limit of vanishing viscosity (this is reexpressed here as a
minimum in order to minimize a suitable action function), to obtain
\begin{equation}
\psi(x,t) = -\min_{y_J}\left [ {(x-y_J)^2 \over 2(t-t_J)}
-\psi(y_J,t_{J-})-F_J(y_J) \right ],
\label{hopfcolebetweenkicks}
\end{equation}
where the index $J$ is such that $t_J<t\leq t_{J+1}$. Repeating this 
step as often as necessary to work our way back to the initial time, 
we obtain
\begin{equation}
\psi(x,t) = -\min_{\{y_j\}_{j_0\leq j\leq J}}\left [ A\left(x,t; \{y_j\} 
\right) -  \psi_0(y_{j_0}) \right ],
\label{mini1}
\end{equation}
where $A$ is an action which has to be minimized,
\begin{equation}
A\left(x,t;\{y_j\}\right) = {(x-y_J)^2\over 2(t-t_J)} + \sum_{j=j_0}^{J-1}
\left [ {(y_{j+1}-y_j)^2\over 2(t_{j+1}-t_j)} - F_{j+1}(y_{j+1}) \right ].
\label{defaction}
\end{equation}
There is a similar representation for the case where the forcing  
$f(x,t) = -\partial_x F(x,t)$ is continuously applied, namely 
\begin{equation}
\psi(x,t) = - \min_{y(\cdot)} \left [ A \left( x,t; y(\cdot) \right ) - \psi_0
(y(t_0)) \right ].
\label{minicont}
\end{equation}
The minimum is now taken over continuous curves $y(\cdot)$ such 
that $y(t)=x$, the action being given by
\begin{equation}
A\left(x,t;y(\cdot)\right) = \int_{t_0}^t \left [ {1 \over 2} (\dot y (s))^2
-F(y (s),s) \right ] ds.
\label{defactioncont}
\end{equation}
This representation goes back to work by Oleinik \cite{o57} on general
conservation laws. It can be derived as the continuous limit of the discrete
formulation when letting the time between kicks tend to zero. Many features of
the forced Burgers equation were obtained by E {\it et al.\/} \cite{ekms00}.
As we will see, the key notions introduced by E {\it et al.}, such as
minimizers, global minimizer and main shock, are still valid in the case of
discrete-in-time forcing.

First, we will introduce the notion of minimizing sequence (or minimizer). In
terms of fluid particles trajectories, the minimum representation
(\ref{mini1}) just means that, to obtain the solution at time $t$ and at some
Eulerian location $x$, one has to look at all possible trajectories reaching
$x$, and choose between them those which minimize the action. The sequence for
which the minimum is achieved is, by definition, a minimizer. In general,
there is only one minimizing trajectory arising at a given $x$. But
for a countable set of $x$-values, there are several minimizing
trajectories. These correspond to particles coalescing in a shock.

A minimizer can be explicitly characterized by requiring the vanishing of the
derivatives, with respect to all the  $y_j$'s, of the argument of the minimum
in (\ref{mini1}). A minimizing sequence then has to verify
the following Euler--Lagrange equations:
\begin{eqnarray}
v_{j+1} &=& v_j + f_j(y_j), \label{equationv}\\
y_{j+1} &=& y_j + (t_{j+1} - t_j)\left [v_j + f_j(y_j)\right ],
\label{equationy} 
\end{eqnarray}
where $v_j\equiv (y_j-y_{j-1})/(t_j-t_{j-1})$ is the velocity at the
location $y_j$ just before the kick. These equations have to be supplemented
by the following initial and final conditions:
\begin{eqnarray}
v_{j_0} &=& u_0(y_{j_0}), \label{condinit} \\
x &=& y_J+(t-t_J)v_{J+1}. \label{condfin}
\end{eqnarray}
Note that $v_{J+1}= u(x,t)$. The Euler--Lagrange map is area-preserving. It is
also explicitly invertible, so that for a given $(x,v)$, one can reconstruct
the past history of a particle, except if a shock sits at $x$.

\subsection{Periodic kicks}
\label{subs:periodic}

From now, we will focus on a particular case of forcing which displays
globally the same features as random forcing but is much easier to handle.
Namely, following Ref.~\cite{bfk00}, we consider the case of time-periodic
kicks: the kicking potential is the same at each kick, $F_j(x) = G(x)$ for all
$j$, and the time interval is constant, $t_{j+1}-t_j = 1$, for convenience. The
force can then be written
\begin{equation}
f(x,t) = \sum_{j>j0} g(x) \delta(t-j),
\label{kickper}
\end{equation}
where $g = -dG/dx$.
\begin{figure}[h]
\iffigs
\centerline{\includegraphics[width=8cm]{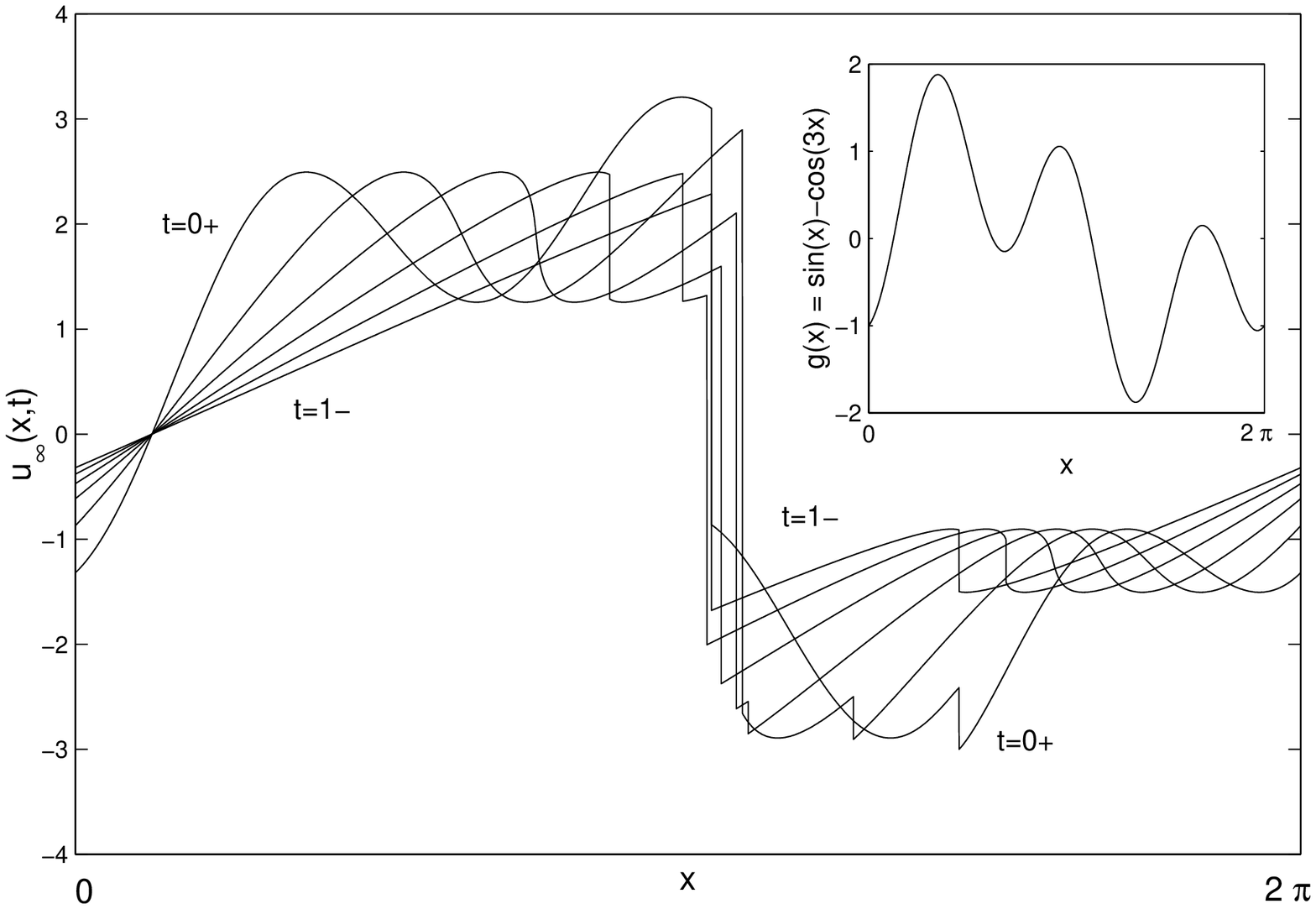}}
\fi
\caption{Snapshots over one time period of the velocity for the limiting
solution corresponding to the kicking force $g(x) = \sin x - \cos(3x)$ on the
space period $[0,2\pi[$ (see upper inset). The main shock is located around
$x=\pi$; the global minimizer, here a fixed point, is the point of vanishing
velocity common to all curves. Notice that during each period,
two new shocks are born and two mergers occur.}
\label{figevol}
\end{figure}
\begin{figure}[h]
\iffigs
\centerline{\includegraphics[width=8cm]{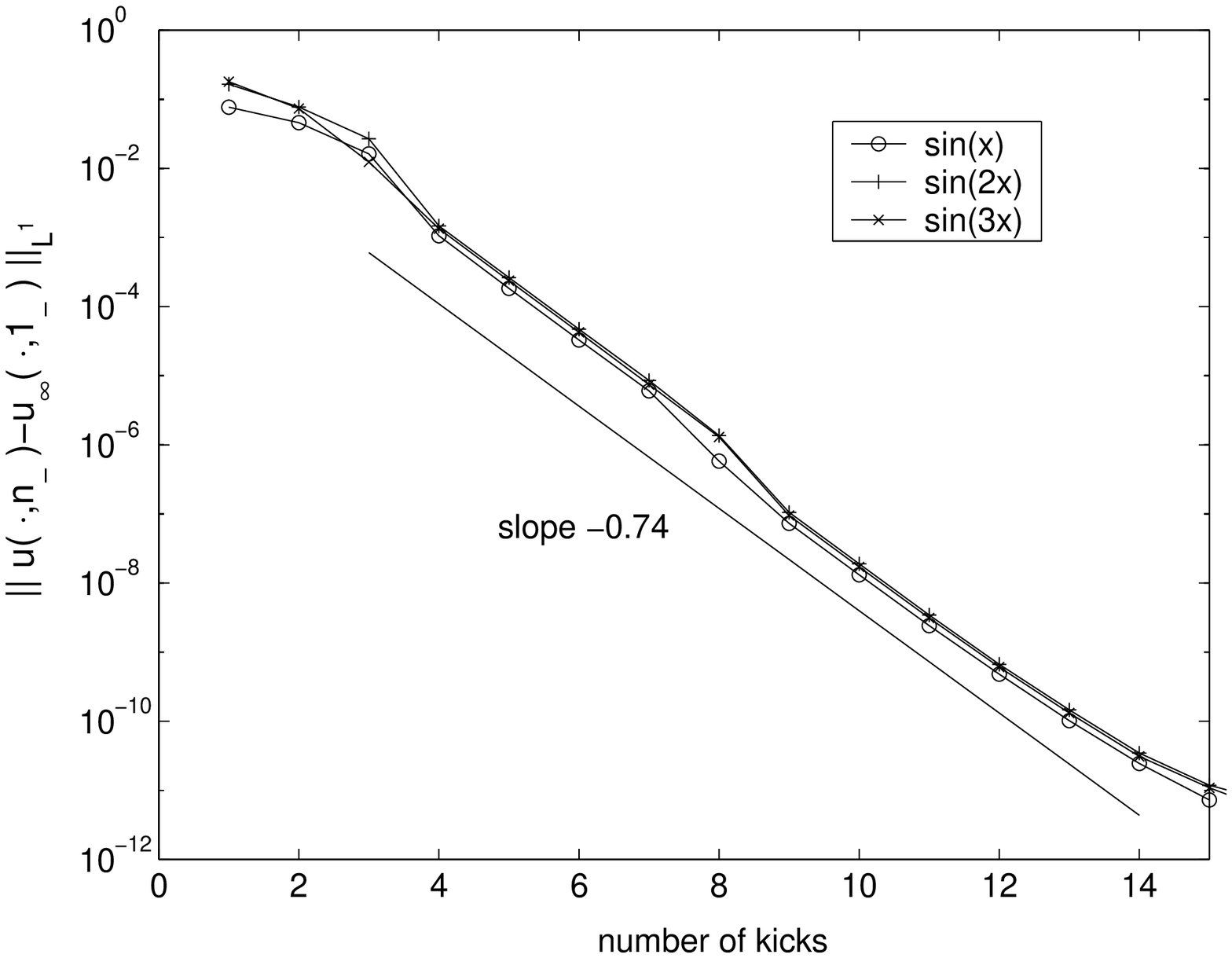}}
\fi
\caption{Exponential relaxation to a time-periodic solution for the same
forcing as in Fig.~\ref{figevol}, with three different
initial conditions, as labeled. $\int_0^{2\pi} |
u(x,n_-) - u_\infty(x,1_-) | dx/(2\pi)$ is plotted vs the number of kicks.}
\label{figrelaxexpo}
\end{figure}

We now show, following Ref.~\cite{bfk00}, that the solution to the Burgers
equation with this kind of forcing converges exponentially fast in time to a
periodic solution $u_\infty(x,t)$. Snapshots of the time-periodic solution for
one instance of kicking are shown in Fig.~\ref{figevol};
Fig.~\ref{figrelaxexpo} shows the exponential relaxation to $u_\infty(x,t)$.
 
\begin{figure}[h]
\iffigs
\centerline{\includegraphics[width=3cm,angle=-90]{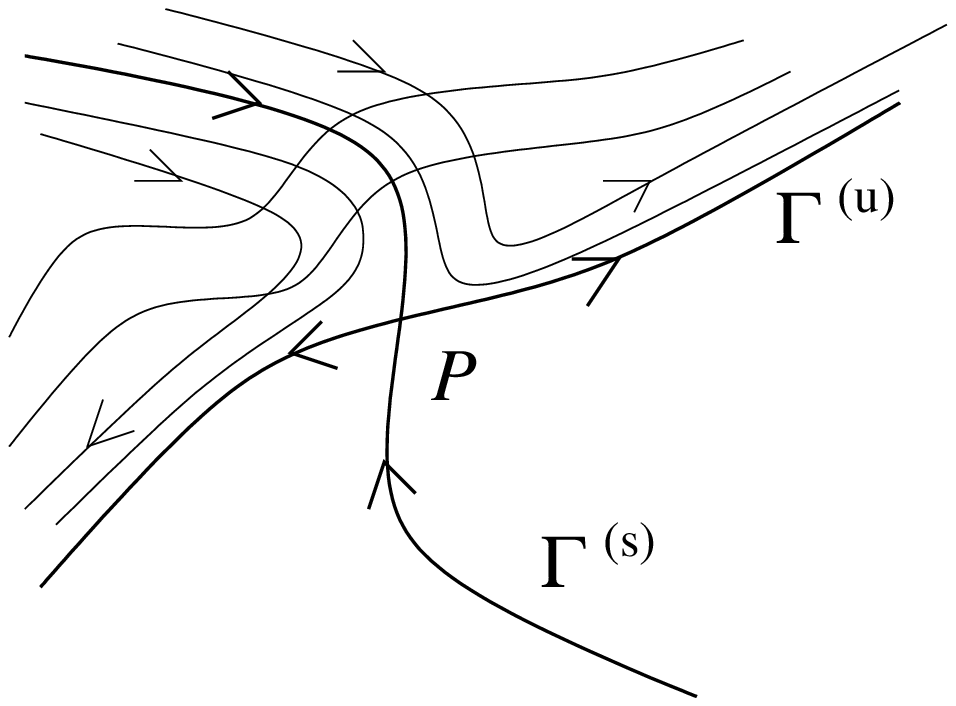}}
\fi
\caption{Sketch of a hyperbolic fixed point $P$ with its stable
  ($\Gamma^{(s)}$) and its unstable ($\Gamma^{(u)}$) manifolds. A
curve, which intersects $\Gamma^{(s)}$, will eventually converge to 
$\Gamma^{(u)}$  under iteration of the map.}
\label{figstabexpo}
\end{figure}

Actually, the convergence to a unique solution at long times is related to
properties near a fixed point of the two-dimensional dynamical system defined
by the Euler--Lagrange map which reads here
\begin{eqnarray}
v_{j+1} &=& v_j + g(y_j), \label{eulagper1} \\
y_{j+1} &=& y_j + v_j + g(y_j). \label{eulagper2}
\end{eqnarray}
A fixed point $(y_\star,v_\star)$, obviously, satisfies $v_\star=0$ and
$g(y_\star) = 0$. The latter expresses that the kicking potential achieves an
extremum at $x= y_\star$. Let $P = (x_c,0)$ be the particular fixed point of
the map (\ref{eulagper1})-(\ref{eulagper2}), which corresponds to the location
where the forcing potential achieves its maximum over the space period.  This
point is hyperbolic because the linearized system in its neighborhood
has two  real eigenvalues $\lambda>1$ and $1/\lambda$, where 
\begin{equation}
\lambda = 1+c+\sqrt{c^2+2c}, \quad c=-{1\over2}{d^2\over
dx^2}G(x_c).
\label{eqlambda}
\end{equation}
In the phase space $(x,v)$, two globally invariant curves are
associated to the corresponding eigendirections. These are (i) the
stable manifold $\Gamma^{(s)}$, associated to $1/\lambda$, which is
the set of points $(x,v)$ converging to the fixed point under
iteration of the map (because the eigenvalue is less than one), and
(ii) the unstable manifold $\Gamma^{(s)}$, associated to $\lambda$,
and generated by inverse iteration (see Fig.~\ref{figstabexpo}).  An
arbitrary continuous curve in the $(x,v)$ plane which intersects the
stable manifold will, under iteration, converge exponentially fast to
the unstable manifold at the rate $1/\lambda$.

\begin{figure}[h]
\iffigs
\centerline{{\bf (a)}\includegraphics[width=4cm]{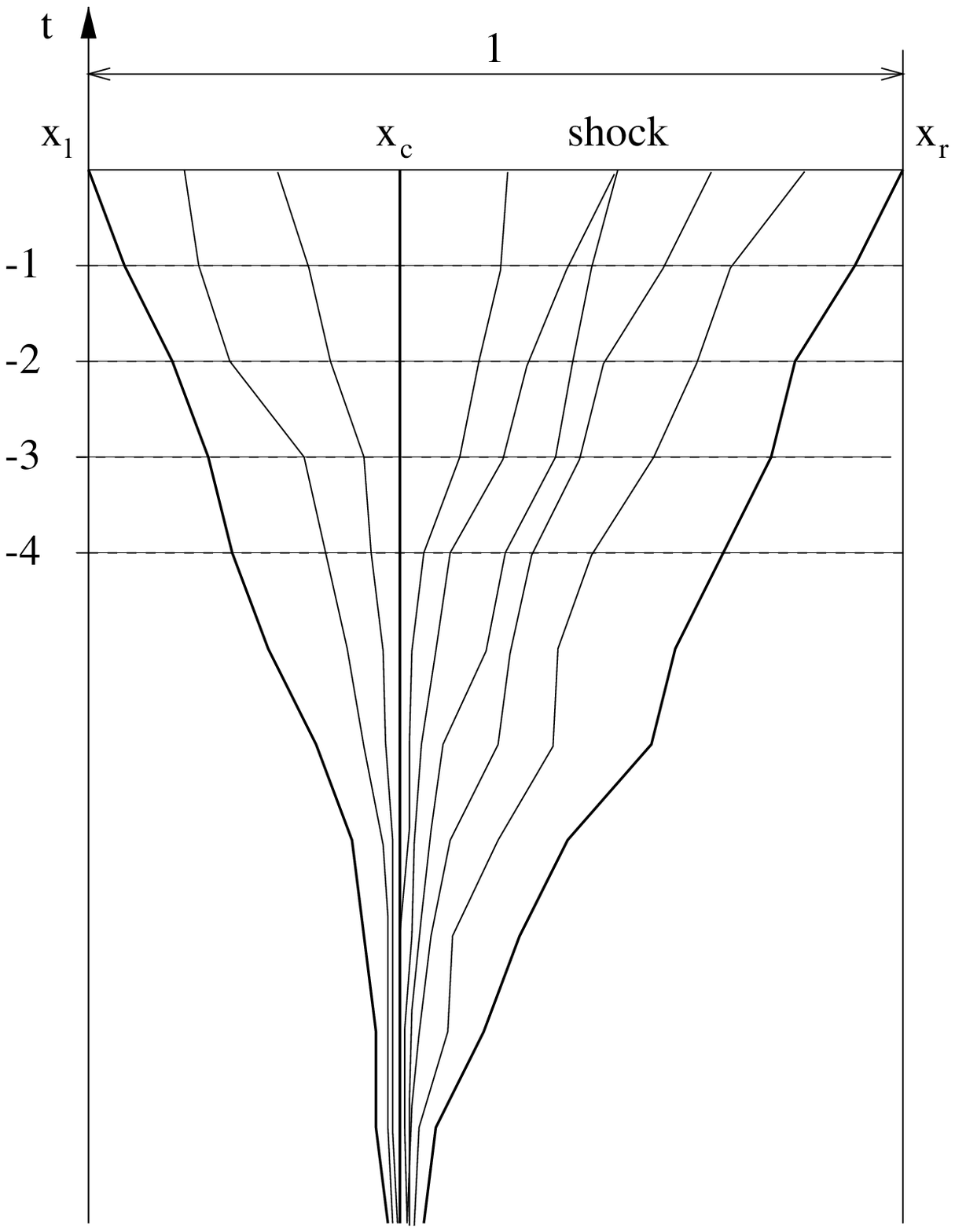}~{\bf
      (b)}\includegraphics[width=6cm]{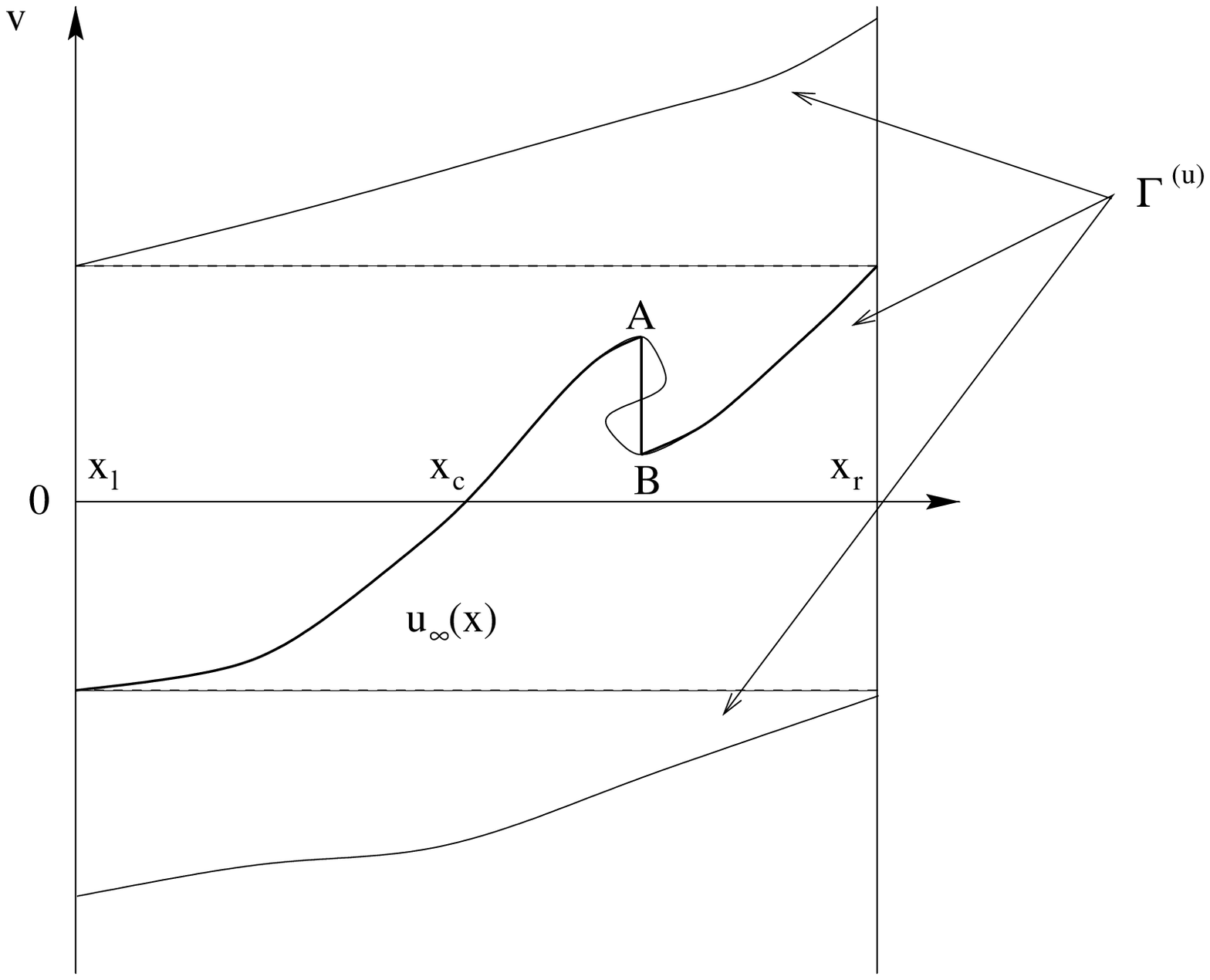}}
\fi
\caption{{\bf (a)} Minimizers on the $(x,t)$ cylinder; initial time $t_0$ is
  taken at $-\infty$. Shock locations are characterized by having two
  minimizers (an instance is at $x_1$); the fat line $x=x_c$ is the global
  minimizer.  {\bf (b)} Unstable manifold $\Gamma^{(u)}$ on the $(x,v)$
  cylinder which passes through the fixed point $P=(x_c,0)$; the bold line is
  the graph of the limiting periodic solution. The main shock is located at
  $x_l=x_r$, and another shock at $x_1$ corresponds to a local zig-zag of
  $\Gamma^{(u)}$ between A and B.}
\label{globalminiandunst}
\end{figure}

In the language of Burgers dynamics, the curve in the $(x,v)$ plane defined by
an initial condition $u_0(x)$ will be mapped after some kicks into a curve
very close to the unstable manifold. To understand this mechanism of
convergence, let us take an initial time $t_0$ tending to $-\infty$ and look
at the behavior of the solution at time $t=0$. The trajectory of the
hyperbolic fixed point $P$ corresponds to the so-called global minimizer. The
global minimizer is the trajectory of a fluid particle never to be absorbed by
a shock. Such a global minimizer is unique, and every minimizing
trajectory converges exponentially fast to the global minimizer as
$t\to-\infty$ \cite{ekms00}. This is illustrated in
Fig.~\ref{globalminiandunst}a.
By definition of the unstable manifold, each point $(y_j,v_j)$ of a minimizer
belongs thus to $\Gamma^{(u)}$ and every regular part of the graph of the
limiting velocity belongs to the unstable manifold. Now we turn to the
construction of the main shock. Since $\lambda>0$, every minimizing trajectory
starting from a point on the right (resp. left) of the global minimizer
approaches it as $t\to-\infty$ from the right (resp. left). Hence, there
exists $x_r$ (resp. $x_l$), the rightmost (resp. leftmost) location from
which a minimizer approaches the global minimizer from the right
(resp. left). By periodicity in space and uniqueness of the global minimizer,
these two points are actually the same: $x_r=x_l \mbox{ mod }1$.  If we shift
the periodicity interval to $[x_l,x_r]$, we can draw $\Gamma^{(u)}$ on the
$(x,v)$-cylinder. The regular parts of the limiting solution belong to this
graph. By construction there is thus a shock at $x=x_l=x_r$ (see
Fig.~\ref{globalminiandunst}b). This is the {\em main shock}, the unique shock
which exists for an infinite time. In Burgers dynamics, shocks are born and
then they may merge.  The main shock is a shock which has always existed when
letting the initial time tend to $-\infty$.  The other shocks are associated to
the regions where $\Gamma^{(u)}$ is multi-valuated in $x$.  Their locations are
determined by requiring that the action be the same at points such as A and B
in Fig.~\ref{globalminiandunst}b.

\subsection{Connections with Aubry--Mather theory}
\label{subs:aubrymather}

So far, we have exclusively considered zero-mean-value initial conditions.
Let us briefly consider the case where 
\begin{equation}
\int_0^1 u(x,t) dx = \int_0^1 u_0(x) dx = a > 0.
\label{nonzeromean}
\end{equation}
The Burgers problem is then in exact correspondence with the description of
equilibrium states of the Frenkel--Kontorova model \cite{kf38}. In the latter,
one has a one-dimensional chain of atoms connected by elastic springs in the
presence of a space-periodic potential.  The potential energy, which must be
minimized to obtain the (classical) ground state, is
\begin{equation}
H(\{y_j\}) = \sum_j {1\over 2} \left (y_{j+1} - y_j -a \right)^2 -G(x),
\label{hamilfrenkel}
\end{equation}
where $a$ is the unstretched distance between atoms. This problem was
investigated by Aubry \cite{a83} and Mather \cite{m82}. The representation
(\ref{hamilfrenkel}) matches the action minimizing representation for Burgers
equation with a mean velocity $a$. The connection between the forced 
Burgers equation and Aubry--Mather theory  was investigated by
Jauslin {\it et al.} \cite{jkm99}, E \cite{e99} and Sobolevski \cite{s99}.

\begin{figure}[h]
\iffigs
\centerline{\includegraphics[width=7cm]{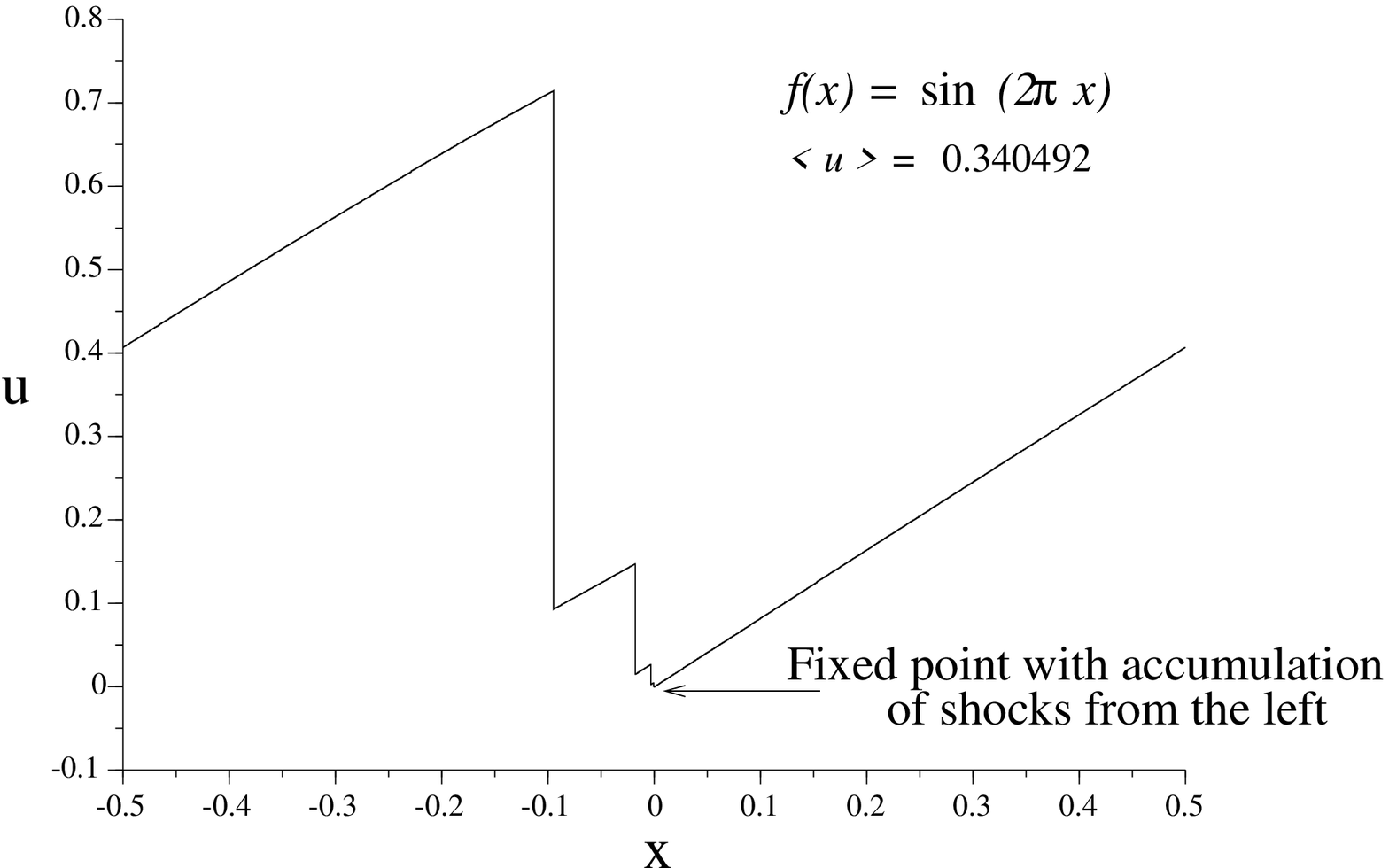}}
\fi
\caption{Velocity profile at the first transition to a rational nonvanishing
rotation number when increasing the mean velocity $a =\la u\ra$ from
0. Note the accumulation of shocks.}
\label{figtrans}
\end{figure}
For $a=0$, the global minimizer is a trivial ground state, associated to a
fixed point, but for $a\neq 0$, it is much more complex. Within some 
intervals of the parameter $a$, the global
minimizer lives on a periodic orbit associated to a rational rotation number
$\rho$ (asymptotic slope of the trajectory when $t\to-\infty$). The graph of
$\rho$ as a function of $a$ is actually a Devil's staircase. The transitions
between the intervals of the mean velocity corresponding to rational rotation
numbers display interesting phenomena, such as accumulations of shocks (see
Fig.~\ref{figtrans}).

\vspace{53mm}
\noindent {\bf Acknowledgments} We are grateful to 
V.~Arnold, M.~Blank, I.A.~Bogaevski, G.~Eyink, J.D.~Fournier, 
W.~E, K.~Khanin, J.~Lukovich, R.~Mohayaee,
Ya.~Sinai, M.R.~Rahimi~Tabar,
E.~Vanden~Eijnden, A.~Sobolevski, M.~Vergassola,  B.~Villone
for useful comments.  Part of this work was done while JB and UF were
visiting the Center of Nonlinear Studies (Los Alamos, USA) and the Isaac
Newton Institute (Cambridge, UK); their support is gratefully
acknowledged. This work was also supported by the European Union under
contract HPRN-CT-2000-00162.


\begin{thebibliography}{99}
\bibitem{h50}
  E.~Hopf, {\it The partial differential equation $u_t +uu_x = u_{xx}$},
  Comm.\ Pure Appl.\ Math.\ {\bf 3} (1950), 201--230. 
\bibitem{c51}
  J.D.~Cole, {\it On a quasi-linear parabolic equation occurring in 
    aerodynamics}, Quart.\ Appl.\ Math.\ {\bf 9} (1951), 225--236.
\bibitem{gms91}
  S.N.~Gurbatov, A.N.~Malakhov and A.I.~Saichev, {\it Non-linear Random
  Waves and Turbulence in Nondispersive Media: Waves, Rays, Particles},
  Manchester University Press, Manchester, 1991. 
\bibitem{z70}
  Y.B.~Zel'dovich, {\it Gravitational instability: an approximate
    theory for large density perturbations}, Astron.\ Astrophys.\ {\bf 5}
  (1970), 84--89.
\bibitem{gs84}
  S.N.~Gurbatov and A.I.~Saichev, {\it Probability distribution and spectra
  of potential hydrodynamic turbulence}, Radiophys.\ Quant.\ Electr.\ {\bf
  27} (1984), 303--313.
\bibitem{saf92}
  Z.~She, E.~Aurell and U.~Frisch, {\it The inviscid Burgers equation
    with initial data of Brownian type}, Comm.\ Math.\ Phys.\
  {\bf  148} (1992), 623--641.
\bibitem{vdfn94}
  M.~Vergassola, B.~Dubrulle, U.~Frisch and A.~Noullez, {\it Burgers'
    equation, Devil's staircases and the mass distribution for large-scale 
    structures}, Astron.\ Astrophys.\ {\bf 289} (1994), 325--356.
\bibitem{p93}
  P.J.E.~Peebles, {\it Principles of Physical Cosmology}, Princeton
  University Press, Princeton, 1993.
\bibitem{cl95}
  P.~Coles and F.~Lucchin, {\it Cosmology: the Origin and Evolution of Cosmic
  Structures}, J.~Wiley and sons, Chichester, 1995.
\bibitem{asz82}
  V.I.~Arnold, S.F.~Shandarin and Y.B.~Zel'dovich, {\it The large-scale
  structure of the Universe I. General Properties. One and two-dimensional
  models}, Geophys.\ Astrophys.\ Fluid Dynam.\ {\bf 20} (1982), 111--130.
\bibitem{kbgn94}
  L.~Kofman, E.~Bertschinger, J.~Gelb and A.~Nusser {\it Evolution of
  one-point distributions from Gaussian initial fluctuations}, Astrophys.\ J.\
  {\bf 420} (1994), 44--57.
\bibitem{css00}
  D.~Chowdhury, L.~Santen and A.~Schadschneider, {\it Statistical physics
    of vehicular traffic and some related systems}, Phys.\ Rep.\ {\bf  329}
  (2000), 199--329.
\bibitem{kpz86}
  M.~Kardar, G.~Parisi and Y.-C.~Zhang, {\it Dynamical scaling of growing
    interfaces}, Phys.\ Rev.\ Lett.\ {\bf 56} (1986), 889-892.
\bibitem{bs95}
  A.L.~Barab\'asi and H.E.~Stanley, {\it Fractal Concepts in Surface
  Growth}, Cambridge University Press, Cambridge, 1995. 
\bibitem{kz87}
 M.~Kardar and Y.-C. Zhang, {\it Scaling of directed polymers in random
 media}, Phys.\ Rev.\ Lett.\ {\bf 58} (1987), 2087--2090.
\bibitem{bmp95}
  J.P.~Bouchaud, M.~M\'ezard and G.~Parisi, {\it Scaling and intermittency in
    Burgers turbulence}, Phys.\ Rev.\ E {\bf 52} (1995), 3656--3674.
\bibitem{o77}
  S.A.~Orszag, {\it Statistical Theory of Turbulence}, in Fluid Dynamics, Les
  Houches 1973, 237--374, eds.~R.~Balian \& J.L.~Peube, Gordon and Breach, 
  New York, 1977.
\bibitem{rs78}
  H.A.~Rose and P.L.~Sulem, {\it Fully developed turbulence and statistical
  mechanics}, J.\ Phys.\ France {\bf 39} (1978), 441--484.
\bibitem{livreufcup95}
  U.~Frisch, {\it Turbulence: The Legacy of A.N.~Kolmogorov}, Cambridge
  University Press, Cambridge, 1995.
\bibitem{l97}
  M.~Lesieur, {\it Turbulence in Fluids}, Fluid Mechanics and Its Applications,
  Vol. 40, Kluwer, 1997.
\bibitem{c75}
  F.~Calogero, {\it Exactly sovable one-dimensional many-body problems},
  Lett.\ Nuovo Cimento {\bf 13}, (1975), 411--416.
\bibitem{fm81}
  U.~Frisch and R.~Morf, {\it Intermittency in nonlinear dynamics and 
  singularities at complex times}, Phys.\ Rev.\ A {\bf 23} (1981), 2673--2705.
\bibitem{ff83}
  J.D.~Fournier and U.~Frisch, {\it L'\'equation de Burgers d\'eterministe
    et statistique}, J.\ M\'ec.\ Th\'eor.\ Appl.\ {\bf 2} (1983), 699--750.
\bibitem{lax}
 P.-D.~Lax,  {\it Hyperbolic systems of conservation laws II},
    Comm.\ Pure Appl.\ Math.\ {\bf 10} (1957), 537--566.
\bibitem{fb70}
  U.~Frisch and R.~Bourret, {\it Parastochastics}, J.\ Math.\ Phys.\ {\bf 11}
  (1970) 364--390. 
\bibitem{ll80}
  L.D.~Landau and E.M.~Lifshitz, {\it Statistical Physics},
  Butterworth-Heinemann, Oxford, 1980.
\bibitem{b99}
  I.A.~Bogaevski, {\it Singularities of convex hulls as fronts of Legendre 
    varieties}, in Geometry and Topology of Caustics -- Caustics '98, Banach
  Center publications {\bf 50} (1999) 61--74, Institute of Mathematics, Polish
  Academy of Sciences (Warsaw).
\bibitem{a00}
  F.~Aicardi, {\it On the classification of generic phenomena in thermodynamic
  binary mixtures},  Physica D (2001), submitted.
\bibitem{nv94}
  A.~Noullez and M.~Vergassola, {\it A fast algorithm for discrete Legendre
  transforms}, J.\ Sci.\ Comput.\ {\bf 9} (1994), 259--281. 
\bibitem{bfk00}
  J.~Bec, U.~Frisch and K.~Khanin, {\it Kicked Burgers turbulence}, 
  J.\ Fluid Mech.\ {\bf 416} (2000), 239--267.
\bibitem{kh38}
  T.~von~K\'arm\'an and L.~Howarth, {\it On the statistical theory of
  isotropic turbulence}, Proc.\ R.\ Soc.\ Lond.~A {\bf 164} (1938), 192--215.
\bibitem{k41decay}
 A.N.~Kolmogorov, {\it On degeneration (decay) of isotropic turbulence in an
  incompressible viscous liquid}, Dokl.\ Akad.\ Nauk SSSR {\bf 31} (1941),
  538--540.
\bibitem{pr54}
  I.~Proudman and W.H.~Reid, {\it On the decay of a normally distributed and
  homogeneous turbulent velocity field}, Phil.\ Trans.\ R.\ Soc.\ Lond.~A {\bf
  247} (1954), 163--189.
\bibitem{burgatov}
  S.N.~Gurbatov, S.I.~Simdyankin, E.~Aurell, U.~Frisch and G.~Toth, {\it On
    the decay of Burgers turbulence}, J.\ Fluid Mech.\ {\bf 344} (1997),
  339--374.
\bibitem{livreburgers}
  J.M.~Burgers, {\it The Nonlinear Diffusion Equation}, D.~Reidel, Dordrecht
  1974.
\bibitem{k79}
  S.~Kida, {\it Asymptotic properties of Burgers turbulence}, J.\ Fluid
  Mech.\ {\bf 93} (1979), 337--377.
\bibitem{msw95}
 S.A.~Molchanov, D.~Surgailis and  W.A.~Woyczynski, {\it Hyperbolic asymptotics
 in Burgers' turbulence and extremal processes}, Comm.\ Math.\ Phys.\ 
 {\bf 168} (1995), 209--226.
\bibitem{ex00}
 G.L.~Eyink and J.~Xin, {\it Self-similar decay in the Kraichnan 
 model of a passive scalar}, J.\ Stat.\ Phys.\ {\bf 100} (2000), 679--741.
\bibitem{et00}
 G.L.~Eyink and D.J.~Thomson, {\it Free decay of turbulence and breakdown of
 self-similarity}, Phys.\ Fluids {\bf 12} (2000), 477--479.
\bibitem{ol00}
 S.~Ossia and M.~Lesieur, {\it Energy backscatter in LES of 3D incompressible
 isotropic turbulence}, J.\ Turbulence~{\bf 1} (2000), 007.
\bibitem{cefv00}
 M.~Chaves, G.~Eyink, U.~Frisch and M.~Vergassola, {\it Universal decay of
 scalar turbulence}, Phys.\ Rev.\ Lett.\~{\bf 86} (2000), 2305--2308.
\bibitem{s92}
 Ya.~Sinai, {\it Statistics of shocks in solutions of inviscid Burgers 
   equation}, Comm.\ Math.\ Phys.\ {\bf 148} (1992), 601--622.
\bibitem{fellervol2}
  W.~Feller, {\it An Introduction to Probability Theory and its Applications},
  vol.~2, J.~Wiley and sons, Chichester, 1995.
\bibitem{ekms97} 
  W.~E, K.~Khanin, A.~Mazel and Ya.~Sinai,
  {\it Probability distribution functions for the random forced Burgers
    equation}, Phys.\ Rev.\ Lett.\ {\bf 78} (1997), 1904--1907.
\bibitem{cy95}
 A.~Chekhlov and V.~Yakhot, {\it Kolmogorov turbulence in a
random-force-driven Burgers equation: anomalous scaling and
probability density functions}, Phys.~Rev.~E {\bf 52} (1995), 5681--5684.
\bibitem{p95}
  A.M.~Polyakov, {\it Turbulence without pressure}, Phys.\ Rev.\ E {\bf 52}
  (1995), 6183--6188.
\bibitem{b97}
  S.A.~Boldyrev, {\it Velocity-difference probability density functions for
    Burgers turbulence}, Phys.\ Rev.\ E {\bf 55} (1997), 6907--6910.
\bibitem{gk98}
  T.~Gotoh and R.H.~Kraichnan, {\it Steady-state Burgers turbulence with
    large-scale forcing}, Phys.\ Fluids {\bf 10} (1998), 2859--2866.
\bibitem{k99}
  R.H.~Kraichnan, {\it Note on forced Burgers turbulence}, Phys.\ Fluids {\bf
  11} (1999), 3738--3742.
\bibitem{eve99}
  W.~E and E.~Vanden~Eijnden, {\it Asymptotic theory for the probability
    density functions in Burgers turbulence}, Phys.\ Rev.\ Lett.\ {\bf 83}
  (1999), 2572--2575. 
\bibitem{eve00}
  W.~E and E.~Vanden~Eijnden, {\it Statistical theory for the stochastic
    Burgers equation in the inviscid limit}, Comm.\ Pure Appl.\ Math.\ {\bf
    53} (2000), 852--901.
\bibitem{bf00}
  J.~Bec and U.~Frisch, {\it Pdf's of derivatives and increments for decaying
    Burgers turbulence}, Phys.\ Rev.\ E {\bf 61} (2000), 1395--1402.
\bibitem{fbv00}
  U.~Frisch, J.~Bec and B.~Villone, {\it Singularities and the distribution
  of density in the Burgers/adhesion model}, Physica D (2000), in press,
  cond-mat/9912110.
\bibitem{js98}
  J.V.~Jos\'e and E.J.~Saletan, {\it Classical Dynamics: A Contemporary
  Approach}, Cambridge University Press, Cambridge, 1998. 
\bibitem{a83}
  S.~Aubry, {\it The twist map, the extended Frenkel--Kontorova model and the
    devil's staircase}, Physica D {\bf 7} (1983), 240--258.
\bibitem{m82}
  J.N.~Mather, {\it Existence of quasi-periodic orbits for twist
    homeomorphisms of the annulus}, Topology {\bf 21} (1982), 457--467. 
\bibitem{o57}
  O.~Oleinik, {\it Discontinuous solutions of nonlinear differential
    equations}, Uspekhi Mat.\ Nauk {\bf 12} (1957), no~3, 3--73. (Russ.\
  Math.\ Survey, Amer.\ Math.\ Transl.\ Series 2 {\bf 26}, 95--172.)
\bibitem{ekms00} 
  W.~E, K.~Khanin, A.~Mazel and Ya.~Sinai,
  {\it Invariant measures for Burgers equation with stochastic forcing},
  Ann.\ Math.\ {\bf 151} (2000), 877--960.
\bibitem{kf38}
  T.~Kontorova and Y.I.~Frenkel, Zh. Eskp. \&
  Teor. Fiz. {\bf 8} (1938), 1340. 
\bibitem{jkm99}
  H.R.~Jauslin, H.O.~Kreiss and J.~Moser, {\it On the forced Burgers
    equation with periodic boundary conditions}, Proc.\ Symp.\ Pure Math. {\bf
    65} (1999), 133--153.
\bibitem{e99}
  W.~E, {\it Aubry--Mather theory and periodic solutions for the forced
  Burgers equation}, Comm.\ Pure Appl.\ Math.\ {\bf 52} (1999), 811--828.
\bibitem{s99}
  A.N.~Sobolevski, {\it Periodic solutions of the Hamilton--Jacobi equation
    with a periodic non-homogeneous term and Aubry--Mather theory}, Sbornik
  Math.\ {\bf 190} (1999), 1487--1504.

\end{thebibliography}
\end{document}